\documentclass[english,12pt]{article}
\usepackage[T1]{fontenc}
\usepackage[latin1]{inputenc}
\usepackage{amsmath}
\usepackage{babel}%
\usepackage{amsfonts}%
\usepackage{amssymb}

\renewcommand{\mathbf}[1]{\mbox{\boldmath$#1$}}


\begin{document}

\vspace{1cm}

\title{{\Large Generalized oscillator representations \\
for generalized Calogero Hamiltonians}}

\author{I.V. Tyutin\thanks{e-mail: tyutin@lpi.ru} and
B.L. Voronov\footnote{e-mail:voronov@lpi.ru}}
\date{}
\maketitle

\begin{center}
\textit{P.N. Lebedev Physical Institute of the Russian
Academy of Science,\\
Leninsky prospect 53, Moscow, 119991, Russia}
\end{center}

\begin{abstract}
This paper is a natural continuation of the previous paper \cite{TyuVo13}
where generalized oscillator representations for Calogero Hamiltonians with
potential $V(x)=\alpha/x^2$, $\alpha\geq-1/4$, were constructed. In this
paper, we present generalized oscillator representations for all
generalized Calogero Hamiltonians with potential $V(x)=g_{1}/x^2+g_{2}x^2$,
$g_{1}\geq-1/4$, $g_{2}>0$. These representations are generally highly
nonunique, but there exists an optimum representation for each Hamiltonian,
representation that explicitly determines the ground state and the
ground-state energy. For generalized Calogero Hamiltonians with coupling
constants $g_1<-1/4$ or $g_2<0$, generalized oscillator representations do
not exist in agreement with the fact that the respective Hamiltonians are
not bounded from below.
\end{abstract}

\vspace{0.5cm}


\vspace{-13cm}
\begin{flushright}
FIAN-TD-2013-17 \hspace{1.2cm}{}~\\
\end{flushright}

\vspace{13cm}

\bigskip

\section{Introduction. Formulation of problem}

Let $\check{H}$ be the self-adjoint (s.a. in what follows) generalized
Calogero differential operation on the positive real semiaxis,
\begin{equation}
\check{H}=-d_x^2+g_1x^{-2}+g_2x^2,\,d_x=\frac{d}{dx},\;
x\in\mathbb{R}_{+},\,\label{1.1}%
\end{equation}
the arbitrary real parameters $g_{1}$ and $g_{2}$ are called the coupling
constants, $g_{1}$ is dimensionless and $g_{2}$ is of dimension of the
fourth degree of inverse length. By definition, the generalized Calogero
Hamiltonians\footnote{A remark about our terminology. In the paper, we use
the term ``generalized'' in different senses, a specific sense is clear from
a context.

The term ``generalized'' in this context is conventionally used in the case of
$g_{2}\neq0$; in the case of $g_{2}=0$, this term is omitted, and we
conventionally speak about the ``Calogero differential operation'' and
``Calogero Hamiltonians'' respectively, the coupling constant $g_{1}$ is then
usually denoted by $\alpha$.} commonly symbolized by\footnote{For specific
Hamiltonians, the index $\mathfrak{e}$ is replaced by more informative
indices.} $\hat{H}_{\mathfrak{e}}$ are s.a. differential operators in
$L^{2}(\mathbb{R}_{+})$ associated with this differential
operation\footnote{By definition, a differential operator $\hat{f}$ is called
associated with a differential operation $\check{f}=\check{f}(x,d_{x})$ if
the operator $\hat{f}$ acts on its domain $D_{f\text{ }}$ by $\check{f}$:
$\hat{f}\psi(x)=\check{f}\psi(x),\forall\psi(x)\in D_f$.}. All generalized
Calogero Hamiltonians with arbitrary coupling constants were constructed in
\cite{GitTyV12}, sec. 8.4, and their spectra and (generalized)
eigenfunctions\footnote{The term ``generalized'' in this context means
distributions.} were evaluated including inversion formulas.

By construction, each $\hat{H}_{\mathfrak{e}}$ with fixed coupling constants
$g_{1}$ and $g_{2}$ is a s.a. extension of the initial symmetric operator
$\hat{H}$ associated with $\check{H}$ and defined on the subspace
$\mathcal{D}(\mathbb{R}_{+})$ of smooth functions with a compact support in
$(0,\infty)$. If the coupling constant $g_{1}\geq3/4$, the extension is
unique, so that for each $g_{1}\geq3/4$ and any fixed $g_{2}$,
$-\infty<g_{2}$ $<\infty$, there is a unique generalized Calogero
Hamiltonian $\hat{H}_1$ (in notation of \cite{GitTyV12}). If the coupling
constant $g_{1}<3/4$, the extension is defined nonuniquely, so that for
each $g_{1}<3/4$ and any fixed $g_{2}$, there is a one-parameter family
$\{\hat{H}_{\nu},\nu\in\lbrack-\pi/2,\pi/2],\,-\pi/2\sim\pi/2\}$ of
generalized Calogero\ Hamiltonians\footnote{For brevity, we here use a
single uniform index $\nu$ for labelling generalized Calogero Hamiltonians
with $g_{1}<3/4$, this notation is a condensed one in comparison with
\cite{GitTyV12} where more specific double indices distinguishing different
regions of $g_{1}$ were used. We introduce a more detailed indexing of
generalized Calogero Hamiltonians with $g_{1}<3/4$ in sec. 3 below as
needed.} which are specified by asymptotic s.a. boundary conditions at the
origin. If $g_{1}\geq-1/4$ and $g_{2}\geq0$, each $\hat{H}_{\mathfrak{e}}$
is bounded from below. From the general standpoint, this is a consequence of
that the initial symmetric operator $\hat{H}$ with $g_{1}\geq-1/4$,
$g_{2}\geq0$ is nonnegative and therefore, all its s.a. extensions are
bounded from below \cite{AkhGl81, Naima69}.If $g_{1}<-1/4$, each
$\hat{H}_{\nu}$ is unbounded from below (which is known as ``a fall to the
center'' in the case of $g_{2}\geq0$); if $g_{2}<0$, all
$\hat{H}_{\mathfrak{e}}$ are unbounded from below (``a fall to infinity'').

In this paper, our interest is with the possibility of
representing generalized Calogero Hamiltonians in the generalized oscillator form%

\begin{equation}
\hat{H}_{\mathfrak{e}}=\hat{c}^{+}\hat{c}-u\,\hat{I},\label{1.2}%
\end{equation}
where $\hat{c}$ and $\hat{c}^{+}$ is a pair of closed mutually adjoint
first-order differential operators,$\ \hat{c}^{+}=(\hat{c})^{+},\,\hat{c}$
$=\overline{\hat{c}}$$=(\hat{c}^{+})^{+}$,$\,\hat{I}$ is the identity
operator and $u$ is a real parameter, $\operatorname{Im}u=0$, of dimension
of inverse length squared. This problem was comprehensively considered for
Calogero Hamiltonians, $g_{2}=0$, in \cite{TyuVo13}. In what follows, we
therefore consider the case of $g_{2}\neq0$, the generalized Calogero
Hamiltonians proper.

The parameter $u$ in representation (\ref{1.2}) is not unique and is not
arbitrary. Each Hamiltonian $\hat{H}_{\mathfrak{e}}$ allowing representation
(\ref{1.2}) allows its own region of admissible values of the
parameter $u$ determined by the spectrum of the Hamiltonian, we make this
region more precise just below. Accordingly, the symbols $\hat{c},\hat{c}^{+}
$ of the operator pairs in (\ref{1.2}) implicitly contain $u$ as an argument.
Moreover, generalized oscillator representations with given $u$ for a given
$\hat{H}_{\mathfrak{e}}$ may allow different mutually adjoint pairs
$\hat{c},\hat{c}^{+}$ parametrized by an additional parameter, which then
appears as an additional argument in the symbols $\hat{c},\hat{c}^{+}$.

The representation (\ref{1.2}) is equivalent to the representation%

\begin{equation}
\hat{H}_{\mathfrak{e}}=\hat{d}\,\hat{d}^{+}-u\,\hat{I},\label{1.3}%
\end{equation}
where $\hat{d}$ and $\hat{d}^{+}$ is a pair of closed mutually adjoint
first-order differential operators, it is sufficient to make the
identifications $\hat{c}=\hat{d}^{+},\,\hat{c}^{+}=\hat{d}$.

The problem we are interested in really concerns the cases of $g_1\geq-1/4$
together with $g_{2}>0$ because representations (\ref{1.2}), or
(\ref{1.3}), for a Hamiltonian $\hat{H}_{\mathfrak{e}}$ imply that
$\hat{H}_{\mathfrak{e}}$ is bounded from below, such that its spectrum is
bounded from below by $-u$ (for brevity, we will say that Hamiltonians
$\hat{H}_{\mathfrak{e}}$ (\ref{1.2}), or (\ref{1.3}), are bounded from below
by $-u$). More specifically, if $E_{0}=E_{0}(g_{1},g_{2},\nu)$ is the lower
boundary of the spectrum of the Hamiltonian $\hat{H}_{\mathfrak{e}}$ with
given coupling constants $g_{1},g_{2}$ and extension parameter $\nu$
(if $g_{1}<3/4$), then $E_{0}\geq-u$, so that the region of admissible $u$ for
a given $\hat{H}_{\mathfrak{e}}$ is determined by the condition $u\geq-E_{0}$.
And if the kernel of the operator $\hat{c}$ in representation (\ref{1.2}), or
the operator $\hat{d}^{+}$ in representation (\ref{1.3}), with $u=-E_{0}$ is
nontrivial, \textrm{ker}\,$\hat{c}\neq\{0\}$, or \textrm{ker}\,$\hat{d}^{+}%
\neq\{0\}$, then \textrm{ker}\,$\hat{c}$, or \textrm{ker}\,$\hat{d}^{+}$, is
the ground space (ground state) of the Hamiltonian and $E_{0}$ is its
ground-state energy. Conversely, if a given Hamiltonian
$\hat{H}_{\mathfrak{e}}$ allows representation (\ref{1.2}), or
representation (\ref{1.3}), with a certain $u$ and
\textrm{ker}\,$\hat{c}\neq\{0\}$, or \textrm{ker}$\,\hat{d}^{+}\neq\{0\}$,
then \textrm{ker}\,$\hat{c}$, or \textrm{ker}\,$\hat{d}^{+}$, is the ground
space (ground state) of the Hamiltonian and $u=-E_{0}$, the ground-state
energy with opposite sign. In such a case, we will say that the
representation (\ref{1.2}), or (\ref{1.3}), is an optimum one. We recall
that all the generalized Calogero Hamiltonians with $g_{2}>0$ have a
discrete spectrum, so that we expect that
if a generalized Calogero Hamiltonian with given coupling constants $g_{1}%
\geq-1/4$, $g_{2}>0\,$allows generalized oscillator representations, then
there exists an optimum representation.

It follows from the aforesaid that when examining the possibility of
generalized oscillator representations (\ref{1.2}) or (\ref{1.3}) for
generalized Calogero Hamiltonians with given coupling constants $g_{1}%
\geq-1/4$, $g_{2}>0$, we can predetermine the range of the parameter $u$ by
the condition%
\begin{equation*}
u\geq u_{0}=u_{0}(g_{1},g_{2}),
\end{equation*}
where $u_{0}$ is a quantity opposite in sign to the lower-boundary value
$E_{0}(g_{1},g_{2})$ of the spectrum of a unique Hamiltonian $\hat{H}_1$
with given coupling constants $g_{1}\geq3/4,g_{2}>0$
or to the maximum in $\nu$ of the lower boundaries $E_{0}%
(g_{1},g_{2},\nu)\,$of the spectrum of all the Hamiltonians $\hat{H}_{\nu}$
with given coupling constants $g_{1}\in\lbrack-1/4,3/4),g_{2}>0$,%
\begin{equation*}
u_{0}=u_{0}(g_{1},g_{2})=\left\{
\begin{array}
[c]{l}%
-E_{0}(g_{1},g_{2}),\,g_{1}\geq3/4,\mathrm{\ }g_{2}>0\mathrm{\ }\\
-\max_{\nu}\{E_{0}(g_{1},g_{2},\nu)\},\,g_{1}\in\lbrack-1/4,3/4),\mathrm{\ }%
g_{2}>0
\end{array} \right.  .
\end{equation*}
These boundaries are written down in full in \cite{GitTyV12}, according to
which we have%
\begin{equation}
u_{0}=u_{0}(g_{1},g_{2})=-2\sqrt{g_{2}}(1+\sqrt{1/4+g_{1}}).\label{1.5}%
\end{equation}

The range of admissible $u$ in the generalized oscillator representation for a
specific generalized Calogero Hamiltonian $\hat{H}_{\mathfrak{e}}$ belongs to
the semiaxis $[u_{0},\infty)$\textrm{. }Because the generalized oscillator
representations for a given $\hat{H}_{\mathfrak{e}}$ is generally highly
nonunique, the question arises about an optimum representation.

As to generalized oscillator representations for generalized Calogero
Hamiltonians with coupling constants $g_{1}<-1/4$ or $g_{2}<0$, there are no
such representations and can not be because these Hamiltonians are not bounded
from below.

In solving the problem of generalized oscillator representations for
generalized Calogero Hamiltonians proper with $g_{1}\geq-1/4$
and$\mathrm{\ }g_{2}>0$, we follow ideas and methods in \cite{TyuVo13} where a
similar problem was completely solved in positive for Calogero Hamiltonians,
$g_{2}=0$. We outline them here to give an insight into a content and a
structure of the paper.

The basic idea is independently constructing generalized Calogero Hamiltonians
as s.a. extensions of the initial symmetric operator $\hat{H}$ directly in the
generalized oscillator forms (\ref{1.2}) or (\ref{1.3}) and then comparing
these constructions with the known Hamiltonians in \cite{GitTyV12}.

A starting point is constructing a generalized oscillator representation for
the respective differential operation $\check{H}$ that is a representation of
the form%
\begin{equation}
\check{H}=\check{b}\check{a}-u,\;\,\check{a}=\ \check{b}^{\ast},\,\check
{b}=\check{a}^{\ast},\text{ }\operatorname{Im}u=0,\label{1.6}%
\end{equation}
where $\check{a}$ and $\check{b}$ are finite-order differential operations
mutually adjoint by Lagrange (the superscript $\ast$ denotes the Lagrange
adjoint), see \cite{AkhGl81}, $u$ is a real numerical parameter.

In principle, the parameter $u$ is a variable parameter restricted by the
condition $\ u\geq u_{0}=u_{0}(g_{1},g_{2})$ (\ref{1.5}),
so that we actually deal with a family $\{\check{a}(u),$ $\check{b}(u)\}$ of
differential operations mutually adjoint by Lagrange. Moreover, a given
$\check{H}$ may allow a family of different pairs $\check{a},\check{b}$
parametrized by an additional parameter, let it be\ $\mu$, in representation
(\ref{1.6}) with given $u$. In such a situation we would actually have a
two-parameter family $\{\check{a}(\mu,u),$ $\check{b}(\mu,u)\}$ of different
mutually adjoint pairs $\check{a},\check{b}$ providing the desired
representation (\ref{1.6}) for a given $\check{H}$.

We note that in the case of Calogero Hamiltonians, the coupling constant
$g_{2}=0$, the lower boundary $u_{0}$ of admissible values of the parameter $u$
in representation (\ref{1.2}), or (\ref{1.3}), is zero\footnote{That is why
the parameter $u$ was denoted in \cite{TyuVo13} as $u^{2}=(sk_{0})^{2}$, the
variable dimensionless parameter $s\geq0$, the fixed parameter $k_{0}>0$ is of
dimension of inverse length.}, $u\geq u_{0}=0$. The same holds for the
parameter $u$ in representation (\ref{1.6}) for $\check{H}$ with $g_{2}=0$. A
specific feature of generalized Calogero Hamiltonians with coupling constant
$g_{2}>0$ is that the lower boundary of admissible values of the parameter $u$
is $u_{0}<0$ and $|u_{0}|$ indefinitely grows with $g_{2}$. The same holds for
the parameter $u$ in representation (\ref{1.6}) for $\check{H}$ with $g_{2}>0$.

Let differential operation $\check{H}$ (\ref{1.1}) with given coupling
constants $g_{1}$, $g_{2}\in\mathbb{R}^{2}$ allow generalized oscillator
representation (\ref{1.6}) with some $\check{a}=\check{a}(u)$ and $\check{b}=
$ $\check{b}(u)$,
\begin{equation}
\check{H}=\check{b}(u)\check{a}(u)-u,\;\,\check{a}(u)=\ \check{b}^{\ast
}(u),\,\check{b}(u)=\check{a}^{\ast}(u),\text{ }\label{1.6a}%
\end{equation}
such that $\check{a}(u):$ $\mathcal{D}(\mathbb{R}_{+})\rightarrow
\mathcal{D}(\mathbb{R}_{+})$, $\check{b}(u):$ $\mathcal{D}(\mathbb{R}%
_{+})\rightarrow\mathcal{D}(\mathbb{R}_{+})$; the pair $\check{a}(u)$,
$\check{b}(u)$ of mutually adjoint by Lagrange differential operations may
be defined nonuniquely for given $u\in\mathbb{R}$. If we introduce a pair of
initial differential operators $\hat{a}(u)$ and $\hat{b}(u)$ in
$L^{2}(\mathbb{R}_{+} )$\textrm{\thinspace}defined on
$\mathcal{D}(\mathbb{R}_{+})$ and associated with the respective
differential operations $\check{a}(u)$ and $\check{b}(u)$,
then the initial symmetric operator $\hat{H}$ is evidently represented as%

\begin{equation*}
\hat{H}=\hat{b}(u)\,\hat{a}(u)-u\hat{I}.
\end{equation*}
Let $\hat{c}(u)$ and $\hat{c}^{+}(u)$ be a pair of closed mutually adjoint
operators that are closed extentions of the respective initial operators
$\hat{a}(u)$ and $\hat{b}(u)$, $\hat{a}(u)\subset\hat{c}(u)$, $\hat
{b}(u)\subset\hat{c}^{+}(u)$, the first candidate for $\hat{c}(u)$ is the
closure of $\hat{a}(u)$, $\hat{c}(u)=$ $\overline{\hat{a}}(u)$, then
$\hat{c}^{+}(u)=$ $\hat{a}^{+}(u)$. The operator
\begin{equation}
\hat{H}_{\mathfrak{e},\,c}=\hat{c}^{+}(u)\hat{c}(u)-u\hat{I}\label{1.8}%
\end{equation}
is an evident extension of $\hat{H}$, $\hat{H}\subset
\hat{H}_{\mathfrak{e},c}$. By the von Neumann theorem \cite{Neuma50} (for a
proof, see also \cite{AkhGl81}), operator $\hat{H}_{\mathfrak{e},c}$
(\ref{1.8}) is s.a., which means that $\hat{H}_{\mathfrak{e},c}$ is a
certain generalized Calogero Hamiltonian represented in the generalized
oscillator form (\ref{1.2}) and bounded from below by $-u$, and if the
kernel of the corresponding operator $\hat{c}(u)$ is nontrivial,
\textrm{ker}\,$\hat{c}(u)\neq \{0\}$, then \textrm{ker}\,$\hat{c}(u)$ is the
ground space (ground state) of $\hat{H}_{\mathfrak{e},c}$ and $E_{0\text{
}}=-u\,$is its least eigenvalue (ground-state energy).

Similarly, let $\hat{d}(u)$ and $\hat{d}^{+}(u)$ be a pair of closed mutually
adjoint operators that are closed extensions of the respective initial
operators $\hat{b}(u)$ and $\hat{a}(u)$, $\hat{b}(u)\subset\hat{d}(u)$,
$\hat{a}(u)\subset$ $\hat{d}^{+}(u)$, the first candidate for $\hat{d}(u)$ is
the closure of $\hat{b}(u)$, $\hat{d}(u)=\overline{\hat{b}}(u)$, then $\hat
{d}^{+}(u)=\hat{b}^{+}(u)$. The operator
\begin{equation}
\hat{H}_{\mathfrak{e},d}=\hat{d}(u)\,\hat{d}^{+}(u)-u\hat{I}\label{1.9}%
\end{equation}
is a certain generalized Calogero Hamiltonian represented in the generalized
oscillator form (\ref{1.3}) and having the properties similar to those of the
operator $\hat{H}_{\mathfrak{e},c}$ (\ref{1.8}) with $\hat{d}^{+}(u)$ in place
of $\hat{c}(u)$. Constructing a pair $\hat{c}(u)\supset$ $\hat{a}(u)$,
$\hat{c}^{+}(u)\supset$ $\hat{b}(u)$ or a pair $\hat{d}(u)\supset$ $\hat
{b}(u)$, $\,\hat{d}^{+}(u)\supset$ $\hat{a}(u)$ is a matter of convenience: we
can start with extending $\hat{a}(u)$ to its closure or with extending
$\hat{b}(u)$ to its closure thus obtaining generally different s.a. extensions
of $\hat{H}$.

Varying the parameter $u$ in (\ref{1.6a}) in the admissible region, finding
total families of admissible differential operation pairs $\check{a}%
(u),\check{b}(u)$ in this representation and involving all possible
mutually-adjoint extensions of the initial differential operators
$\hat{a}(u) $ and $\hat{b}(u)$, we can hope to construct generalized
oscillator representations (\ref{1.2}) or (\ref{1.3}) for all generalized
Calogero Hamiltonians with $g_{1}\geq-1/4$ and $g_{2}>0$ in the form
(\ref{1.8}) or (\ref{1.9}). We show below that this hope is justified. An
identification of the Hamiltonians $\hat{H}_{\mathfrak{e},c}$ and
$\hat{H}_{\mathfrak{e},d}$ with the known generalized Calogero Hamiltonians
$\hat{H}_{\mathfrak{e}}$ in \cite{GitTyV12} is straightforward for
$g_{1}\geq3/4$ because the Hamiltonian $\hat{H}_{\mathfrak{e}}=\hat{H}_1$
with given $g_{1}\geq3/4,\,g_{2}>0$ is unique, while for
$g_{1}\in\lbrack-1/4,3/4),\,g_{2}>0$, an identification is achieved by
evaluating the asymptotic behavior of functions belonging to the domains of
$\hat{H}_{\mathfrak{e},c}$ and $\hat{H}_{\mathfrak{e},d}$\ at the origin and
comparing it with the asymptotic s.a. boundary conditions specifying
different generalized Calogero Hamiltonians
$\hat{H}_{\mathfrak{e}}=\hat{H}_\nu$ with given
$g_{1}\in\lbrack-1/4,3/4),\,g_{2}>0$.

We say in advance that generalized oscillator representation (\ref{1.2}), or
(\ref{1.3}), for a given generalized Calogero Hamiltonian is generally highly
nonunique; in fact, there exists a one-, or even two-, parameter family of
generalized oscillator representations for each Hamiltonian, among which there
exists an optimum representation.

In conclusion, we note that the ideas and methods used in this paper, as well
as in \cite{TyuVo13}, can be applied to constructing generalized oscillator
representations for other Hamiltonians associated with s.a. second order
differential operations.

\section{Generalized Calogero Hamiltonians in generalized oscillator form}

We proceed to solving the problem of constructing generalized Calogero
Hamiltonians in generalized oscillator form in accordance with the program
presented above.

\subsection{Basics of constructing generalized oscillator representations
for differential operation $\check{H}$}

We begin with looking into the possibility of representing generalized
Calogero differential operation $\check{H}$ (\ref{1.1}) with coupling
constants $g_{1}\geq-1/4$ and $g_{2}>0$ in generalized oscillator form
(\ref{1.6a}) with $u\geq u_{0}$ given by
(\ref{1.5}).

Directly extending the arguments in \cite{TyuVo13} concerning the Calogero
differential operation, $g_{2}=0$, to the generalized Calogero differential
operation proper, we can assert that the differential operation $\check{H}$
allows generalized oscillator representation (\ref{1.6a}) iff the homogeneous
differential equation
\begin{equation}
-\phi^{\prime\prime}(x)+(g_{1}x^{-2}+g_{2}x^{2}+u)\phi(x)=0,\label{2.1.3}
\end{equation}
or the eigenvalue problem
\begin{equation}
\check{H}\phi(x)=-\phi^{\prime\prime}(x)+(g_{1}x^{-2}+g_{2}x^{2}%
)\phi(x)=-u\,\phi(x) \label{2.1.4}%
\end{equation}
(which can be considered a stationary Schr\"{o}dinger equation \ with
''energy'' $E=-u$), has a real-valued positive solution $\phi(u;x)$,%
\[
\operatorname{Im}\phi(u;x)=0,\;\phi(u;x)>0,\;x>0,
\]
and in this case, $\check{a}(u)$ and $\check{b}(u)$ are first-order
differential operations of the form
\begin{equation}\label{2.1.6}
\begin{array}{lll}
& \check{a}(u)=d_{x}-h(u;x)=\phi(u;x)d_{x}\frac{1}{\phi(u;x)}=\check{b}^{\ast
}(u),\;\\
& \check{b}(u)=-d_{x}-h(u;x)=-\frac{1}{\phi(u;x)}d_{x}\phi(u;x)=\check
{a}^{\ast}(u),\\
& h(u;x)=\phi^{\prime}(u;x)/\phi(u;x)=-\phi(u;x)\left(  \frac{1}{\phi
(u;x)}\right)  ^{\prime}.%
\end{array}
\end{equation}
It is evident that the real-valued $h(u;x)$ is smooth in $(0,\infty)$ as a
function of $x$ because the real-valued $\phi(u;x)$ is smooth and positive, so
that $\check{a}(u):$ $\mathcal{D}(\mathbb{R}_{+})\rightarrow
\mathcal{D}(\mathbb{R}_{+})$, $\check{b}(u):$ $\mathcal{D}(\mathbb{R}%
_{+})\rightarrow\mathcal{D}(\mathbb{R}_{+})$. It is also evident
that the function $\phi(u;x)$ in (\ref{2.1.6}) is defined up to a positive
constant factor\footnote{The differential operations in (\ref{2.1.6}) are
evidently defined up to arbitrary phase factors, $\check{a}\rightarrow$
$e^{i\theta(x)}\check{a}$, $\check{b}\rightarrow$ $\check{b}e^{-i\theta(x)}$.
These factors are irrelevant because they trivially cancel in the product
$\check{b}\,\check{a} $; their choosing is a matter of convenience, we here
choose $\theta(x)=0$.}.

If eq. (\ref{2.1.3}), or (\ref{2.1.4}), with given coupling constants
$g_{1}$, $g_{2}$ and external real parameter $u$ has no real-valued positive
solution, there exists no generalized oscillator representation with given
$u$ for $\check{H}$ with given $g_{1}$ and $g_{2}$.

If eq. (\ref{2.1.3}) with given $u$ has a unique, up to a positive constant
factor, real-valued positive solution $\phi(u;x)$, there exists a unique
generalized oscillator representation with given $u$ for given $\check{H}$.
It may happen that eq. (\ref{2.1.3}) with different $u$ has desired
solutions $\phi(u;x)$, then we get a one-parameter family
$\{\check{a}(u),\check {b}(u)\}$ of different pairs
$\check{a}(u),\check{b}(u)$ of differential operations (\ref{2.1.6}) with
the respective different functions $\phi(u;x)$ providing the desired
generalized oscillator representation for given $\check{H}$.

It may happen that eq. (\ref{2.1.3}) with given $u$ has two linearly
independent real-valued positive solutions $\phi_{1}(u;x)$ and $\phi_{2}%
(u;x)$, then the general real-valued positive solution $\phi(u;x)$ of this
equation, defined modulo a positive constant factor, is of the form
$\phi(\mu,u;x)=\sin\mu\,\phi_{1}(u;x)+\cos\mu\,\phi_{2}(u;x)$ with $\mu$
belonging to a certain interval in $[0,2\pi]$ such that the positive functions
$\phi(\mu,u;x)$ with different $\mu$ are pairwise linearly independent; of
course, this interval contains the segment $[0,\pi/2]$. In such a case, we get
a two-parameter family $\{\check{a}(\mu,u),\check{b}(\mu,u)\}$ of different
pairs $\check{a}(\mu,u),\check{b}(\mu,u)$ of differential operations
(\ref{2.1.6}) with the respective different functions $\phi(\mu,u;x)$
providing the desired generalized oscillator representation for
given $\check{H}$.

For completeness, we first show that generalized Calogero differential
operation $\check{H}$ (\ref{1.1}) with the coupling constant $g_{1}<-1/4$ or
with the coupling constant $g_{2}<0$ does not allow generalized oscillator
representation with whatever $u$, which is in complete agreement with the fact
that the respective generalized Calogero Hamiltonians $\hat{H}_{\mathfrak{e}}%
$\ with such coupling constants are not bounded from below and therefore
cannot be represented in generalized oscillator form.

It is sufficient to prove that eq. (\ref{2.1.3}) with $g_{1}<-1/4$ or with
$g_{2}<0$ and any real $u$ has no real-valued positive solution.

Let $g_{1}=-1/4-\sigma^{2},\,\sigma>0$. In this case, eq. (\ref{2.1.3}) with
any $g_{2\text{ }}$and $u$ has a solution $\phi_{1}(x)$ whose asymptotic
behavior at the origin is given by%
\begin{equation*}
\phi_{1}(x)=(k_{0}x)^{1/2+i\sigma}[1+O(x^{2})],\,x\rightarrow0,
\end{equation*}
where $k_{0}$ is an arbitrary, but fixed, parameter of dimension of inverse
length. The linearly independent solution $\phi_{2}(x)$ is the complex
conjugate of $\phi_{1}(x)$, $\phi_{2}(x)=\overline{\phi_{1}(x)}$. The
general real-valued
solution of eq. (\ref{2.1.3}) is of the form $\phi(x)=A\,\phi_{1}%
(x)+\overline{A}\,\overline{\phi_{1}(x)},$ $A=|A|\,e^{i\varphi}$ is an
arbitrary constant, its asymptotic behavior at the origin is given by%
\begin{equation*}
\phi(x)=2|A|(k_{0}x)^{1/2}[1+O(x^{2})]\cos[\sigma\ln(k_{0}x)+\varphi
+O(x^{2})],\,x\rightarrow0,
\end{equation*}
which demonstrates an infinite number of zeroes of $\phi(x)$ accumulated at
the origin.

Let $g_{2}=-\omega^{2},\,\omega>0$. In this case, eq. (\ref{2.1.3}) with any
$g_{1\text{ }}$and $u$ has a solution $\phi_{1}(x)$ whose asymptotic behavior
at infinity is given by
\begin{equation*}
\phi_{1}(x)=(\omega x^{2})^{-\frac{1}{4}(1-i\frac{u}{\omega})}\,e^{-i\frac{1}%
{2}\omega x^{2}}[1+O(1/x^{2})],\,x\rightarrow \infty.
\end{equation*}

Again, the linearly independent solution $\phi_{2}(x)$ is the complex
conjugate of $\phi_{1}(x)$, $\phi_{2}(x)=\overline{\phi_{1}(x)}$, and the
general real-valued
solution of eq. (\ref{2.1.3}) is of the form $\phi(x)=A\,\phi_{1}%
(x)+\overline{A}\,\overline{\phi_{1}(x)},$ $A=|A|\,e^{i\varphi}$. The
asymptotic behavior of $\phi(x)$ at infinity is given by%
\begin{equation*}
2|A|(\omega x^{2})^{-1/4}[1+O(x^{2})]\cos[\frac{1}{2}\omega x^{2}%
-\frac{u}{4\omega}\ln(\omega x^{2})-\varphi+O(x^{2})],\,x\rightarrow
\infty,
\end{equation*}
which demonstrates an infinite number of zeroes of $\phi(x)$ accumulated at infinity.

We are now coming to generalized oscillator representations for generalized
Calogero differential operations $\check{H}$ (\ref{1.1}) with coupling
constants $g_{1}>-1/4$ and $g_{2}>0$, which are of our main interest.

In finding the general solution of eq. (\ref{2.1.3}), it is convenient to go
from the old parameters $g_{1}$, $g_{2}$, $u$, $u_{0}$ to new parameters
$\varkappa$, $\upsilon$, $w$, $w_{0}$ defined by%

\begin{equation}\label{2.1.11}
\begin{array}{llll}
& g_{1}=-1/4+\varkappa^{2},\,\varkappa=\sqrt{g_{1}+1/4}\geq0,\\
& g_{2}=\upsilon^{4},\,\upsilon=\sqrt[4]{g_{2}}>0,\,\\
& u=4\sqrt{g_{2}}w,\,w=u/4\upsilon^{2}\geq w_{0},\\
& w_{0}=\frac{u_{0}}{4\upsilon^{2}}=-\frac{1}{2}(1+\varkappa)<0,%
\end{array}
\end{equation}
$\upsilon$ is of dimension of inverse length, $\varkappa$, $w$ and $w_{0}$
are dimensionless, and from the old space variable $x$ to a new dimensionless
variable
\begin{equation}
\rho=(\upsilon x)^{2}\geq0.\label{2.1.12}%
\end{equation}
The Ansatz%
\begin{equation}
\phi(x)=\mathrm{e}^{-\rho/2}\rho^{1/4+\varkappa/2}\tilde{\phi}(\rho
)\label{2.1.13}%
\end{equation}
reduces eq. (\ref{2.1.3}) to the equation
\begin{equation}\label{2.1.14}
\begin{array}{ll}
& \rho\frac{d^{2}\tilde{\phi}(\rho)}{d\rho^{2}}+(\beta-\rho)\frac{d\tilde
{\phi}(\rho)}{d\rho}-\alpha\tilde{\phi}(\rho)=0,\\
& \alpha=1/2(1+\varkappa)+w=\beta/2+w\geq0,\,\beta=1+\varkappa\geq1,%
\end{array}
\end{equation}
which is the so-called confluent hypergeometric equation. A great body of
information on its solutions, which are called confluent hypergeometric
functions, can be found in \cite{BatEr53}, Ch. 6 and in \cite{GraRy94}, Ch.9.2.

When considering the fundamental systems of solutions and the respective
representations for the general solution of confluent hypergeometric equation
(\ref{2.1.14}), we have to distinguish two cases: the case of $\alpha>0$ and
the case of $\alpha=0$.

As the fundamental system of solutions of eq. (\ref{2.1.14}) with $\alpha>0$,
we can take the two standard confluent hypergeometric functions, the function
\begin{equation}
\Phi(\alpha,\beta;\rho)=\sum_{k=0}^{\infty}\frac{(\alpha)_{k}}{(\beta)_{k}%
}\frac{\rho^{k}}{k!}=1+\frac{\alpha}{\beta}\rho+\frac{\alpha(\alpha+1)}%
{\beta(\beta+1)}\frac{\rho^{2}}{2!}+...,\label{2.1.15}%
\end{equation}
where
\begin{equation*}
\,(\alpha)_{k}=\frac{\Gamma(\alpha+k)}{\Gamma(\alpha)}=\left\{
\begin{array}
[c]{l}%
1,\,k=0,\\
\alpha(\alpha+1)(\alpha+k-1),\,k=1,2,3,...
\end{array}
\right.  ,
\end{equation*}
is the Pochhammer symbol, and the function%

\begin{equation}
\Psi(\alpha,\beta;\rho)=\frac{\Gamma(1-\beta)}{\Gamma(\alpha-\beta+1)}%
\Phi(\alpha,\beta;\rho)+\frac{\Gamma(\beta-1)}{\Gamma(\alpha)}\rho^{1-\beta
}\Phi(\alpha-\beta+1,2-\beta;\rho).\label{2.1.17}%
\end{equation}
The indeterminacy of the r.h.s in (\ref{2.1.17}) at integers $\beta=n+1,$
$n=0,1,2,...$, is resolved by the\ passage to the limit $\beta\rightarrow n+1
$, which produces the term with logarithmic factor, in particular,%
\begin{align}
& \Psi(\alpha,1;\rho)=\frac{1}{\Gamma(\alpha)}\left\{  \Phi(\alpha
,n+1;\rho)\ln\frac{1}{\rho}\right.  +\nonumber\\
& \left.  +\sum_{r=0\,}^{\infty}\frac{(\alpha)_{r}}{(n+1)_{r}}\left[
2\psi(1+r)-\psi(\alpha+r)\right]  \frac{\rho^{r}}{r!}\right\}
,\label{2.1.17a}%
\end{align}
where $\psi$ is the logarithmic derivative of the Euler $\Gamma$ function,
$\psi(z)$ $=\Gamma^{\prime}(z)/\Gamma(z)$. The function $\Psi(\alpha
,\beta;\rho)$ also allows the representation
\begin{align}
& \Psi(\alpha,\beta;\rho)=\frac{1}{\Gamma(\alpha)}\int_{0}^{\infty
}dt\,t^{\alpha-1}(1+t)^{\beta-\alpha-1}e^{-\rho t}=\nonumber\\
& \,=\frac{\rho^{1-\beta}}{\Gamma(\alpha)}\int_{0}^{\infty}dt\,t^{\alpha
-1}(\rho+t)^{\beta-\alpha-1}e^{-t}.\label{2.1.18}%
\end{align}

The general solution of eq. (\ref{2.1.14}) with $\alpha>0$ is then given by%
\begin{equation}
\tilde{\phi}(\rho)=A\,\Phi(\alpha,\beta;\rho)+B\,\Psi(\alpha,\beta
;\rho),\,\alpha>0,\label{2.1.19}%
\end{equation}
where $A$ and $B$ are arbitrary complex coefficients.

In what follows, we need the asymptotic behavior of the functions $\Phi
(\alpha,\beta;\rho)$ and $\Psi(\alpha,\beta;\rho)$ as a functions of $\rho$ at
the origin and at infinity, which we present in the form sufficient for our
purposes. The asymptotic behavior of these functions at the origin is
respectively given by%
\begin{equation}
\Phi(\alpha,\beta;\rho)=1+O(\rho),\;\rho\rightarrow0,\label{2.1.19a}%
\end{equation}
and
\begin{equation}
\Psi(\alpha,\beta;\rho)=\left\{
\begin{array}
[c]{l}%
O(\rho^{1-\beta}),\;\beta\geq2\text{ }\\
\frac{\Gamma(\beta-1)}{\Gamma(\alpha)}\rho^{1-\beta}(1+O(\rho))+\\
+\frac{\Gamma(1-\beta)}{\Gamma(\alpha-\beta+1)}(1+O(\rho)),\;\beta\in(1,2)\\
-\frac{1}{\Gamma(\alpha)}\ln\rho\,(1+O(\rho))+\\
+[2\psi(1)-\psi(\alpha)](1+O(\rho)),\,\beta=1
\end{array}
\right.  \rightarrow\infty,\;\rho\rightarrow0,\,\label{2.1.19b}%
\end{equation}
while their asymptotic behavior at infinity is respectively given by%
\begin{equation}
\Phi(\alpha,\beta;\rho)=\frac{\Gamma(\beta)}{\Gamma(\alpha)}\rho^{\alpha
-\beta}e^{\rho}\left(  1+O(1/\rho)\right)  \rightarrow\infty,\;\rho
\rightarrow\infty,\label{2.1.19c}%
\end{equation}
and%
\begin{equation}
\Psi(\alpha,\beta;\rho)=\rho^{-\alpha}(1+O(1/\rho))\rightarrow0,\;\,\rho
\rightarrow\infty.\label{2.1.19d}%
\end{equation}

The case of $\alpha=0$ is the exceptional case because $\Phi(0,\beta
;\rho)=\,\Psi(0,\beta;\rho)$ $=1$. As the fundamental system of solutions of
eq. (\ref{2.1.14}) with $\alpha=0$, we can take the functions%
\begin{equation*}
\Phi(\beta;\rho)=\Phi(0,\beta;\rho)=1 
\end{equation*}
and%
\begin{equation}
\Psi(\beta;\rho)=\int_{a^{2}}^{\rho}d\tau\tau^{-\beta}e^{\tau},\label{2.1.21}%
\end{equation}
where $a>0$ is a certain fixed number.

The general solution of eq. (\ref{2.1.14}) with $\alpha=0$ is then given by%
\begin{equation}
\tilde{\phi}(\rho)=A\,+B\,\Psi(\beta;\rho),\,\alpha=0.\label{2.1.22}%
\end{equation}

Returning to eq. (\ref{2.1.3}), we use the notation introduced in
(\ref{2.1.11}), (\ref{2.1.12}) and (\ref{2.1.14}), where $u=4\upsilon^{2}w$,
and have to distinguish the region $w>w_{0}=-\frac{1}{2}(1+\varkappa)$
$(\alpha>0)$ and the point $w=w_{0}$ $(\alpha=0)$.

The general solution of eq. (\ref{2.1.3}) with $g_{1}%
\geq-1/4,\,g_{2}>0$, and $w>w_{0}$ is obtained by combining (\ref{2.1.13}) and
(\ref{2.1.19}). But to get a suitable form of the asymptotic behavior of the
solution at the origin, we renormalize the coefficient $B$ in (\ref{2.1.19})
as follows\footnote{This is equivalent to an evident change of the fundamental
system of solutions of eq. (\ref{2.1.14}).}:
\begin{equation*}
B\rightarrow\left\{
\begin{array}
[c]{l}%
B\frac{\Gamma(\alpha)}{\Gamma(\beta-1)},\,g_{1}>-1/4\;(\varkappa>0)\\
B\Gamma(\alpha),\,g_{1}=-1/4\;(\varkappa=0)
\end{array}
\right.  .
\end{equation*}
Under this convention, the general solution of eq. (\ref{2.1.3}) with
$w>w_{0}$ is given by
\begin{align}
& \phi(w;x)=A\phi_{1}(w;x)+B\phi_{2}(w;x),\,w>w_{0}=-\frac{1}{2}%
(1+\varkappa), \label{2.1.23}\\
& \phi_{1}\left(  w;x\right)  =\mathrm{e}^{-\rho/2}\rho^{1/4+\varkappa/2}%
\Phi(\alpha,\beta;\rho),\nonumber\\
& \phi_{2}\left(  w;x\right)  =\left\{
\begin{array}
[c]{l}%
\mathrm{e}^{-\rho/2}\rho^{1/4+\varkappa/2}\frac{\Gamma(\alpha)}{\Gamma
(\varkappa)}\Psi(\alpha,\beta;\rho),\;g_{1}>-1/4\;(\varkappa>0)\\
\mathrm{e}^{-\rho/2}\rho^{1/4}\Gamma(\alpha)\Psi(\alpha,1;\rho),\;g_{1}%
=-1/4\;(\varkappa=0)
\end{array}
\right.  \nonumber,%
\end{align}
where the functions $\Phi(\alpha,\beta;\rho)$ and $\Psi(\alpha,\beta;\rho)$ are
given by respective (\ref{2.1.15}) and (\ref{2.1.17}), (\ref{2.1.17a}) or
(\ref{2.1.18}), $\ A$ and $B$ are arbitrary complex numbers. It is remarkable
that the both functions $\phi_{1}$ and $\phi_{2}$ are real-valued and positive.

According to (\ref{2.1.13}) and (\ref{2.1.22}), the general solution of eq.
(\ref{2.1.3}) with $w=w_{0}$ is given by%

\begin{equation}
\phi(w_{0};x)=\mathrm{e}^{-\rho/2}\rho^{1/4+\varkappa/2}(A\,+B\Psi(\beta
;\rho)\,),\,w=w_{0}=-\frac{1}{2}(1+\varkappa),\label{2.1.24}%
\end{equation}
where the function $\Psi(\beta;\rho)$ is given by (\ref{2.1.21}).

Accordingly, we have to consider separately the region
$w>w_{0}=-1/2(1+\varkappa)$ and the point $w=w_{0}$ in future treatment. In
addition, as it follows from \cite{GitTyV12}, in the both cases
$w>w_{0}$ and $w=w_{0}$, the point $g_{1}=-1/4$ $(\varkappa=0)$ is naturally
distinguished by a specific behavior of the functions involved at the origin,
which is an essential point in the analysis. We begin with the region
$g_{1}>-1/4$ $(\varkappa>0)$\ and $w>w_{0}$.

\subsection{Region $g_{1}>-1/4$ ($\varkappa>0$),\ $w>w_{0}$
$=-1/2(1+\varkappa)$ ($\alpha>0$)}

\subsubsection{Generalized oscillator representations for $\check{H}$,
differential operations $\check{a}$ and $\check{b}$}

In this region of parameters, the general solution of eq. (\ref{2.1.3}) is
given by (\ref{2.1.23}) with real-valued and positive linearly-independent
functions $\phi_{1}$ and $\,\phi_{2}$. In addition, the function $\Phi
(\alpha,\beta;\rho)$ increases monotonically from $1$ to $\infty$ as
$\rho=(\upsilon x)^{2}$ together with $x$ ranges from $0$ to $\infty$, see
(\ref{2.1.15}), while the function $\Psi(\alpha,\beta;\rho)$ decreases
monotonically from $\infty$ to $0$ because the integrand of the first
integral in the r.h.s. of (\ref{2.1.18}) is a decreasing function of
$\rho$, is nonintegrable at $\rho=0$\ and vanishes as
$\rho\rightarrow\infty$. It follows by the arguments in the previous
subsection, see the text after (\ref{2.1.6}), that the general real-valued
positive solution of eq. (\ref{2.1.3}) defined modulo a positive constant
factor is given by (\ref{2.1.23}) with $A=\sin\mu$, $B=\cos\mu$,
$\mu\in\lbrack0,\pi/2]$,
\begin{align}
& \phi(\mu,w;x)=\mathrm{e}^{-\rho/2}\rho^{1/4+\varkappa/2}\left[  \Phi
(\alpha,\beta;\rho)\sin\mu+\frac{\Gamma(\alpha)}{\Gamma(\varkappa)}\Psi
(\alpha,\beta;\rho)\cos\mu\right]  ,\label{2.2.1.2}\\
& \,\mu\in\lbrack0,\pi/2],\,\alpha=1/2(1+\varkappa)+w>0,\,\beta=1+\varkappa
,\,\varkappa>0,\rho=(\upsilon x)^{2},\nonumber%
\end{align}
which implies that we have the two-parameter family $\{\check{a}(\mu
,w),\check{b}(\mu,w);\;\mu\in\lbrack0,\pi/2],\;w\in(w_{0},\infty
),\;\varkappa>0\}$ of different pairs of mutually adjoint first-order
differential operations $\check{a}(\mu,w)$ and $\check{b}(\mu,w)$ given by
(\ref{2.1.6})\ with the evident substitutions $\check{a}(u)\rightarrow
\check{a}(\mu,w)$, $\check{b}(u)\rightarrow$ $\check{b}(\mu,w)$, and
$\phi(u;x)\rightarrow\phi(\mu,w;x)$:%
\begin{equation}\label{2.2.1.3}
\begin{array}{lll}
& \check{a}(\mu,w)=\phi(\mu,w;x)d_{x}\frac{1}{\phi(\mu,w;x)}=\check{b}^{\ast
}(\mu,w),\\
& \check{b}(\mu,w)=-\frac{1}{\phi(\mu,w;x)}d_{x}\phi(\mu,w;x)=\check{a}^{\ast
}(\mu,w),\\
& \mu\in\lbrack0,\pi/2],\,w\in(w_{0},\infty),\,\varkappa\in(0,\infty),%
\end{array}
\end{equation}
and providing a two-parameter family\footnote{As a rule, we indicate the
ranges of parameters $\mu$ and $w$ in formulas to follow only if they
differ from the whole ranges, here these are $[0,\pi/2]$ for $\mu$ and
($w_{0},\infty)$ for $w$, the range of $\varkappa$ is clear from the title
of section, subsection or subsubsection, here this is $(0,\infty)$. As to
the main resulting formulas, we indicate the ranges of all the parameters
including $\varkappa$.} of different generalized oscillator representations
(\ref{1.6a}) for generalized Calogero differential operation
$\check{H}\,$\ (\ref{1.1}) with $g_{1}>-1/4$ and $g_{2}>0$,%
\begin{align}
& \check{H}=-d_{x}^{2}+g_{1}x^{-2}+g_{2}x^{2}=\check{b}(\mu,w)\check{a}%
(\mu,w)-4\upsilon^{2}w,\label{2.2.1.4}\\
& \mu\in\lbrack0,\pi/2],\,w\in(w_{0},\infty).\nonumber%
\end{align}

In an analysis to follow, we need the asymptotic behavior of the functions
$\phi(\mu,w;x)$ and $1/\phi(\mu,w;x)$ at the origin and at infinity. According
to (\ref{2.1.19a}), (\ref{2.1.19b}) and (\ref{2.1.19c}), (\ref{2.1.19d}), the
asymptotic behavior of these functions at the origin is respectively given by%
\begin{align}
& \phi(\mu,w;x)=\nonumber\\
& \,=\left\{
\begin{array}
[c]{l}%
\left\{
\begin{array}
[c]{l}%
O(x^{1/2-\varkappa}),\;\varkappa\geq1\\
\widetilde{A}\,(\upsilon x)^{1/2+\varkappa}+\widetilde{B}\,(\upsilon
x)^{1/2-\varkappa}+O(x^{5/2-\varkappa}),\;\varkappa\in(0,1)
\end{array}
\right. ,\,\mu\in\lbrack0,\pi/2)\\
O(x^{1/2+\varkappa}),\,\,\mu=\pi/2,\,\forall\varkappa>0,
\end{array}
\right. ,\;x\rightarrow0, \label{2.2.1.5}\\
& \widetilde{A}=\widetilde{A}(\mu,w)=\sin\mu-\cos\mu\frac{\Gamma
(1-\varkappa)\Gamma\left(  \frac{1}{2}(1+\varkappa)+w\right)  }{\Gamma
(1+\varkappa)\Gamma\left(  \frac{1}{2}(1-\varkappa)+w\right)  },\,\nonumber\\
& \widetilde{B}=\,\widetilde{B}(\mu,w)\,=\cos\mu,\,\nonumber%
\end{align}
and%
\begin{align}
& 1/\phi(\mu,w;x)=\nonumber\\
& \,=\left\{
\begin{array}
[c]{l}%
\left\{
\begin{array}
[c]{l}%
O(x^{\varkappa-1/2}),\;\varkappa\geq1\\
\frac{1}{\widetilde{B}(\mu,w)}\frac{(\upsilon x)^{\varkappa-1/2}%
}{1+(\widetilde{A}(\mu,w)/\widetilde{B}(\mu,w))(\upsilon x)^{2\varkappa}%
}+O(x^{3/2+\varkappa}),\;\varkappa\in(0,1)
\end{array}
\right.  ,\,\mu\in\lbrack0,\pi/2)\\
O(x^{-1/2-\varkappa}),\,\,\mu=\pi/2,\,\forall\varkappa>0,
\end{array}
\right.  ,x\rightarrow0,\label{2.2.1.6}%
\end{align}
while their asymptotic behavior at infinity is respectively given by%
\begin{align}
& \phi(\mu,w;x)=\nonumber\\
& =\left\{
\begin{array}
[c]{l}%
\sin\mu\frac{\Gamma(1+\varkappa)}{\Gamma\left(  \frac{1}{2}(1+\varkappa
)+w\right)  }(\upsilon x)^{-1/2+2w}e^{\frac{1}{2}(\upsilon x)^{2}}%
(1+O(x^{-2})),\,\\
\mu\in(0,\pi/2],\,\forall\varkappa>0\\
\frac{\Gamma\left(  \frac{1}{2}(1+\varkappa)+w\right)  }{\Gamma(\varkappa
)}(\upsilon x)^{-1/2-2w}e^{-\frac{1}{2}(\upsilon x)^{2}}(1+O(x^{-2}%
)),\,\mu=0,\,\forall\varkappa>0
\end{array}
\right.  ,\;x\rightarrow\infty,\label{2.2.1.7}%
\end{align}
and%
\begin{align}
& 1/\phi(\mu,w;x)=\nonumber\\
& \,=\left\{
\begin{array}
[c]{l}%
\frac{1}{\sin\mu}\frac{\Gamma\left(  \frac{1}{2}(1+\varkappa)+w\right)
}{\Gamma(1+\varkappa)}(\upsilon x)^{+1/2-2w}e^{-\frac{1}{2}(\upsilon x)^{2}%
}(1+O(x^{-2})),\\
\mu\in(0,\pi/2],\,\forall\varkappa>0\\
\frac{\Gamma(\varkappa)}{\Gamma(\frac{1}{2}(1+\varkappa)+w)}(\upsilon
x)^{1/2+2w}e^{\frac{1}{2}(\upsilon x)^{2}}(1+O(x^{-2})),\,\mu=0,\,\forall
\varkappa>0
\end{array}
\right.  ,\,x\rightarrow\infty.\label{2.2.1.8}%
\end{align}

\subsubsection{Initial operators $\hat{a}$ and $\hat{b}$}

We introduce the pairs of initial differential operators $\hat{a}(\mu,w)$
and $\hat{b}(\mu,w)\,$in $L^{2}(\mathbb{R}_+)$\textrm{\thinspace}defined on
the subspace $\mathcal{D}(\mathbb{R}_{+})$ of smooth compactly
supported functions,
$D_{a(\mu,w)}=D_{b(\mu,w)}=\mathcal{D}(\mathbb{R}_{+})$, and associated
with each pair of the respective differential operations $\check
{a}(\mu,w)$ and $\check{b}(\mu,w)$ (\ref{2.2.1.3}). These operators have the
property%
\begin{equation}
\left(  \psi,\hat{a}(\mu,w)\xi\right)  =\left(  \hat{b}(\mu,w)\psi,\xi\right)
,\;\forall\psi(x),\xi(x)\in\mathcal{D}(\mathbb{R}_{+}),\label{2.2.2.1}%
\end{equation}
which is easily verified by integration by parts. According to (\ref{2.2.1.4}),
the initial symmetric operator $\hat{H}$ associated with $\check{H}$ and
defined on $\mathcal{D}(\mathbb{R}_{+})$ allows the representation%
\begin{equation}
\hat{H}=\hat{a}(\mu,w)\hat{b}(\mu,w)\,-4\upsilon^{2}w\,\hat{I},\,\forall\mu
\in\lbrack0,\pi/2],\,\forall\,w>w_{0},\forall\varkappa>0,\label{2.2.2.2}%
\end{equation}
which, in particular, implies that $\hat{H}$ with $g_{1}>-1/4,g_{2}>0$ is
bounded from below by $-4\upsilon^{2}w$,%
\begin{align}
& \left(  \xi,\hat{H}\xi\right)  =\left(  \xi,(\hat{b}(\mu,w)\,\hat{a}(\mu
,w)-4\upsilon^{2}w\,\hat{I})\,\xi\right)  =\nonumber\\
& =\left(  \hat{a}(\mu,w)\xi,\hat{a}(\mu,w)\xi\right)  -4\upsilon
^{2}w\,\left(  \xi,\xi\right)  \geq-4\upsilon^{2}w\,\left(  \xi,\xi\right)
,\;\forall\xi(x)\in\mathcal{D}(\mathbb{R}_{+}),\nonumber
\end{align}
and taking the infimum of $w$, which is $w_{0}$, we obtain that $\hat{H}$ with
$g_{1}>-1/4,$ $g_{2}>0\,$is bounded from below by $-4\upsilon^{2}%
w_{0}=2\upsilon^{2}(1+\varkappa)$. These representations provide a
basis for constructing s.a. generalized Calogero Hamiltonians
$\hat{H}_{\mathfrak{e}}$ in generalized oscillator form as s.a. extensions of
$\hat{H}$ (\ref{2.2.2.2}) $\,$in accordance with the program formulated in
sec.1. Namely, we should construct all possible extensions of each pair
$\hat{a}(\mu,w)$, $\hat{b}(\mu,w)$ of initial operators with given$\,\mu$ and
$w$ to a pair of closed mutually adjoint operators $\hat{c}(\mu,w)$,
$\hat{c}^{+}(\mu,w)$, $\hat{a}(\mu,w)\subset\hat{c}(\mu,w)$, $\hat{b}%
(\mu,w)\subset\hat{c}^{+}(\mu,w)$, beginning from the closure $\overline
{\hat{a}}(\mu,w)$ of $\hat{a}(\mu,w)$ or to a pair of closed mutually adjoint
operators $\hat{d}(\mu,w)$, $\hat{d}^{+}(\mu,w)$, $\hat{a}(\mu,w)\subset
\hat{d}^{+}(\mu,w)$, $\hat{b}(\mu,w)\subset\hat{d}(\mu,w)$, beginning from the
closure $\overline{\hat{b}}(\mu,w)$ of $\hat{b}(\mu,w)$. These extensions
produce the respective s..a. operators
\begin{equation}
\hat{H}_{\mathfrak{e\,}c(\mu,w)}=\hat{c}^{+}(\mu,w)\,\hat{c}(\mu
,w)-4\upsilon^{2}w\,\hat{I},\label{2.2.2.4}%
\end{equation}
and
\begin{equation}
\hat{H}_{\mathfrak{e\,}d(\mu,w)}=\hat{d}(\mu,w)\,\hat{d}^{+}(\mu
,w)-4\upsilon^{2}w\,\hat{I},\label{2.2.2.5}%
\end{equation}
which are certain generalized Calogero Hamiltonians in generalized
oscillator form. It then remains to identify
$\hat{H}_{\mathfrak{e\,}c(\mu,s)}$ and $\hat{H}_{\mathfrak{e\,}d(\mu,s)}$
with the known generalized Calogero Hamiltonians. It should be noted that
the operators $\hat{H}_{\mathfrak{e\,}c(\mu,w)}$ and
$\hat{H}_{\mathfrak{e\,}d(\mu,w)}$ are not necessarily different even if
$c(\mu,w)\neq\hat{d}^{+}(\mu,w)$, and what is more, the l.h.s in
(\ref{2.2.2.4}) or in (\ref{2.2.2.5}) may not depend on $\mu$ and $w$.

We proceed to constructing all possible extensions of an arbitrary pair of
initial operators $\hat{a}(\mu,w)$, $\hat{b}(\mu,w)$\ to a pair of closed
mutually adjoint operators.

\subsubsection{Adjoint operators $\hat{a}^{+}$ and $\hat{b}^{+}$, closed
operators $\overline{\hat{a}}$ and $\overline{\hat{b}}$}

Because all the operators $\hat{a}(\mu,w)$ and $\hat{b}(\mu,w)$ are densely
defined, they have the adjoins, the respective $\hat{a}^{+}(\mu,w)$ and
$\hat{b}^{+}(\mu,w)$. The defining equation for $\hat{a}^{+}(\mu,w)$, i.e.,
the equation for pairs $\psi(x)\in D_{a^{+}(\mu,w)}$ and $\eta(x)=\hat{a}^{+}%
(\mu,w)\psi(x)$ forming the graph of the operator $\hat{a}^{+}(\mu,w)$, see
\cite{GitTyV12}, sec. 2.6, reads%
\begin{equation}
(\psi,\hat{a}(\mu,w)\xi)=(\eta,\xi),\ \ \forall\xi(x)\in\mathcal{D}%
(\mathbb{R}_{+}).\label{2.2.3.1}%
\end{equation}
The equality (\ref{2.2.2.1}) then implies that $\hat{b}(\mu,w)$
$\subset\hat{a}^{+}(\mu,w)$: eq. (\ref{2.2.3.1}) has solutions%
\begin{equation*}
\psi(x)=\zeta(x),\ \ \eta(x)=\hat{b}(\mu,w)\zeta(x),\ \ \forall\zeta
(x)\in\mathcal{D}(\mathbb{R}_{+}).
\end{equation*}
It follows that $\hat{a}^{+}(\mu,w)$\thinspace is densely defined and in
turn has the adjoint $(\hat{a}^{+}(\mu,w))^{+}$, while the operator
$\hat{a}$ $(\mu,w)$ has a closure $\overline{\hat{a}}(\mu,w)=(\hat{a}^{+}%
(\mu,w))^{+}\subseteq\hat{b}^{+}(\mu,w)$ and $\left(  \overline{\hat{a}}%
(\mu,w)\right)  ^{+}=\hat{a}^{+}(\mu,w)$. Similarly, we obtain that
$\hat{a}(\mu,w)\subset\hat{b}^{+}(\mu,w)$, and therefore, there exists the
adjoint $(\hat{b}^{+}(\mu,w))^{+}$ of $\hat{b}^{+}(\mu,w)$,the operator
$\hat{b}(\mu,w)$ has a closure $\overline{\hat{b}}(\mu,w)=(\hat{b}^{+}%
(\mu,w))^{+}\subseteq\hat{a}^{+}(\mu,w)$ and $\left(  \overline{\hat{b}}%
(\mu,w)\right)  ^{+}=\hat{b}^{+}(\mu,w)$. We thus obtain the chains
of inclusions%
\begin{equation}\label{2.2.3.3}
\begin{array}{ll}
\hat{a}(\mu,w)  & \subset\overline{\hat{a}}(\mu,w)=(\hat{a}^{+}(\mu
,w))^{+}\subseteq\hat{b}^{+}(\mu,w),\,\\
\hat{b}(\mu,w)  & \subset\overline{\hat{b}}\,(\mu,w)=\,(\hat{b}^{+}%
(\mu,w))^{+}\subseteq\hat{a}^{+}(\mu,w).%
\end{array}
\end{equation}

\subsubsection{Domains of operators $\hat{a}^{+}$, $\hat{b}^{+}$,
$\overline{\hat{a}}$ and $\overline{\hat{b}}$}

In evaluating the operators $\hat{a}^{+}(\mu,w),\ \hat{b}^{+}(\mu
,w),\ \overline{\hat{a}}(\mu,w),$ and$\ \overline{\hat{b}}(\mu,w)$, we follow
\cite{TyuVo13} where the case of $g_{2}=0$ was considered. The operators
$\hat{a}^{+}(\mu,w)$ and $\overline{\hat{b}}(\mu,w)$ are associated with the
differential operation $\check{b}(\mu,w)=\check{a}^{\ast}(\mu,w)$, while the
operators $\hat{b}^{+}(\mu,w)$ and $\overline{\hat{a}}(\mu,w)$
are associated with the differential operation $\check{a}(\mu,w)=\check
{b}^{\ast}(\mu,w)$. It is therefore sufficient to evaluate the domains of the
operators involved, which either coincide with or belong to the natural
domains for the respective differential operations\footnote{We recall that the
natural domain $D_{\check{f}}^{n} \subset L^{2}(\mathbb{R}_{+}%
)$ for a given differential operation $\check{f}$ is the
maximum possible domain for operators associated with $\check{f}$, see
\cite{GitTyV12}.}.

i) The domain $D_{a^{+}(\mu,w)}$ of the operator $\hat{a}^{+}(\mu,w)$ is the
natural domain for $\check{b}(\mu,w)$:%
\begin{align}
&  D_{a^{+}(\mu,w)}=D_{\check{b}(\mu,w)}^{n}=\{\psi(x):\psi
(x)\;\text{\textrm{is a.c. in} }\mathbb{R}_{+};\,\,\nonumber\\
&  \psi(x),\check{b}(\mu,w)\psi(x)=-\frac{1}{\phi(\mu,w;x)}\frac{d}{dx}\left(
\phi(\mu,w;x)\psi(x)\right)=\eta(x)\in L^{2}(\mathbb{R}_{+}%
)\},\label{2.2.4.1}%
\end{align}
the symbol ``a.c.'' is a contraction of ``absolutely continuous''.

According to (\ref{2.2.4.1}), a generic function $\psi(x)$ belonging to
$D_{a^{+}(\mu,w)}$ can be considered as the general solution of the
inhomogeneous differential equation $\check{b}_{(\mu,w)}\psi(x)=\eta(x)$ under
the additional conditions that the both $\psi(x)$ and $\eta(x)$ are square
integrable on $\mathbb{R}_{+}$. It follows with taking estimates
(\ref{2.2.1.5}), (\ref{2.2.1.6}) and (\ref{2.2.1.7}),\ (\ref{2.2.1.8}) into
account that a generic $\psi(x)\in$ $D_{a^{+}(\mu,w)}$ allows the representation%

\begin{equation}
\psi(x)=\frac{1}{\phi(\mu,w;x)}\left[  C-\int_{x_{0}}^{x}dy\,\phi(\mu
,w;y)\eta(y)\right]  ,\,\,\eta(x)=\check{b}_{(\mu,w)}\psi(x)\in L^{2}%
(\mathbb{R}_{+}),\label{2.2.4.3}%
\end{equation}
where the point $x_{0}$ and constant $C$ depend on the values of $\mu$ and
$\varkappa$ as follows:%
\begin{equation}\label{2.2.4.4}
\begin{array}{lll}
\mu & =0:\left\{
\begin{array}
[c]{l}%
x_{0}\in(0,\infty)\mathrm{\,}\text{\textrm{for} }\varkappa\geq1\\
x_{0}=0,\mathrm{\,}\text{\textrm{for} }\varkappa\in(0,1)
\end{array}
\right. \text{\textrm{\ and }}C=\int_{x_0}^{\infty}dy\phi(0,w;y)\eta(y),\\
\mu & \in(0,\pi/2),\,\left\{
\begin{array}
[c]{l}%
x_{0}\in(0,\infty)\,\mathrm{\,}\text{\textrm{for }}\varkappa\geq1\\
x_{0}=0,\,\text{\textrm{for }}\varkappa\in(0,1)
\end{array}
\right.  \text{\textrm{\ and }}C\,\text{\textrm{is an arbitrary constant}},\\%
\mu & =\pi/2:\,x_{0}=0\text{\textrm{\ and }}C=0\,\,\text{\textrm{for} }%
\forall\varkappa>0. \\ %
\end{array}
\end{equation}
A subtlety is that for $\mu\in\lbrack0,\pi/2)$, the constant $C$ in
(\ref{2.2.4.3}) can take arbitrary values, but for $\mu\in(0,\pi/2)$,
the constant $C$ is independent of the function $\eta(x)$, while for $\mu=0$,
it is uniquely related to $\eta$, so that the representation
(\ref{2.2.4.3}), (\ref{2.2.4.4}) with $\mu=0$ for $\psi(x)\in$
$D_{a^{+}(0,w)}$ is equivalent to
\begin{align}
& \psi(x)=\frac{1}{\phi(0,w;x)}\int_{x}^{\infty}dy\,\phi(0,w;y)\eta
(y),\,\,\eta(x)=\check{b}_{(0,w)}\psi(x)\in L^{2}(\mathbb{R}_{+}),\label{2.2.4.5}\\
& \mu=0,\,\forall\varkappa>0.\nonumber%
\end{align}

Using estimates (\ref{2.2.1.5})-(\ref{2.2.1.8}) and estimating the integral
terms in (\ref{2.2.4.3}) and (\ref{2.2.4.5}) with the Cauchy--Bunyakovskii
inequality, we obtain that the asymptotic
behavior of functions $\psi(x)\in D_{a^{+}(\mu,w)}$ at the origin is given by%
\begin{equation}
\psi(x)=\left\{
\begin{array}
[c]{l}%
O(x^{1/2}),\,\forall\mu,\,\varkappa>1\text{\textrm{\ or }}\mu=\pi
/2,\,\varkappa\in(0,1]\\
O(x^{1/2}\sqrt{\ln\frac{1}{x}}),\,\,\mu\in\lbrack0,\pi/2),\,\varkappa=1\\
\frac{C}{\widetilde{B}(\mu,w)}(\upsilon x)^{\varkappa-1/2}[1+O(x^{2\varkappa
})]+O(x^{1/2}),\,\\
\,\mu\in\lbrack0,\pi/2),\,\varkappa\in(0,1)
\end{array}
\right.  ,\,x\rightarrow0,\label{2.2.4.6}%
\end{equation}
while at infinity, $\psi(x)\in D_{a^{+}(\mu,w)}$ vanishes,
\begin{equation}
\psi(x)\rightarrow0,\;x\rightarrow\infty,\;\forall\mu,\forall\varkappa
>0.\label{2.2.4.7}%
\end{equation}

We note that for $\varkappa\in(0,1)$, the domain $D_{a^{+}(\mu,w)}$ of the
operator $\hat{a}^{+}(\mu,w)$ with $\mu\in\lbrack0,\pi/2)$ can be represented
as a direct sum of the form %
\begin{equation*}
D_{a^{+}(\mu,w)}=\{C\psi_{0}(\mu,w;x)\}+\tilde{D}_{a^{+}(\mu,w)},\,\,\mu
\in\lbrack0,\pi/2),\,\varkappa\in(0,1),
\end{equation*}
where the function $\psi_{0}$ $(\mu,w;x)\,$belonging to $D_{a^{+}(\mu,w)}$ is
given by%
\begin{equation*}
\psi_{0}(\mu,w;x)=\frac{1}{\phi(\mu,w;x)}\zeta\left(  x\right)  ,\mathrm{\,}%
\text{\textrm{so that}}\ \check{b}(\mu,w)\psi_{0}(\mu,w;x)=-\frac{1}{\phi
(\mu,w;x)}\zeta^{\prime}\left(  x\right)  ,
\end{equation*}
$\zeta\left(  x\right)  $ is a fixed smooth function with a compact support
and equal to $1$ in a neighborhood of the origin, and $\tilde
{D}_{a^{+}(\mu,w)}$ is the subspace of functions belonging to $D_{a^{+}%
(\mu,w)}$ and vanishing at the origin:%
\begin{equation}
\tilde{D}_{a^{+}(\mu,w)}=\left\{  \psi(x)\in D_{a^{+}(\mu,w)}:\,\psi
(x)=O(x^{1/2}),\,x\rightarrow0\right\}  ,\,\mu\in\lbrack0,\pi/2),\;\varkappa
\in(0,1).\label{2.2.4.10}%
\end{equation}

ii) The domain $D_{b^{+}(\mu,w)}$ of the operator $\hat{b}^{+}(\mu,w)$ is
described quite similarly. It is the natural domain for $\check{a}%
(\mu,w)=\check{b}^{\ast}(\mu,w)$:
\begin{align}
& D_{b^{+}(\mu,w)}=D_{\check{a}(\mu,w)}^{n}=\{\chi(x):\chi
(x)\;\text{\textrm{is a.c. in} }\mathbb{R}_{+};\,\nonumber\\
& \chi(x),\,\check{a}(\mu,w)\chi(x)=\phi(\mu,w;x)d_{x}\left(  \frac{1}%
{\phi(\mu,w;x)}\chi(x)\right)  =\eta(x)\in L^{2}(\mathbb{R}_{+}%
)\}.\label{2.2.4.12}%
\end{align}

Using arguments similar to those in the previous item i) for a function
$\psi(x)$ belonging to $D_{a^{+}(\mu,w)}$ (\ref{2.2.4.1}) with the natural
interchange $\phi(\mu,w)\leftrightarrow1/\phi(\mu,w)$, we establish that a
generic function $\chi(x)$ belonging to $D_{b^{+}{(\mu,w)}}$ allows the representation%

\begin{equation}\label{2.2.4.13}
\begin{array}{ll}
& \chi(x)=\phi(\mu,w;x)\left[  D+\int_{x_{0}}^{x}dy\frac{1}{\phi(\mu,w;y)}%
\eta(y)\right]  \\
& \eta(x)=\check{a}(\mu,w;x)\chi(x)\in L^{2}(\mathbb{R}_{+}),%
\end{array}
\end{equation}
where the point $x_{0}$ and constant $D$ depend on the values of $\mu$ and
$\varkappa$ as follows:%
\begin{equation}\label{2.2.4.14}
\begin{array}{lll}
& \mu=0:x_{0}=0\text{\textrm{\ and }}\left\{
\begin{array}
[c]{l}%
D=0\,\text{\textrm{for} }\varkappa\geq1\\
D\,\,\text{\textrm{is an arbitrary constant for }}\varkappa\in(0,1)
\end{array} \right.  ,\\
& \mu\in(0,\pi/2),\,x_{0}=0\,\text{\textrm{and }}\left\{
\begin{array}[c]{l}%
D=\int_{0}^{\infty}dy\frac{1}{\phi(\mu,w;y)}\eta(y)=0\,\text{\textrm{for}
}\varkappa\geq1\\
D=-\int_{0}^{\infty}dy\frac{1}{\phi(\mu,w;y)}\eta(y),\;\varkappa\in(0,1)
\end{array} \right.  ,\\
& \mu=\pi/2:\,0<x_{0}<\infty\text{\textrm{\ and }}D=-\int_{x_{0}}^{\infty
}dy\frac{1}{\phi(\pi/2,w;y)}\eta(y),\,\forall\varkappa>0.%
\end{array}
\end{equation}
A subtlety is that for $\mu\in\lbrack0,\pi/2),\varkappa\in(0,1)$ and for
$\mu=\pi/2,$ $\forall\varkappa>0$, the constant $D$ in (\ref{2.2.4.13}) can
take arbitrary values, but for $\mu=0$, the constant $D$ is independent of the
function $\eta(x)$, while for $\mu\in(0,\pi/2),\varkappa\in(0,1)\,$and for
$\mu=\pi/2,$ $\forall\varkappa>0$, it is uniquely related to $\eta$.

All this implies that representation (\ref{2.2.4.13}), (\ref{2.2.4.14}) for
$\chi(x)\in D_{b^{+}(\mu,w)}$ with $\mu\in(0,\pi/2]$, $\forall\varkappa>0$, is
equivalent to
\begin{align}
& \chi(x)=-\phi(\mu,w;x)\int_{x}^{\infty}dy\,\frac{1}{\phi(\mu,w;y)}%
\eta(y),\,\,\eta(x)=\check{a}_{(\mu,w)}\chi(x)\in L^{2}(\mathbb{R}%
_{+}),\label{2.2.4.15}\\
& \,\mu\in(0,\pi/2],\forall\varkappa>0.\nonumber%
\end{align}

The asymptotic behavior of functions $\chi(x)\in$ $D_{b^{+}(\mu,w)}$ at the
origin and at infinity estimated using (\ref{2.2.1.5})-(\ref{2.2.1.8}) and the
Cauchy--Bunyakovskii inequality for integral terms in (\ref{2.2.4.13}) and
(\ref{2.2.4.15}) is respectively given by%
\begin{equation}
\chi(x)=\left\{
\begin{array}
[c]{l}%
O(x^{1/2}),\text{\textrm{\ }}\forall\mu,\;\varkappa\geq1\text{\textrm{\ or }%
}\mu=\pi/2,\;\,\varkappa\in(0,1)\\
D\widetilde{B}(\mu,w)(\upsilon x)^{1/2-\varkappa}[1+O(x^{2\varkappa
})]+O(x^{1/2}),\\
\ \mu\in\lbrack0,\pi/2),\;\varkappa\in(0,1)
\end{array}
\right.  \,,\,x\rightarrow0,\label{2.2.4.16}%
\end{equation}
and by%
\begin{equation}
\chi(x)\rightarrow0,\,x\rightarrow\infty,\text{ }\forall\mu,\,\forall
\varkappa>0.\label{2.2.4.17}%
\end{equation}

As it follows from (\ref{2.2.4.13}), (\ref{2.2.4.14}) with $\eta(x)=0$, the
kernel of $\hat{b}^{+}(\mu,w)$ \ is nontrivial only for $\mu=0$, $\varkappa
\in(0,1)$:%
\begin{equation}
\ker\hat{b}^{+}(\mu,w)=\left\{
\begin{array}
[c]{l}%
\{0\},\,\forall\mu,\,\varkappa\geq1\;\mathrm{or\;\,}\mu\in(0,\pi
/2],\varkappa\in(0,1)\\
\{c\left(  \phi(0,w;x)=\mathrm{e}^{-(\upsilon x)^{2}/2}(\upsilon
x)^{1/2+\varkappa\,}\frac{\Gamma(\alpha)}{\Gamma(\varkappa)}\Psi(\alpha
,\beta;(\upsilon x)^{2})\right)  \},\\
\,\mu=0,\varkappa\in(0,1)
\end{array}
\right.  .\label{2.2.4.17b}%
\end{equation}
A simple reason is that $\hat{b}^{+}(\mu,w)\chi(x)=\check{a}(\mu,w)
\chi(x)=0\Rightarrow\chi(x)=c\phi(\mu,w;x)\,$\ and $\phi(\mu,w;x)\,$%
(\ref{2.2.1.2}) is square integrable on $\mathbb{R}_{+}$ only for $\mu=0$ and
$\varkappa\in(0,1)$, see (\ref{2.2.1.5}) and (\ref{2.2.1.7}).

For $\varkappa\in(0,1)$, the domain $D_{b^{+}(\mu,w)}$ of the operator
$\hat{b}^{+}(\mu,w)$ with $\mu\in\lbrack0,\pi/2)$ can be represented as a
direct sum of the form
\begin{equation}
D_{b^{+}(\mu,w)}=\{D\chi_{0}(\mu,w;x)\}+\tilde{D}_{b^{+}(\mu,w)},\,\mu
\in\lbrack0,\pi/2),\,\varkappa\in(0,1),\label{2.2.4.18}%
\end{equation}
where the function $\chi_{0}(\mu,w;x)\,$belonging to $D_{b^{+}(\mu,w)}$ is
given by%
\begin{equation*}
\chi_{0}(\mu,w;x)=\phi(\mu,w;x)\zeta\left(  x\right)  ,\mathrm{\,}%
\text{\textrm{so that}}\ \ \check{a}(\mu,w)\chi_{0}(\mu,w;x)=\phi
(\mu,w;x)\zeta^{\prime}\left(  x\right)  ,
\end{equation*}
$\zeta\left(  x\right)  $ is a fixed smooth function with a compact support
equal to $1$ in a neighborhood of the origin, and $\tilde{D}_{b^{+}(\mu,w)}$
is the subspace of functions belonging to $D_{b^{+}(\mu,w)}$ and vanishing at
the origin:%
\begin{equation}
\tilde{D}_{b^+(\mu,w)}=\left\{\chi(x)\in D_{b^+(\mu,w)}:\,
\chi(x)=O(x^{1/2}),\;x\rightarrow0\right\},\,\mu\in[0,\pi/2),\,
\varkappa\in(0,1).\label{2.2.4.20}%
\end{equation}

iii) The operator $\overline{\hat{a}}(\mu,w)$ is evaluated in accordance with
(\ref{2.2.3.3}): as a restriction of $\hat{b}^{+}(\mu,w)$, this operator is
associated with $\check{a}(\mu,w)\,$and its domain belongs to or
coincides with $D_{b^{+}(\mu,w)}$, while the defining equation for
$\overline{\hat{a}}$ $(\mu,w)\,$as $(\hat{a}^{+}(\mu,w))^{+}$,%
\begin{equation*}
(\chi,\hat{a}^{+}(\mu,w)\psi)-(\overline{\hat{a}}(\mu,w)\chi,\psi
)=0,\,\chi(x)\in D_{\bar{a}(\mu,w)},\text{\thinspace}\forall\psi(x)\in
D_{a^{+}(\mu,w)},
\end{equation*}
is reduced to the equation for $D_{\bar{a}(\mu,w)}$, i.e., for the functions
$\chi(x)\in D_{\bar{a}(\mu,w)}\subseteq D_{b^{+}(\mu,w)}$, of the form%
\begin{equation}
(\chi,\check{b}(\mu,w)\psi)-(\check{a}(\mu,w)\chi,\psi)=0,\,\chi(x)\in
D_{\bar{a}(\mu,w)}\subseteq D_{b^{+}(\mu,w)},\,\forall\psi(x)\in D_{a^{+}%
(\mu,w)}.\label{2.2.4.22}%
\end{equation}
Integrating by parts in $(\check{a}(\mu,w)\chi,\psi)$ and taking estimates
(\ref{2.2.4.6}), (\ref{2.2.4.7}) and (\ref{2.2.4.16}), (\ref{2.2.4.17}) into
account, we establish that for $\mu\in\lbrack0,\pi/2)$ $\varkappa\geq1$ and
for $\mu=\pi/2,\,\forall\varkappa>0$, eq. (\ref{2.2.4.22}) holds identically
for all$\,\chi(x)\in$ $D_{b^{+}(\mu,w)}$, while for $\mu\in\lbrack0,\pi/2),$
$\,\varkappa\in(0,1)$, eq. (\ref{2.2.4.22}) is reduced to
\begin{equation*}
\overline{D}C=0,\text{ }\forall C,
\end{equation*}
which requires that $D=0$.

We finally obtain that
\begin{equation}\label{2.2.4.23}
\begin{array}{ll}
&  \overline{\hat{a}}(\mu,w)=\hat{b}^{+}(\mu,w),\,D_{\bar{a}(\mu,w)}=
\mathrm{\,}D_{b^{+}(\mu,w)}\text{\textrm{(\ref{2.2.4.12})}},\\
&  \text{\textrm{for} }\forall\mu\,,\;\varkappa\geq1
\text{\textrm{\ and\thinspace}}\mu=\pi/2,\;\varkappa\in(0,1),%
\end{array}
\end{equation}
and
\begin{equation}
\overline{\hat{a}}(\mu,w)\subset\hat{b}^{+}(\mu,w),\,D_{\bar{a}(\mu,w)}=
\tilde{D}_{b^{+}(\mu,w)}\,\text{\textrm{(\ref{2.2.4.20})},}\,\,
\text{\textrm{for} }\mu\in\lbrack0,\pi/2),\;
\varkappa\in(0,1),\label{2.2.4.24}%
\end{equation}
which, in particular, implies that for $\varkappa\in(0,1)$, \ the asymptotic
behavior at the origin of the functions belonging to the domain of any
operator $\overline{\hat{a}}(\mu,w)$ is estimated as $O(x^{1/2})$,%
\begin{equation}
D_{\bar{a}(\mu,w)}\ni\chi(x)=O(x^{1/2}),\,x\rightarrow0,\,\mathrm{for\,}%
\forall\mu,\forall w>w_{0},\,\varkappa\in(0,1),\label{2.2.4.24a}%
\end{equation}
for $\mu=\pi/2$ this follows from (\ref{2.2.4.23})\thinspace and
(\ref{2.2.4.16}), while for $\mu\in\lbrack0,\pi/2)\,$\ this follows
from (\ref{2.2.4.24})\thinspace and (\ref{2.2.4.20}).

The kernel of any operator $\overline{\hat{a}}(\mu,w)$\thinspace is trivial,%
\begin{equation}
\ker\overline{\hat{a}}(\mu,w)=\{0\},\,\forall\mu,\,\forall\,w>w_{0}%
,\forall\varkappa>0,\label{2.2.4.24b}%
\end{equation}
for $\varkappa\geq1$ and $\mu=\pi/2,\;\varkappa\in(0,1)$ this follows from
(\ref{2.2.4.23})\thinspace and (\ref{2.2.4.17b}), while for
$\mu\in[0,\pi/2),\,\varkappa\in(0,1)$, this follows from that $\overline{\hat{a}}%
(\mu,w)\chi(x)=$ $\check{a}(\mu,w)\chi(x)=0\Rightarrow\chi(x)=c\,\phi
(\mu,w;x)$, but the function $\phi(\mu,w;x)$ is estimated at the origin as
$O(x^{1/2-\varkappa})$, see (\ref{2.2.1.5}), and therefore cannot belong to
$D_{\bar{a}(\mu,w)}$ according to (\ref{2.2.4.24a})
(in addition, $\phi(\mu,w;x)$ with $\mu\in(0,\pi/2)$ is not square
integrable on $\mathbb{R}_{+}$, see (\ref{2.2.1.7})).

iv) Quite similarly, we find
\begin{equation}
\overline{\hat{b}}(\mu,w)=\hat{a}^{+}(\mu,w),\,D_{\bar{b}(\mu,w)}%
=D_{a^{+}(\mu,w)}\,\text{\textrm{(\ref{2.2.4.1})}},
\text{\textrm{for}}\,\forall\mu,\;\varkappa\geq1 \text{\textrm{\
and\thinspace}}\,\mu=\pi/2,\;
\varkappa\in(0,1),\label{2.2.4.25}%
\end{equation}
and
\begin{equation}
\overline{\hat{b}}(\mu,w)\subset\hat{a}^{+}(\mu,w),\,D_{\bar{b}(\mu,w)}%
=\tilde{D}_{a^{+}(\mu,w)}\mathrm{\,(\ref{2.2.4.10})},\text{ }\,\mu\in
\lbrack0,\pi/2),\varkappa\in(0,1),\label{2.2.4.26}%
\end{equation}
which, in particular, implies that for $\varkappa\in(0,1)$, \ the asymptotic
behavior at the origin of the functions belonging to the domain of any
operator $\overline{\hat{b}}(\mu,w)$ with $\mu\in\lbrack0,\pi/2)\,$is
estimated as $O(x^{1/2})$,%
\begin{equation}
D_{\bar{b}(\mu,w)}\ni\psi(x)=O(x^{1/2}),\,x\rightarrow0,\,\mathrm{for\,}\mu
\in\lbrack0,\pi/2),\;w\in(w_{0},\infty),\;\,\varkappa\in
(0,1).\label{2.2.4.26a}%
\end{equation}

We note that equality (\ref{2.2.4.25}) and inclusion (\ref{2.2.4.26}) directly
follow from the respective previous equality (\ref{2.2.4.23}) and inclusion
(\ref{2.2.4.24}) by taking the adjoints, and only the domain $D_{\bar{b}%
(\mu,w)}$ in the last case has to be evaluated.

It is easy to prove that there are no other pairs of closed mutually
adjoint operators that are extensions of each pair $\hat{a}(\mu,w),\,\hat
{b}(\mu,w)$. Indeed, let $\hat{g},\,\hat{g}^{+}$ be such a pair, then because
$\overline{\hat{a}}(\mu,w)$ and $\overline{\hat{b}}(\mu,w)$ are minimum closed
extensions of the respective $\hat{a}(\mu,w)$ and $\,\hat{b}(\mu,w)$, we have%
\[
\hat{a}(\mu,w)\subset\overline{\hat{a}}(\mu,w)\subseteq\hat{g}=\overline
{\hat{g}}=(\,\hat{g}^{+})^{+},\,\hat{b}(\mu,w)\subset\overline{\hat{b}}%
\,(\mu,w)\subseteq\hat{g}^{+}.
\]
It follows by taking the adjoints of these inclusions that
\[
\hat{g}^{+}\subseteq\hat{a}^{+}(\mu,w),\,\hat{g}\subseteq\hat{b}^{+}(\mu,w),
\]
so that we finally have
\[
\overline{\hat{a}}(\mu,w)\subseteq\hat{g}\subseteq\hat{b}^{+}(\mu,w),
\]
in particular,
\[
D_{\overline{a}(\mu,w)}\subseteq D_{g}\subseteq D_{b^{+}(\mu,w)}.
\]
It then directly follows from (\ref{2.2.4.23}) that $\hat{g}=\overline
{\hat{a}}(\mu,w)=\hat{b}^{+}(\mu,w)$ and therefore $\hat{g}^{+}=\overline
{\hat{b}}(\mu,w)=\hat{a}^{+}(\mu,w)$ for $\mu\in\lbrack0,\pi/2),\,\varkappa
\geq1$ and for$\,\mu=\pi/2,\,\forall\,\varkappa>0$, while for %
$\mu\in\lbrack0,\pi/2),$ $\varkappa\in(0,1)$, it follows from (\ref{2.2.4.24}%
), (\ref{2.2.4.18}) that the domains $D_{b^{+}(\mu,w)}$ and $D_{\overline
{a}(\mu,s)}=\tilde{D}_{b^{+}(\mu,s)}$ differ by a one-dimensional subspace, so
that either $\hat{g}=\overline{\hat{a}}(\mu,w)$, and therefore $\hat{g}^{+}%
=\hat{a}^{+}(\mu,w)$, or $\hat{g}=\hat{b}^{+}(\mu,w)$, and therefore
$\hat{g}^{+}=\overline{\hat{b}}(\mu,w)$.

We thus show that each pair $\check{a}(\mu,w)$, $\check{b}(\mu,w) $,
$\mu\in\lbrack0,\pi/2]$, $w\in(w_{0},\infty)$,$\ w_{0}=-\frac{1}%
{2}(1+\varkappa)$, $\varkappa>0$, of mutually adjoint\ by Lagrange
differential operations (\ref{2.2.1.3}) providing generalized oscillator
representation (\ref{2.2.1.4}) with $\,w\in(w_{0},\infty)$ for $\check{H}%
$\ (\ref{1.1}) with $g_{1}>-1/4\,$ and $g_{2}>0$ generates a unique %
pair $\overline{\hat{a}}(\mu,w)=\hat{b}^{+}(\mu,w)$, $\ \hat{a}^{+}%
(\mu,w)=\overline{\hat{b}}(\mu,w)$ of closed mutually
adjoint\textrm{\thinspace} operators for $\,\mu\in\lbrack0,\pi/2),\,\varkappa
\geq1$ and for $\mu=\pi/2,\,\forall\,\varkappa>0$, while for $\mu\in
\lbrack0,\pi/2)$,$\,\ \varkappa\in(0,1)$, each pair $\check{a}(\mu,w)$,
$\check{b}(\mu,w)\,$generates two different pairs $\overline{\hat{a}}%
(\mu,w),\ \hat{a}^{+}(\mu,w)$ and $\,\hat{b}^{+}(\mu,w)$, $\overline{\hat{b}%
}(\mu,w)$ of closed mutually adjoint\textrm{\thinspace} operators such that
$\overline{\hat{a}}(\mu,w)\subset$ $\hat{b}^{+}(\mu,w)$ and $\overline{\hat
{b}}(\mu,w)\subset$ $\hat{a}^{+}(\mu,w)$. The operators $\overline
{\hat{a}}(\mu,w) $\ and $\hat{b}^{+}(\mu,w)$ are extensions of the initial
operator $\hat{a}(\mu,w)$, they are associated with $\check{a}(\mu,w)$, and
their domains are given by the respective (\ref{2.2.4.23}),(\ref{2.2.4.24})
and (\ref{2.2.4.12}). The operators $\overline{\hat{b}}(\mu,w)$\ and
$\hat{a}^{+}(\mu,w)$\ are extensions of the initial operator $\hat{b}(\mu,w)$,
they are associated with $\check{b}(\mu,w)$, and their domains are given by
the respective (\ref{2.2.4.25}),(\ref{2.2.4.26}) and (\ref{2.2.4.1}).

\subsection{Region $g_{1}>-1/4$ ($\varkappa>0$),\ $w=w_{0}=-\frac{1}%
{2}(1+\varkappa)\,$ ($\alpha=0$)}

A consideration for this region of parameters follows the standard scheme
presented in the previous subsec. 2.2 where $w>w_{0}$. The
distinctive feature is that here, we encounter a unique pair of
differential operations $\check{a}(u)$ and $\check{b}(u)$ providing
generalized oscillator representation (\ref{1.6a}) with $u=u_{0}=4\upsilon
^{2}w_{0}$ for $\check{H}$ (\ref{1.1}) with $g_{1}>-1/4$ and $g_{2}>0$.

\subsubsection{Generalized oscillator representations for $\check{H}$,
differential operations $\check{a}$ and $\check{b}$}

In this region of parameters, the general solution of differential equation
(\ref{2.1.3}) with $u=u_{0}=4\upsilon^{2}w_{0}$ is given by (\ref{2.1.24}).
The function $\Psi(\beta,\rho)$ given by (\ref{2.1.21}) increases
monotonically from $-\infty$ to $\infty$ as $\rho=(\upsilon x)^{2}$ together
with $x$ ranges from $0$ to $\infty$. It follows that eq.
(\ref{2.1.3}) with these values of parameters has a unique, up to a positive
constant factor, real-valued positive solution
\begin{equation}
\phi(w_{0};x)=\mathrm{e}^{-\rho/2}\rho^{1/4+\varkappa/2},\,\rho=(\upsilon
x)^{2},\label{2.3.1.1}%
\end{equation}
which implies, see the text after (\ref{2.1.6}), that in this case, we have a
unique\footnote{Up to irrelevant phase factors.} pair of mutually adjoint
first-order differential operations $\check{a}(w_{0})$ and $\check{b}(w_{0})$
given by (\ref{2.1.6}) with the evident substitutions $\check{a}%
(u)\rightarrow\check{a}(w_{0})$, $\check{b}(u)\rightarrow$ $\check{b}(w_{0}),$
$h(u;x)\rightarrow h(w_{0};x)$ and $\phi(u;x)\rightarrow\phi(w_{0};x)$:%
\begin{equation}\label{2.3.1.2}
\begin{array}{ll}
\check{a}(w_{0})=\phi(w_{0};x)d_{x}\frac{1}{\phi(w_{0};x)}=
\check{b}^{\ast}(w_{0}),\\
\check{b}(w_{0})=-\frac{1}{\phi(w_{0};x)}d_{x}\phi(w_{0};x)=
\check{a}^{\ast}(w_{0}),\;\varkappa>0,%
\end{array}
\end{equation}
and providing unique generalized oscillator representation (\ref{1.6a})
with $u=4\upsilon^{2}w_{0}$ for generalized Calogero differential operation
$\check{H}\,$\ (\ref{1.1}) with $g_{1}>-1/4$ and $g_{2}>0$:%
\begin{equation}
\check{H}=-d_{x}^{2}+g_{1}x^{-2}+g_{2}x^{2}=\check{b}(w_{0})\check{a}%
(w_{0})-4\upsilon^{2}w_{0},\,\varkappa>0.\label{2.3.1.3}%
\end{equation}

The asymptotic behavior of the functions $\phi(w_{0};x)$ and $1/\phi(w_{0};x)$
at the origin and at infinity is evident from (\ref{2.3.1.1}).

We note that all the results to follow in this subsection can be obtained
from the corresponding results of the previous subsection for $\mu=\pi/2$ in
the limit $w\rightarrow w_{0}$.

\subsubsection{Initial operators $\hat{a}$ and $\hat{b}$}

Following subsubsec. 2.2.2, we introduce the pair of initial differential
operators $\hat{a}(w_{0})$ and $\hat{b}(w_{0})\,$in $L^{2}(\mathbb{R}_+)$
associated with respective differential operations $\check{a}(w_{0})$ and
$\check
{b}(w_{0})$ (\ref{2.3.1.2}) and defined on the subspace $\mathcal{D}%
(\mathbb{R}_{+})$ of smooth compactly supported functions,
$D_{a(w_{0})}=D_{b(w_{0})}=\mathcal{D}(\mathbb{R}_{+})$. These operators
satisfy the relation%
\begin{equation}
\left(  \psi,\hat{a}(w_{0})\xi\right)  =\left(  \hat{b}(w_{0})\psi,\xi\right)
,\;\forall\psi(x),\xi(x)\in\mathcal{D}(\mathbb{R}_{+}).\label{2.3.2.1}%
\end{equation}
The representation (\ref{2.3.1.3}) for $\check{H}$ provide the
representation
\begin{equation}
\hat{H}=\hat{b}(w_{0})\,\hat{a}(w_{0})-4\upsilon^{2}w_{0}\,\hat{I},\,\,\forall
\varkappa>0,\label{2.3.2.2}%
\end{equation}
for the initial symmetric operator $\hat{H}$ $\,$associated with $\check{H}$
and defined on $\mathcal{D}(\mathbb{R}_{+})$, which, in particular, confirms
that $\hat{H}$ with $g_{1}>-1/4,$ $g_{2}>0$ is bounded from below by
$-4\upsilon^{2}w_{0}=2\upsilon^{2}(1+\varkappa)$,$\,$
\begin{equation*}
\left(  \xi,\hat{H}\xi\right)  =\left(  \hat{a}(w_{0})\xi,\hat{a}(w_{0}%
)\xi\right)  -4\upsilon^{2}w_{0}(\xi,\xi)\geq2\upsilon^{2}\,(1+\varkappa
)\left(  \xi,\xi\right)  ,\;\forall\xi(x)\in\mathcal{D}(\mathbb{R}%
_{+}).
\end{equation*}
This representation is a basis for constructing, maybe new, s.a. generalized
Calogero Hamiltonians $\hat{H}_{\mathfrak{e}}$ as s.a. extensions of $\hat{H}
$ (\ref{2.3.2.2}) in generalized oscillator form (\ref{1.8}) or (\ref{1.9})
via constructing all possible extensions of the pair $\hat{a}(w_{0})$,
$\hat{b}(w_{0})$ of initial operators to a pair of closed mutually adjoint
operators $\hat{c}(w_{0})$, $\hat{c}^{+}(w_{0})$, $\hat{a}(w_{0}%
)\subset\hat{c}(w_{0})$, $\hat{b}(w_{0})\subset\hat{c}^{+}(w_{0})$, beginning
from the closure $\overline{\hat{a}}(w_{0})$ of $\hat{a}(w_{0})$, or to a pair
of closed mutually adjoint operators $\hat{d}(w_{0})$, $\hat{d}^{+}(w_{0})$,
$\hat{a}(w_{0})\subset\hat{d}^{+}(w_{0})$, $\hat{b}(w_{0})\subset\hat{d}%
(w_{0})$, beginning from the closure $\overline{\hat{b}}(w_{0})$ of $\hat
{b}(w_{0})$.

\subsubsection{Adjoint operators $\hat{a}^{+}$ and $\hat{b}^{+}$,
closed operators $\overline{\hat{a}}$ and $\overline{\hat{b}}$}

By arguments completely similar to those in the previous subsubsecs. 2.2.3 and
2.2.4, we establish that the closures $\overline{\hat{a}}(w_{0})$ and
$\overline{\hat{b}}(w_{0})\,$of the operators $\hat{a}(w_{0})$ and $\hat
{b}(w_{0})$, as well as the adjoints $\hat{a}^{+}(w_{0})$ and $\hat{b}%
^{+}(w_{0})$ of the latters, do exist and the chains of inclusions%
\begin{equation}\label{2.3.3.1}
\begin{array}{ll}
\hat{a}(w_{0})\subset\overline{\hat{a}}(w_{0})=(\hat{a}^{+}(w_{0}))^+
\subseteq\hat{b}^{+}(w_{0}),\\
\hat{b}(w_{0})\subset\overline{\hat{b}}(w_{0})=
(\hat{b}^{+}(w_{0}))^{+}\subseteq\hat{a}^{+}(w_{0}),%
\end{array}
\end{equation}
hold. The operators $\overline{\hat{a}}(w_{0})$ and $\hat{b}^{+}(w_{0})$ are
associated with $\check{a}(w_{0})$, while the operators $\overline{\hat{b}%
}(w_{0})$ and $\hat{a}^{+}(w_{0})$ are associated with $\check{b}(w_{0})$, so
that to specify these operators, it is sufficient to evaluate their domains
following the method in subsubsec. 2.2.4.

\subsubsection{Domains of operators $\hat{a}^{+}$, $\hat{b}^{+}$,
$\overline{\hat{a}}$ and $\overline{\hat{b}}$}

i) The domain of the operator $\hat{a}^{+}$ $(w_{0})$ is the natural domain
for $\check{b}(w_{0})$,$\,$which is given by a copy of (\ref{2.2.4.1}) with
the substitution $\check{b}(\mu,w)\rightarrow\check{b}(w_{0})$:
\begin{align}
& D_{a^{+}(w_{0})}=D_{\check{b}(w_{0})}^{n}=\{\psi(x):\psi\;\text{\textrm{is
a.c. in} }\mathbb{R}_{+},\nonumber\\
& \psi(x),\check{b}(w_{0})\psi(x)=-\frac{1}{\phi(w_{0};x)}\frac{d}{dx}\left(
\phi(w_{0};x)\psi(x)\right)  =\eta(x)\in L^{2}(\mathbb{R}_{+}%
)\}.\label{2.3.4.1}%
\end{align}
By arguments similar to those in the item i), subsubsec. 2.2.4, it follows
that a generic function $\psi(x)$ belonging to $D_{a^{+}(w_{0})}$ allows
the representations%
\begin{equation}\label{2.3.4.2}
\begin{array}{ll}
\psi(x)=-\frac{1}{\phi(w_{0};x)}\int_{0}^{x}dy\,\phi(w_{0};y)
\eta(y)=\frac{1}{\phi(w_{0},x)}\int_{x}^{\infty}dy\,\phi(w_{0};y)
\eta(y),\\
\eta(x)=\check{b}_{(w_{0})}\psi(x)\in L^{2}(\mathbb{R}_{+}),\
\text{\textrm{and\ }}\int_{0}^{\infty}dy\,\phi(w_{0};y)\eta(y)=0.%
\end{array}
\end{equation}
The latter equality in (\ref{2.3.4.2}) means that the range $R_{a^{+}(w_{0})}%
$of the operator $\hat{a}^{+}(w_{0})$ is orthogonal to the one-dimensional
subspace $\{c\phi(w_{0};x)\}\subset L^{2}(\mathbb{R}_{+}),$ $R_{a^{+}(w_{0}%
)}\bot\{c\phi(w_{0};x)\}$, and thereby its closure $\overline{R_{a^{+}(w_{0}%
)}}\,$can not be the whole $L^{2}(\mathbb{R}_{+})$, which in turn implies
that the kernel of the adjoint operator $(\hat{a}^{+}(w_{0}))^{+}=\overline
{\hat{a}}(w_{0})$ is not trivial, $\{c\phi(w_{0};x)\}\subseteq\ker
\overline{\hat{a}}(w_{0})\neq\{0\}$, see below.

Estimating the integral terms in (\ref{2.3.4.2}) with the
Cauchy--Bunyakovskii inequality, we obtain that the asymptotic
behavior of functions $\psi(x)\in$ $D_{a^{+}(w_{0})}$ at the origin and at
infinity is respectively is given by%
\begin{equation}\label{2.3.4.3}
\begin{array}{ll}
\psi(x)  & =O(x^{1/2}),\,x\rightarrow0,\\
\psi(x)  & \rightarrow0,\;x\rightarrow\infty,\,\forall\varkappa>0.
\end{array}
\end{equation}
\
ii) The domain $D_{b^{+}(w_{0})}$ of the operator $\hat{b}^{+}(w_{0})$ is the
natural domain for $\check{a}(w_{0})$, which is given by a copy of
(\ref{2.2.4.12}) with the substitution $\check{a}(\mu,w)\rightarrow\check
{a}(w_{0})$:
\begin{align}
& D_{b^{+}(w_{0})}=D_{\check{a}(w_{0})}^{n}=\{\chi(x):\chi\;\text{\textrm{is
a.c. in} }\mathbb{R}_{+},\,\nonumber\\
& \chi(x),\,\check{a}(w_{0})\chi(x)=\phi(w_{0};x)d_{x}\left(  \frac{1}%
{\phi(w_{0};x)}\chi(x)\right)  =\eta(x)\in L^{2}(\mathbb{R}_{+}%
)\},\label{2.3.4.5}%
\end{align}
whence it follows that a generic function $\chi(x)\in$ $D_{b^{+}(w_{0})}$
allows the representation%
\begin{equation}\label{2.3.4.6}
\chi(x)=\phi(w_0;x)[D+\int_{x_0}^xdy\frac{1}{\phi(w_{0};y)}\eta(y)],
\;\eta(x)=\check{b}(w_{0})\chi(x)\in L^{2}(\mathbb{R}_{+}),
\end{equation}
where $x_{0}\in(0,\infty)$ and $D$ is an arbitrary constant, and its
asymptotic behavior at the origin and at infinity estimated using the
Cauchy--Bunyakovskii inequality is respectively given by
\begin{equation}\label{2.3.4.7}
\begin{array}{ll}
\chi(x)  & =O(x^{1/2}),\;x\rightarrow0,\\
\chi(x)  & \rightarrow0,\;x\rightarrow\infty,\text{ }\forall\varkappa>0.%
\end{array}
\end{equation}

As it follows from (\ref{2.3.4.6}) with $\eta(x)=0$ and (\ref{2.3.1.1}), the
kernel of the operator $\hat{b}^{+}(w_{0})$ is nontrivial,%
\begin{equation*}
\ker\hat{b}^{+}(w_{0})=\{c\left(  \phi(w_{0};x)=
(\upsilon x)^{1/2+\varkappa}\mathrm{e}^{-(\upsilon x)^{2}/2}\right)
  \},\,\forall\varkappa>0.
\end{equation*}

iii) The domain $D_{\bar{a}(w_{0})}$ of the operator $\overline{\hat{a}}%
(w_{0})$ is evaluated in accordance with (\ref{2.3.3.1}) using arguments
similar to those in the item iii), subsubsec. 2.2.4: the defining equation for
\textrm{\thinspace}$\overline{\hat{a}}(w_{0})$ as $(\hat{a}^{+})^{+}(w_{0})$,
which is a restriction of $\hat{b}^{+}(w_{0})$, is reduced to the equation for
$D_{\bar{a}(w_{0})}\subseteq$ $D_{b^{+}(w_{0})}$ of the form%
\begin{equation}
(\chi,\check{b}(w_{0})\psi)-(\check{a}(w_{0})\chi,\psi)=0,\,\chi(x)\in
D_{\bar{a}(w_{0})}\subseteq D_{b^{+}(w_{0})},\,\forall\psi(x)\in
D_{a^{+}(w_{0})}.\label{2.3.4.9}%
\end{equation}
Integrating by parts in $(\check{a}(w_{0})\chi,\psi)$ and taking asymptotic
estimates (\ref{2.3.4.3}) and (\ref{2.3.4.7}) into account, we establish
that eq. (\ref{2.3.4.9}) holds identically for all$\,\chi(x)\in
D_{b^{+}(w_{0})}$, which implies that
\begin{equation}
\overline{\hat{a}}(w_{0})=\hat{b}^{+}(w_{0}),\,\text{\textrm{\ }%
}D_{\bar{a}(w_{0})}=\mathrm{\,}D_{b^{+}(w_{0})}\,\text{\textrm{(\ref{2.3.4.5}%
)}},\text{\textrm{\thinspace}}\forall\varkappa>0,\label{2.3.4.10}%
\end{equation}
in particular, the asymptotic behavior at the origin of the functions
$\chi(x)$ belonging to the domain of the operator $\overline{\hat{a}}(w_{0})$
is estimated by a copy of (\ref{2.3.4.7}),%
\begin{equation}
\chi(x)=O(x^{1/2}),\;x\rightarrow0,\,\forall\varkappa>0,\label{2.3.4.10a}%
\end{equation}
\ \ and
\begin{equation}
\ker\overline{\hat{a}}(w_{0})=\{c\left(  \phi(w_{0};x)=(\upsilon
x)^{1/2+\varkappa}\mathrm{e}^{-(\upsilon x)^{2}/2}\right)  \},\,\forall
\varkappa>0.\label{2.3.4.10b}%
\end{equation}

iv) It directly follows from (\ref{2.3.4.10}) by taking the adjoints that
\begin{equation}
\overline{\hat{b}}(w_{0})=\hat{a}^{+}(w_{0}),\,D_{\bar{b}(w_{0})}%
=D_{a^{+}(w_{0})}\mathrm{\,}\text{\textrm{(\ref{2.3.4.1})}},\forall
\varkappa>0.\label{2.3.4.11}%
\end{equation}

By arguments similar to those in the end of subsubsec. 2.2.4, it is easy to
prove that there is no other pair $\hat{g}=\overline{\hat{g}}$ and
$\,\hat{g}^{+}$of closed mutually adjoint operators that are extensions of the
respective $\hat{a}(w_{0})$ and $\hat{b}(w_{0})$, $\hat{a}(w_{0})\subset
\hat{g}\,,\hat{b}(w_{0})\subset\hat{g}\,^{+}$.

We thus show that the pair $\check{a}(w_{0}),\check{b}(w_{0})$ of mutually
adjoint\ by Lagrange differential operations (\ref{2.3.1.2}) providing
unique generalized oscillator representation (\ref{2.3.1.3}) for
$\check{H}$\ (\ref{1.1}) with $g_{1}>-1/4$, $g_{2}>0$ generates a unique
pair $\overline{\hat{a}}(w_{0})=\hat{b}^{+}(w_{0})$, $\ \hat{a}^{+}(w_{0})$
$=\overline{\hat{b}}(w_{0})$ of closed mutually adjoint operators. The
operator $\overline{\hat{a}}(w_{0})=$ $\hat{b}^{+}(\mu,w)$ is an extension
of the initial operator $\hat{a}(w_{0})$, it is associated with
$\check{a}(w_{0})$, and its domain is given by
(\ref{2.3.4.5}). The operator $\overline{\hat{b}}(w_{0})=\hat{a}^{+}(w_{0}%
)$\ is an extension of the initial operator $\hat{b}(w_{0})$, it is associated
with $\check{b}(w_{0})$, and its domain is given by (\ref{2.3.4.1}).

\subsection{Region $g_{1}=-1/4$ ($\varkappa=0$), $w>w_{0}=-1/2$
\ ($\alpha>0$)}

A consideration for this region of parameters literally follows the scheme of
subsec. 2.2, we even do not change the notation having in mind that here
$\varkappa=0,$ $\beta=1$.

The only distinction is in a specific asymptotic behavior of the real-valued
positive functions $\phi_{1}$ and $\,\phi_{2}$ (\ref{2.1.23}) with $\beta=1
$, the fundamental solutions of eq. (\ref{2.1.3}) with
$g_{1}=-1/4\,(\varkappa =0)$, at the origin.

\subsubsection{Generalized oscillator representations for $\check{H}$,
differential operations $\check{a}$ and $\check{b}$}

The general real-valued positive solution of eq. (\ref{2.1.3}) with
$g_{1}=-1/4$ $(\varkappa=0)\,$and $u=4\upsilon^{2}w>-2\upsilon^{2}$%
,$\,$defined modulo a positive constant factor, is given by
\begin{align}
& \phi(\mu,w;x)=\,\mathrm{e}^{-\rho/2}\rho^{1/4}\left[  \Phi(\alpha
,1;\rho)\sin\mu+\Gamma(\alpha)\Psi(\alpha,1;\rho)\,\cos\mu\right]
,\,\label{2.4.1.1}\\
& \mu\in\lbrack0,\pi/2],\,\alpha=1/2+w>0,\ \rho=(\upsilon x)^{2}%
,\nonumber%
\end{align}
where $\,\Phi(\alpha,1;\rho)$ and $\Psi(\alpha,1;\rho)$ are given by the
respective (\ref{2.1.15}) with $\beta=1$ and (\ref{2.1.17a}).

This implies that in this case, we have the two-parameter family $\{\check
{a}(\mu,w),\check{b}(\mu,w)$; $\mu\in\lbrack0,\pi/2]$, $w\in(-1/2,\infty)\}$
of different pairs of mutually adjoint first-order differential operations
$\check{a}(\mu,w)$ and $\check{b}(\mu,w)$ given by a copy of (\ref{2.2.1.3})
with $\phi(\mu,w;x)$ (\ref{2.4.1.1}) instead of $\phi(\mu,w;x)$ (\ref{2.2.1.2}%
) and providing a two-parameter family of different generalized oscillator
representations for generalized Calogero differential operation $\check{H}%
\,$\ (\ref{1.1}) with $g_{1}=-1/4,g_{2}>0$,
\begin{align}
\check{H}  & =-d_{x}^{2}-\frac{1}{4}x^{-2}+g_{2}x^{2}=\check{b}(\mu
,w)\check{a}(\mu,w)-4\upsilon^{2}w,\label{2.4.1.2}\\
\mu & \in[0,\pi/2],\,w\in(-1/2,\infty),\nonumber%
\end{align}
which are copies of (\ref{2.2.1.4}) with $\varkappa=0$. According to
(\ref{2.1.19a}), (\ref{2.1.19b}) and (\ref{2.1.19c}), (\ref{2.1.19d}), the
asymptotic behavior of functions $\phi(\mu,w;x)$ (\ref{2.4.1.1}) and
$1/\phi(\mu,w;x)$ at the origin is respectively given by%
\begin{align}
& \phi(\mu,w;x)=\left\{
\begin{array}
[c]{l}%
\tilde{A}\text{ }(\upsilon x)^{1/2}+\tilde{B}\,(\upsilon x)^{1/2}\ln(\upsilon
x)+O(x^{5/2}\ln x),\,\mu\in\lbrack0,\pi/2)\\
(\upsilon x)^{1/2}+O(x^{5/2}),\,\mu=\pi/2
\end{array}
\right.  ,\,x\rightarrow0,\label{2.4.1.3}\\
& \tilde{A}\text{ }=\tilde{A}\text{ }(\mu,w)=\sin\mu+\cos\mu(2\psi
(1)-\psi(\alpha)),\alpha=1/2+w>0,\nonumber\\
& \tilde{B}=\tilde{B}(\mu,w)=-2\cos\mu,\,\,\nonumber%
\end{align}

where $\psi(z)$ $=\Gamma^{\prime}(z)/\Gamma(z)$, and
\begin{equation}
\frac{1}{\phi(\mu,w;x)}=\left\{
\begin{array}
[c]{l}%
\frac{1}{\tilde{B}(\mu,w)}\frac{(\upsilon x)^{-1/2}}{\ln(\upsilon x)}%
\frac{1}{1+(\tilde{A}/\tilde{B}\text{ })/\ln(\upsilon x)}+O(x^{3/2}/\ln
x),\,\,\mu\in\lbrack0,\pi/2)\\
(\upsilon x)^{-1/2}+O(x^{3/2}),\,\mu=\pi/2
\end{array}
\right.  ,\,x\rightarrow0,\label{2.4.1.4}%
\end{equation}
while their asymptotic behavior at infinity is respectively given by
\begin{equation}
\phi(\mu,w;x)=\left\{
\begin{array}
[c]{c}%
\sin\mu\frac{1}{\Gamma(1/2+w)}(\upsilon x)^{-1/2+2w}e^{\frac{1}{2}(\upsilon
x)^{2}}(1+O(x^{-2})),\,\mu\in(0,\pi/2]\\
\Gamma(1/2+w)(\upsilon x)^{-1/2-2w}e^{-\frac{1}{2}(\upsilon x)^{2}}%
(1+O(x^{-2})),\,\mu=0
\end{array}
\right.  ,\,x\rightarrow\infty,\label{2.4.1.5}%
\end{equation}
and
\begin{equation}
\frac{1}{\phi(\mu,w;x)}=\left\{
\begin{array}
[c]{l}%
\frac{\Gamma(1/2+w)}{\sin\mu}(\upsilon x)^{1/2-2w}e^{-\frac{1}{2}(\upsilon
x)^{2}}(1+O(x^{-2})),\,\mu\in(0,\pi/2]\\
\frac{1}{\Gamma(1/2+w)}(\upsilon x)^{1/2+2w}e^{\frac{1}{2}(\upsilon x)^{2}%
}(1+O(x^{-2})),\,\mu=0
\end{array}
\right.  ,\,x\rightarrow\infty.\label{2.4.1.6}%
\end{equation}

\subsubsection{Initial operators $\hat{a}$ and $\hat{b}$}

In perfect analogy to subsubsec. 2.2.2, we introduce the pairs $\hat{a}(\mu
,w)$, $\hat{b}(\mu,w)$ of initial differential operators \textrm{\thinspace
}defined on $\mathcal{D}(\mathbb{R}_{+})$ and associated with the respective
pairs of differential operations $\check{a}(\mu,w)$,$\check{b}(\mu,w)$.
These operators have the property that is the copy of (\ref{2.2.2.1}). The
representations (\ref{2.4.1.2}) imply that the initial symmetric operator
$\hat{H}$ defined on $\mathcal{D}(\mathbb{R}_{+})$ and associated with
$\check{H}$ can be represented as
\begin{equation}
\hat{H}=\hat{b}(\mu,w)\,\hat{a}(\mu,w)-4\upsilon^{2}w\,\hat{I},\,\forall\mu
\in\lbrack0,\pi/2],\forall w>-1/2,\label{2.4.2.1}%
\end{equation}
which, in particular, implies that $\hat{H}$ with $g=-1/4,\,g_{2}>0$ is
bounded from below by $2\upsilon^{2}$.

The representations (\ref{2.4.2.1}) is a basis for constructing s.a.
generalized Calogero Hamiltonians $\hat{H}_{\mathfrak{e}}$ with
$g=-1/4,\,g_{2}>0\,$as s.a. extensions of $\hat{H}$ in generalized oscillator
form copiing (\ref{2.2.2.4}) or (\ref{2.2.2.5}) via constructing all possible
extensions of each pair $\hat{a}(\mu,w)$, $\hat{b}(\mu,w)$ of initial
operators respectively to a pair of closed mutually adjoint operators
$\hat{c}(\mu,w)$, $\hat{c}^{+}(\mu,w)$ beginning from the pair $\overline
{\hat{a}}(\mu,w),\hat{a}^{+}(\mu,w)$, the closure of $\hat{a}(\mu,w)$ and its
adjoint, or to a pair of closed mutually adjoint operators $\hat{d}(\mu,w)$,
$\hat{d}^{+}(\mu,w)$ beginning from the pair$\overline{\hat{b}}(\mu,w),\hat
{b}^{+}(\mu,w)$, the closure of $\hat{b}(\mu,w)$ and its adjoint.

\subsubsection{Adjoint operators $\hat{a}^{+}$ and $\hat{b}^{+}$, closed
operators $\overline{\hat{a}}$ and $\overline{\hat{b}}$}

By arguments repeating those in subsubsecs. 2.2.3 and 2.2.4, we establish that
all these operators, the closures $\overline{\hat{a}}(\mu,w)$, $\overline
{\hat{b}}(\mu,w)$,\ as well as the adjoints $\hat{a}^{+}(\mu,w)$, $\hat{b}%
^{+}(\mu,w)$, do exist and satisfy the chains of inclusions%
\begin{equation}\label{2.4.3.1}
\begin{array}{ll}
\hat{a}(\mu,w)\subset\overline{\hat{a}}(\mu,w)=(\hat{a}^+(\mu,w))^+
\subseteq\hat{b}^{+}(\mu,w),\;\\
\hat{b}(\mu,w)\subset\overline{\hat{b}}(\mu,w)=(\hat{b}^+(\mu,w))^+
\subseteq\hat{a}^{+}(\mu,w),%
\end{array}
\end{equation}
the copies of (\ref{2.2.3.3}).

The operators in the first chain are associated with $\check{a}(\mu,w)$, while
the operators in the second chain are associated with $\check{b}(\mu,w)$, and
to specify these operators, it is sufficient to evaluate their domains
repeating the reasoning in subsubsec. 2.2.4.

\subsubsection{Domains of operators $\hat{a}^{+}$, $\hat{b}^{+}$,
$\overline{\hat{a}}$ and $\overline{\hat{b}}$}

In fact, the content of this subsubsection is a copy of that of subsubsec.
2.2.4 in the part related to the case of $\varkappa\in(0,1)$ with the
substitution of $\phi(\mu,w;x)$ (\ref{2.4.1.1}) ($\varkappa=0$) for $\phi
(\mu,w;x)$ (\ref{2.2.1.2})\ ($\varkappa\in(0,1)$).

i) The domain $D_{a^{+}(\mu,w)}$ of the operator $\hat{a}^{+}(\mu,w)$ is the
natural domain for $\check{b}(\mu,w)$, which is given by a copy of
(\ref{2.2.4.1}). A generic $\psi(x)$ belonging to $D_{a^{+}(\mu,w)}\,$allows
the representation%

\begin{equation}
\psi(x)=\frac{1}{\phi(\mu,w;x)}\left[  C-\int_{0}^{x}dy\,\phi(\mu
,w;y)\eta(y)\right]  ,\,\,\eta(x)=\check{b}_{(\mu,w)}\psi(x)\in L^{2}%
(\mathbb{R}_{+}),\label{2.4.4.1}%
\end{equation}
where the constant $C$ depends on the values of $\mu$ as follows:%
\begin{equation}\label{2.4.4.2}
\begin{array}{lll}
& \mu=0:C=\int_{0}^{\infty}dy\,\phi(\mu,w;y)\eta(y),\\
& \mu\in(0,\pi/2):C\,\text{\textrm{is an arbitrary constant}},\\
& \mu=\pi/2:C=0.%
\end{array}
\end{equation}
A subtlety is that for $\mu\in\lbrack0,\pi/2)$, the constant $C$ in
(\ref{2.4.4.2}) can take arbitrary values, but for $\mu\in(0,\pi/2)$, the
constant $C$ is independent of the function $\eta(x)$, while for $\mu=0$, it
is uniquely related to $\eta$, so that the representation (\ref{2.4.4.1}%
),(\ref{2.4.4.2}) with $\mu=0$ for $\psi(x)\in$ $D_{a^{+}(0,w)}$ is equivalent
to
\begin{equation}
\psi(x)=\frac{1}{\phi(0,w;x)}\int_{x}^{\infty}dy\,\phi(0,w;y)\eta
(y),\,\,\eta(x)=\check{b}_{(0,w)}\psi(x)\in L^{2}(\mathbb{R}_{+}%
),\mu=0.\label{2.4.4.3}%
\end{equation}

The asymptotic behavior of functions $\psi(x)\in$ $D_{a^{+}(\mu,w)}$ at the
origin and at infinity estimated using (\ref{2.4.1.3})-(\ref{2.4.1.6}) and the
Cauchy--Bunyakovskii inequality for integral terms in (\ref{2.4.4.1}) and
(\ref{2.4.4.3}) is respectively given by%
\begin{equation}
\psi(x)=\left\{
\begin{array}
[c]{l}%
\frac{C}{\tilde{B}(\mu,w)}\frac{(vx)^{-1/2}}{\ln(\upsilon x)}\left[  1+O(1/\ln
x)\right]  +O(x^{1/2}),\,\mu\in\lbrack0,\pi/2)\\
O(x^{1/2}),\,\mu=\pi/2
\end{array}
\right.  ,\,x\rightarrow0,\label{2.4.4.4}%
\end{equation}
and by%
\begin{equation}
\psi(x)\rightarrow0,\,x\rightarrow\infty,\text{ }\forall\mu.\label{2.4.4.5}%
\end{equation}

The domain $D_{a^{+}(\mu,w)}$ of the operator $\hat{a}^{+}(\mu,w)$ with
$\mu\in\lbrack0,\pi/2)$ can be represented as a direct sum of the
form %
\begin{equation*}
D_{a^{+}(\mu,w)}=\{C\psi_{0}(\mu,w;x)\}+\tilde{D}_{a^{+}(\mu,w)},\,\,\mu
\in\lbrack0,\pi/2),\,
\end{equation*}
where the function $\psi_{0}(\mu,w;x)\,$belonging to $D_{a^{+}(\mu,w)}$ is
given by%
\begin{equation*}
\psi_{0}(\mu,w;x)=\frac{1}{\phi(\mu,w;x)}\zeta\left(  x\right)  ,\mathrm{\,}%
\text{\textrm{so that}}\ \check{b}(\mu,w)\psi_{0}(\mu,w;x)=-\frac{1}{\phi
(\mu,w;x)}\zeta^{\prime}\left(  x\right)  ,\label{2.4.4.7}
\end{equation*}
$\zeta\left(  x\right)  $ is a fixed smooth function with a compact support
and equal to $1$ in a neighborhood of the origin, and $\tilde
{D}_{a^{+}(\mu,w)}$ is the subspace of functions belonging to $D_{a^{+}%
(\mu,w)}$ and vanishing at the origin:%
\begin{equation}
\tilde{D}_{a^{+}(\mu,w)}=\left\{  \psi(x)\in D_{a^{+}(\mu,w)}:\,\psi
(x)=O(x^{1/2}),\,x\rightarrow0\right\}  ,\,\mu\in\lbrack0,\pi
/2).\label{2.4.4.8}%
\end{equation}

ii) The domain $D_{b^{+}(\mu,w)}$ of the operator $\hat{b}^{+}(\mu,w)$ is the
natural domain for $\check{a}(\mu,w)$, which is given by a copy of
(\ref{2.2.4.12}). A generic $\chi(x)$ belonging to $D_{b^{+}(\mu,w)}\,$allows
the representation%
\begin{equation}
\chi(x)=\phi(\mu,w;x)\left[  D+\int_{x_{0}}^{x}dy\frac{1}{\phi(\mu,w;y)}%
\eta(y)\right]  ,\text{ }\eta(x)=\check{a}(\mu,w;x)\chi(x)\in L^{2}%
(\mathbb{R}_{+}),\label{2.4.4.9}%
\end{equation}
where the point $x_{0}$ and constant $D$ depend on the values of $\mu$ as
follows:%
\begin{equation}\label{2.4.4.10}
\begin{array}{lll}
\mu=0:x_{0}=0\text{\textrm{\ and }}D\,\text{\textrm{is an arbitrary
constant}},\\
\mu\in(0,\pi/2),\,x_{0}=0\,\text{\textrm{and }}D=-\int_{0}^{\infty
}dy\frac{1}{\phi(\mu,w;y)}\eta(y),\\
\mu=\pi/2:\,x_{0}\in(0,\infty)\text{\textrm{\ and\ }}D=-\int_{x_{0}%
}^{\infty}dy\frac{1}{\phi(\pi/2,w;y)}\eta(y).%
\end{array}
\end{equation}

A subtlety is that the constant $D$ can take arbitrary values, but for $\mu
=0$, the constant $D$ is independent of the function $\eta(x)$, while for
$\mu\in(0,\pi/2]$, it is uniquely related to $\eta$, so that representation
(\ref{2.4.4.9}),(\ref{2.4.4.10}) for $\chi(x)\in D_{b^{+}(\mu,w)}$ with
$\mu\in(0,\pi/2]$ is equivalent to
\begin{equation}
\chi(x)=-\phi(\mu,w;x)\int_{x}^{\infty}dy\,\frac{1}{\phi(\mu,w;y)}%
\eta(y),\,\,\eta(x)=\check{a}_{(\mu,w)}\chi(x)\in L^{2}(\mathbb{R}_+),\,\,
 \mu\in(0,\pi/2].\,\label{2.4.4.11}%
\end{equation}

The asymptotic behavior of functions $\chi(x)\in$ $D_{b^{+}(\mu,w)}$ at the
origin and at infinity estimated using (\ref{2.4.1.3})-(\ref{2.4.1.6}) and the
Cauchy--Bunyakovskii inequality for integral terms in (\ref{2.4.4.9}) and
(\ref{2.4.4.11}) is respectively given by%
\begin{equation}
\chi(x)=\left\{
\begin{array}
[c]{l}%
D\,\tilde{B}(\mu,w)(\upsilon x)^{1/2}\ln(\upsilon x)\left[  1+O(1/\ln
x)\right]  +O(x^{1/2}\ln^{1/2}x),\\
\mu\in\lbrack0,\pi/2)\\
O(x^{1/2}\ln^{1/2}x),\,\mu=\pi/2
\end{array}
\right.  ,\,\,x\rightarrow0,\label{2.4.4.12}%
\end{equation}
and by%
\begin{equation}
\chi(x)\rightarrow0,\,x\rightarrow\infty,\text{ }\forall\mu.\label{2.4.4.13}%
\end{equation}

As it follows from (\ref{2.4.4.9}),(\ref{2.4.4.10}), the kernel of $\hat
{b}^{+}(\mu,w)$ \ is nontrivial only for $\mu=0$:%
\begin{equation}
\ker\hat{b}^{+}(\mu,w)=\left\{
\begin{array}
[c]{l}%
\{0\},\mathrm{\,}\mu\in(0,\pi/2]\\
\{c\left(  \phi(0,w;x)=\mathrm{e}^{-(\upsilon x)^{2}/2}(\upsilon
x)^{1/2\,}\Gamma(\alpha)\Psi(\alpha,1;(\upsilon x)^{2})\right)  \},\,\\
\mu=0
\end{array}
\right.  .\label{2.4.4.13a}%
\end{equation}
A simple reason is that $\hat{b}^{+}(\mu,w)\chi(x)=\check{a}(\mu
,w)\chi(x)=0\Rightarrow\chi(x)=c\phi(\mu,w;x)\,$\ and $\phi(\mu,w;x)\,$%
(\ref{2.4.1.1}) is square integrable on $\mathbb{R}_{+}$ only for $\mu=0$,
see (\ref{2.4.1.3}) and (\ref{2.4.1.5}).

The domain $D_{b^{+}(\mu,w)}$ of the operator $\hat{b}^{+}(\mu,w)$ with
$\mu\in\lbrack0,\pi/2)$ can be represented as a direct sum of the
form %
\begin{equation*}
D_{b^{+}(\mu,w)}=\{D\chi_{0}(\mu,w)(x)\}+\tilde{D}_{b^{+}(\mu,w)},\,\,\mu
\in\lbrack0,\pi/2),\,
\end{equation*}
where the function $\chi_{0}(\mu,w;x)\,$belonging to $D_{b^{+}(\mu,w)}$ is
given by%
\begin{equation*}
\chi_{0}(\mu,w;x)=\phi(\mu,w;x)\zeta\left(  x\right)  ,\mathrm{\,}%
\text{\textrm{so that}}\ \check{a}(\mu,w)\chi_{0}(\mu,w;x)=\phi(\mu
,w;x)\zeta^{\prime}\left(  x\right)  ,
\end{equation*}
$\zeta\left(  x\right)  $ is a fixed smooth function with a compact support
and equal to $1$ in a neighborhood of the origin, and $\tilde
{D}_{b^{+}(\mu,w)}$ is the subspace of functions belonging to $D_{b^{+}%
(\mu,w)}$ and vanishing at the origin as $O(x^{1/2}\ln^{1/2}x)$:%
\begin{equation}
\tilde{D}_{b^{+}(\mu,w)}=\left\{  \chi(x)\in D_{b^{+}(\mu,w)}:\,\chi
(x)=O(x^{1/2}\ln^{1/2}x),\,x\rightarrow0\right\}  ,\,\mu\in\lbrack
0,\pi/2).\label{2.4.4.16}%
\end{equation}

iii) The domain $D_{\bar{a}(\mu,w)}$ of the operator $\overline{\hat{a}}$
$(\mu,w)$ is evaluated in accordance with (\ref{2.4.3.1}): the defining
equation for $\overline{\hat{a}}$ $(\mu,w)$ as $(\hat{a}^{+}(\mu,w))^{+}$,
which is a restriction of $\hat{b}^{+}(\mu,w)$, is reduced to the equation for
$D_{\bar{a}(\mu,w)}\subseteq D_{b^{+}(\mu,w)}$ of the form%
\begin{equation}
(\chi,\check{b}(\mu,w)\psi)-(\check{a}(\mu,w)\chi,\psi)=0,\,\chi(x)\in
D_{\bar{a}(\mu,w)}\subseteq D_{b^{+}(\mu,w)},\,\forall\psi(x)\in D_{a^{+}%
(\mu,w)}.\label{2.4.4.17}%
\end{equation}
Integrating by parts in $(\check{a}(\mu,w)\chi,\psi)$ and taking estimates
(\ref{2.4.4.4}),(\ref{2.4.4.5}) and (\ref{2.4.4.12}),(\ref{2.4.4.13}) into
account, we establish that for $\mu=\pi/2$, eq. (\ref{2.4.4.17}) holds
identically for all$\,\chi(x)\in$ $D_{b^{+}(\mu,w)}$, while for $\mu\in
\lbrack0,\pi/2)$, eq. (\ref{2.4.4.17})\thinspace is reduced to
\begin{equation*}
\overline{D}C=0,\text{ }\forall C,
\end{equation*}
which requires that $D=0$.

We finally obtain that
\begin{equation}\label{2.4.4.19}
\overline{\hat{a}}(\pi/2,w)=\hat{b}^{+}(\pi/2,w),\;
D_{\bar{a}(\pi/2,w)}=D_{b^{+}(\pi/2,w)},
\end{equation}
a copy of (\ref{2.2.4.12}) with $\mu=\pi/2$, and%
\begin{equation}
\overline{\hat{a}}(\mu,w)\subset\hat{b}^{+}(\mu,w),\,D_{\bar{a}(\mu,w)}%
=\tilde{D}_{b^{+}(\mu,w)}\,\text{\textrm{(\ref{2.4.4.16})},}\,\mu\in
\lbrack0,\pi/2),\label{2.4.4.20}%
\end{equation}
which, in particular, implies that the asymptotic behavior at the origin of
the functions belonging to the domain of any operator $\overline{\hat{a}}%
(\mu,w)$ is estimated as $O(x^{1/2}\ln^{1/2}x)$,%
\begin{equation}
D_{\bar{a}(\mu,w)}\ni\chi(x)=O(x^{1/2}\ln^{1/2}x),\,x\rightarrow
0,\,\mathrm{for\,}\forall\mu,\forall w>-1/2,\label{2.4.4.20a}%
\end{equation}
for $\mu=\pi/2$ this follows from (\ref{2.4.4.19}) and (\ref{2.4.4.12}), while
for $\mu\in\lbrack0,\pi/2)\,$\ this follows from (\ref{2.4.4.20})\thinspace
and (\ref{2.4.4.16}).

The kernel of any operator $\overline{\hat{a}}(\mu,w)$\thinspace is trivial,%
\begin{equation}
\ker\overline{\hat{a}}(\mu,w)=\{0\},\,\forall\mu,\,\forall
\,w>-1/2,\label{2.4.4.20b}%
\end{equation}
for $\mu=\pi/2$ this follows from (\ref{2.4.4.19})\thinspace and
(\ref{2.4.4.13a}), while for $\mu\in[0,\pi/2)$, this follows from that
$\overline{\hat{a}}(\mu,w)\chi(x)=$ $\check{a}(\mu,w)\chi(x)=0\Rightarrow
\chi(x)=c\,\phi(\mu,w;x)$, but the function $\phi(\mu,w;x)$ is estimated at
the origin as $-2\cos\mu(\upsilon x)^{1/2}\ln(\upsilon x)+O(x^{1/2})$,
see (\ref{2.4.1.3}), and therefore cannot
belong to $D_{\bar{a}(\mu,w)}$ according to (\ref{2.4.4.20a}) (in addition,
$\phi(\mu,w;x)$ with $\mu\in(0,\pi/2)$ is not square integrable on
$\mathbb{R}_{+}$, see (\ref{2.4.1.5})).

iv) Quite similarly, we find
\begin{equation}\label{2.4.4.21}
\overline{\hat{b}}(\pi/2,w)=\hat{a}^{+}(\pi/2,w),\;
D_{\bar{b}(\mu,w)}=D_{a^{+}(\mu,w)},\,\mathrm{a\ copy\ of\ }%
\text{\textrm{(\ref{2.2.4.1}) with }}\mu=\pi/2,\text{ }%
\end{equation}
and
\begin{equation}
\overline{\hat{b}}(\mu,w)\subset\hat{a}^{+}(\mu,w),\,D_{\bar{b}(\mu,w)}%
=\tilde{D}_{a^{+}(\mu,w)}\mathrm{\,(\ref{2.4.4.8})},\text{ }\,\mu\in
\lbrack0,\pi/2),\label{2.4.4.22}%
\end{equation}
which, in particular, implies that the asymptotic behavior at the origin of
the functions belonging to the domain of any operator $\overline{\hat{b}}%
(\mu,w)$ is estimated as $O(x^{1/2})$,%
\begin{equation}
D_{\bar{b}(\mu,w)}\ni\psi(x)=O(x^{1/2}),\,x\rightarrow0,\,\mathrm{for\,}%
\forall\mu,\forall w>-1/2,\label{2.4.4.22a}%
\end{equation}
for $\mu=\pi/2$ this follows from (\ref{2.4.4.21}) and (\ref{2.4.4.4}), while
for $\mu\in\lbrack0,\pi/2)\,$\ this follows from (\ref{2.4.4.22})\thinspace
and (\ref{2.4.4.8}). We note that equality (\ref{2.4.4.21}) and inclusion
(\ref{2.4.4.22}) directly follow from the respective previous equality
(\ref{2.4.4.19}) and inclusion (\ref{2.4.4.20}) by taking the adjoints, and
only the domain $D_{\bar{b}(\mu,w)}$ in the last case has to be evaluated.

By arguments similar to those in the end of subsubsec. 2.2.4, it is easy to
prove that there are no other pairs $\hat{g}=\overline{\hat{g}}$ and
$\,\hat{g}^{+}$ of closed mutually adjoint operators that are extensions of
the respective $\hat{a}(\mu,w)$ and $\,\hat{b}(\mu,w)$, $\hat{a}(\mu
,w)\subset\hat{g}\,,\hat{b}(\mu,w)\subset\hat{g}\,^{+}$.

We thus show that each pair $\check{a}(\mu,w)$, $\check{b}(\mu,w) $,
$\mu\in\lbrack0,\pi/2],\,w\in(-1/2,\infty)$,$\ $of mutually adjoint\ by
Lagrange differential operations given by a copy of (\ref{2.2.1.3}) with
$\phi(\mu,w;x)$ (\ref{2.4.1.1}) and providing a two-parameter family of
different generalized oscillator representations (\ref{2.4.2.1}) for
generalized Calogero differential operation $\check{H}\,$\ (\ref{1.1}) with
$g_{1}=-1/4,g_{2}>0$ generates a unique pair $\overline{\hat{a}}%
(\pi/2,w)=\hat{b}^{+}(\pi/2,w)$,
$\hat{a}^+(\pi/2,w)=\overline{\hat{b}}(\pi/2,w)$ of closed mutually adjoint
operators for $\mu=\pi/2$, while for $\mu\in\lbrack0,\pi/2)$, each pair
$\check{a}(\mu,w)$, $\check{b}(\mu,w)\,$generates two different pairs
$\overline{\hat{a}}(\mu,w),\ \hat{a}^{+}(\mu,w)$ and $\,\hat{b}^{+}(\mu,w)$,
$\overline{\hat{b}}(\mu,w)$ of closed mutually adjoint\textrm{\thinspace}
operators such that $\overline{\hat{a}}(\mu,w)\subset$ $\hat{b}^{+}(\mu,w)$
and $\overline{\hat{b}}(\mu,w)\subset$ $\hat{a}^{+}(\mu,w)$. The operators
$\overline{\hat{a}}(\mu,w) $\ and $\hat{b}^{+}(\mu,w)$ are extensions of the
initial operator $\hat{a}(\mu,w)$, they are associated with $\check{a}(\mu
,w)$, and their domains are given by the respective (\ref{2.4.4.19}%
),(\ref{2.4.4.20}) and a copy of (\ref{2.2.4.12}). The operators
$\overline{\hat{b}}(\mu,w)$\ and $\hat{a}^{+}(\mu,w)$\ are extensions of the
initial operator $\hat{b}(\mu,w) $, they are associated with $\check{b}%
(\mu,w)$, and their domains are given by the respective (\ref{2.4.4.21}%
),(\ref{2.4.4.22}) and a copy of (\ref{2.2.4.1}).

\subsection{Region $g_{1}=-1/4$ ($\varkappa=0$), $w=w_{0}=-1/2$
($\alpha=0$)}

A consideration in this subsection is completely similar to that in subsec.
2.3, and all the results to follow can be obtained from the results of
subsec. 2.3 in the limit $\varkappa\rightarrow0$ $(\beta\rightarrow1)$. We
even do not change the notation having in mind that here $\varkappa=0,$
$\beta=1 $.

On the other hand, all the results of this subsection can be obtained from
the results of the previous subsec. 2.4 related to the case $\mu=\pi/2$ in
the limit $w\rightarrow w_{0}=-1/2$.

\subsubsection{Generalized oscillator representations for $\check{H}$,
differential operations $\check{a}$ and $\check{b}$}

For these values of parameters, the general solution of differential
equation (\ref{2.1.3}) with $u=u_{0}=4\upsilon^{2}w_{0}=-2\upsilon^{2}$ is
given by (\ref{2.1.24}) with $\varkappa=0$, $\beta=1$. The function
$\Psi(1,\rho)$ given by (\ref{2.1.21}) with $\beta=1$ increases
monotonically from $-\infty$ to $\infty$ as $\rho=(\upsilon x)^{2}$
together with $x$ ranges from $0$ to $\infty$. It follows that eq.
(\ref{2.1.3}) with these values of parameters has a unique, up to a
positive constant factor, real-valued positive solution
\begin{equation}
\phi(w_{0};x)=\mathrm{e}^{-\rho/2}\rho^{1/4},\,\rho=(\upsilon x)^{2}%
,\label{2.5.1.1}%
\end{equation}
which is a copy of (\ref{2.3.1.1}) with $\varkappa=0$. This implies that in
this case, we have a unique\footnote{Up to irrelevant phase factors.} pair of
mutually adjoint first-order differential operations $\check{a}(w_{0})$ and
$\check{b}(w_{0})$ given by a copy of (\ref{2.3.1.2}) with $\phi(w_{0};x)$
(\ref{2.5.1.1}) instead of $\phi(w_{0};x)$ (\ref{2.3.1.1}) and providing
unique generalized oscillator representation (\ref{1.6a}) with
$u=-2\upsilon^2$ for generalized Calogero differential operation $\check{H}$
(\ref{1.1}) with $g_{1}=-1/4$, $g_{2}>0$:%
\begin{equation}
\check{H}=-d_{x}^{2}-\frac{1}{4}x^{-2}+g_{2}x^{2}=\check{b}(w_{0})\check
{a}(w_{0})+2\upsilon^{2}.\label{2.5.1.2}%
\end{equation}
This representation is obtained from representation (\ref{2.3.1.3}) in the
limit $g_{1}\rightarrow-1/4$, $\varkappa\rightarrow0$. Accordingly, a
consideration in this subsection is a copy of that in subsec. 2.3 with the
substitution $\phi(w_{0};x)$ (\ref{2.5.1.1}) for $\phi(w_{0};x)$
(\ref{2.3.1.1}).

The asymptotic behavior of the functions $\phi(w_{0};x)$ and $1/\phi(w_{0};x)$
at the origin and at infinity is evident from (\ref{2.5.1.1}).

\subsubsection{Initial operators $\hat{a}$ and $\hat{b}$}

In perfect analogy to subsubsec. 2.3.2, we introduce the pair
$\hat{a}(w_{0}), $ $\hat{b}(w_{0})\,$of initial differential operators in
$L^2(\mathbb{R}_+)$ defined on
$\mathcal{D}(\mathbb{R}_{+})$ and associated with the respective pair of
differential operations $\check{a}(w_{0})$,$\check{b}(w_{0})$. These
operators satisfy the relation that is the copy of (\ref{2.3.2.1}). The
representation (\ref{2.5.1.2}) for $\check{H}$ provide the representation
\begin{equation}
\hat{H}=\hat{b}(w_{0})\,\hat{a}(w_{0})+2\upsilon^{2}\,\hat{I}\label{2.5.2.1}%
\end{equation}
for the initial symmetric operator $\hat{H}$ $\,$associated with $\check{H}$
and defined on $\mathcal{D}(\mathbb{R}_{+})$, which in particular confirms
that $\hat{H}$ with $g_{1}=-1/4,$ $g_{2}>0$ is bounded from below by
$2\upsilon^{2}$.

The representation (\ref{2.5.2.1}) is a basis for constructing, maybe new,
s.a. generalized Calogero Hamiltonians $\hat{H}_{\mathfrak{e}}$ with
$g=-1/4,\,g_{2}>0\,$as s.a. extensions of $\hat{H}$ (\ref{2.5.2.1}) in
generalized oscillator form (\ref{1.8}) or (\ref{1.9}) via constructing all
possible extensions of the pair $\hat{a}(w_{0})$, $\hat{b}(w_{0})$ of initial
operators to a pair of closed mutually adjoint operators $\hat{c}(w_{0})$,
$\hat{c}^{+}(w_{0})$, $\hat{a}(w_{0})\subset\hat{c}(w_{0})$, $\hat{b}%
(w_{0})\subset\hat{c}^{+}(w_{0})$, beginning from the the pair $\overline
{\hat{a}}(w_{0}),$ $\hat{a}^{+}(w_{0})$, or to a pair of closed mutually
adjoint operators $\hat{d}(w_{0})$, $\hat{d}^{+}(w_{0})$, $\hat{a}(w_{0}%
)\subset\hat{d}^{+}(w_{0})$, $\hat{b}(w_{0})\subset\hat{d}(w_{0})$, beginning
from the $\overline{\hat{b}}(w_{0}),$ $\hat{b}^{+}(w_{0})$.

\subsubsection{Adjoint operators $\hat{a}^{+}$ and $\hat{b}^{+}$,
closed operators $\overline{\hat{a}}$ and $\overline{\hat{b}}$}

By arguments similar to those in subsubsecs. 2.2.3 and 2.2.4, we establish
that the closures $\overline{\hat{a}}(w_{0})$ and $\overline{\hat{b}}%
(w_{0})\,$of the respective operators $\hat{a}(w_{0})$ and $\hat{b}(w_{0})$,
as well as the adjoints $\hat{a}^{+}(w_{0})$ and $\hat{b}^{+}(w_{0})$ of
the latters, do
exist and the chains of inclusions%
\begin{equation}\label{2.5.3.1}
\begin{array}{ll}
\hat{a}(w_{0})\subset\overline{\hat{a}}(w_{0})=(\hat{a}^{+}(w_{0}%
))^{+}\subseteq\hat{b}^{+}(w_{0}),\\
\hat{b}(w_{0})\subset\overline{\hat{b}}(w_{0})=(\hat{b}^{+}(w_{0}%
))^{+}\subseteq\hat{a}^{+}(w_{0}),%
\end{array}
\end{equation}
which are the copies of (\ref{2.3.3.1}), hold. The operators $\overline
{\hat{a}}(w_{0})$ and $\hat{b}^{+}(w_{0})$ are associated with $\check
{a}(w_{0})$, while the operators $\overline{\hat{b}}(w_{0})$ and
$\hat{a}^{+}(w_{0})$ are associated with $\check{b}(w_{0})$, so that to
specify these operators, it is sufficient to evaluate their domains actually
repeating the content of subsubsec. 2.3.4 with the substitution $\phi
(w_{0};x)$ (\ref{2.5.1.1}) for $\phi(w_{0};x)$ (\ref{2.3.1.1}).

\subsubsection{Domains of operators $\hat{a}^{+}$, $\hat{b}^{+}$,
$\overline{\hat{a}}$ and $\overline{\hat{b}}$}

i) The domain of the operator $\hat{a}^{+}$ $(w_{0})$ is the natural domain
for $\check{b}(w_{0})$,$\,$which is given by the copy of (\ref{2.3.4.1}). A
generic function $\psi(x)$ belonging to $D_{a^{+}(w_{0})}$ allows the representations%
\begin{equation}\label{2.5.4.1}
\begin{array}{ll}
\psi(x)=-\frac{1}{\phi(w_{0},x)}\int_{0}^{x}dy\,\phi(w_{0},y)\eta
(y)=\frac{1}{\phi(w_{0},x)}\int_{x}^{\infty}dy\,\phi(w_{0},y)\eta(y),\\
\eta(x)=\check{b}_{(w_{0})}\psi(x)\in L^{2}(\mathbb{R}_{+}%
),\,\text{\textrm{and }}\int_{0}^{\infty}dy\,\phi(w_{0},y)\eta
(y)=0.%
\end{array}
\end{equation}
The equality $\int_{0}^{\infty}dy\,\phi(w_{0},y)\eta(y)=0$ means that the
range $R_{a^{+}(w_{0})}$of the operator $\hat{a}^{+}(w_{0})$ is orthogonal to
the one-dimensional subspace $\{c\phi(w_{0};x)\}\subset L^{2}(\mathbb{R}%
_{+}),$ $R_{a^{+}(w_{0})}\bot\{c\phi(w_{0};x)\}$, and thereby its closure
$\overline{R_{a^{+}(w_{0})}}\,$can not be the whole $L^{2}(\mathbb{R}_{+})$,
which in turn implies that the kernel of the adjoint operator $(\hat{a}^{+}%
(w_{0}))^{+}=\overline{\hat{a}}(w_{0})$ is not trivial, $\{c\phi
(w_{0};x)\}\subseteq\ker\overline{\hat{a}}(w_{0})\neq\{0\}$, see below.

Estimating the integral terms in (\ref{2.5.4.1}) with the
Cauchy--Bunyakovskii inequality, we obtain that the asymptotic
behavior of functions $\psi(x)\in$ $D_{a^{+}(w_{0})}$ at the origin and at
infinity is respectively is given by%
\begin{equation}\label{2.5.4.2}
\begin{array}{ll}
\psi(x)=O(x^{1/2}),\,x\rightarrow0,\\
\psi(x)\rightarrow0,\;x\rightarrow\infty.%
\end{array}
\end{equation}

ii) The domain $D_{b^{+}(w_{0})}$ of the operator $\hat{b}^{+}(w_{0})$ is the
natural domain for $\check{a}(w_{0})$, which is given by a copy of
(\ref{2.3.4.5}). A generic function $\chi(x)$ belonging to $D_{b^{+}(w_{0})}$
allows the representation%
\begin{equation}\label{2.5.4.3}%
\chi(x)=\phi(w_{0},x)\left[  D+\int_{x_{0}}^{x}dy\frac{1}{\phi(w_{0},y)}%
\eta(y)\right],\;\eta(x)=\check{b}(w_{0})\chi(x)\in L^{2}(\mathbb{R}_{+}),
\end{equation}
where $x_{0}\in(0,\infty)$ and $D$ is an arbitrary constant, and its asymptotic
behavior at the origin and at infinity estimated using the
Cauchy--Bunyakovskii inequality is respectively given by
\begin{equation}\label{2.5.4.4}
\begin{array}{ll}
\chi(x)=O(x^{1/2}\ln^{1/2}x),\;x\rightarrow0,\\
\chi(x)\rightarrow0,\;x\rightarrow\infty.
\end{array}
\end{equation}

As it follows from (\ref{2.5.4.3}) with $\eta(x)=0$ and (\ref{2.5.1.1}), the
kernel of the operator $\hat{b}^{+}(w_{0})$ is nontrivial,%
\begin{equation}
\ker\hat{b}^{+}(w_{0})=\{c\left(  \phi(w_{0};x)=(\upsilon x)^{1/2}%
\mathrm{e}^{-(\upsilon x)^{2}/2}\right)  \}.\label{2.5.4.4a}%
\end{equation}

iii) The domain $D_{\bar{a}(w_{0})}$ of the operator $\overline{\hat{a}}%
(w_{0})$ is evaluated in accordance with (\ref{2.5.3.1}): the defining
equation for $\overline{\hat{a}}(w_{0})$ as $(\hat{a}^{+}(w_{0}))^{+}$, which
is a restriction of $\hat{b}^{+}(w_{0})$, is reduced to the equation for
$D_{\bar{a}(w_{0})}\subseteq$ $D_{b^{+}(w_{0})}$ of the form%
\begin{equation}
(\chi,\check{b}(w_{0})\psi)-(\check{a}(w_{0})\chi,\psi)=0,\,\chi(x)\in
D_{\bar{a}(w_{0})}\subseteq D_{b^{+}(w_{0})},\,\forall\psi(x)\in
D_{a^{+}(w_{0})}.\label{2.5.4.5}%
\end{equation}
Integrating by parts in $(\check{a}(w_{0})\chi,\psi)$ and taking asymptotic
estimates (\ref{2.5.4.2}) and (\ref{2.5.4.4}) into account, we establish that
eq. (\ref{2.5.4.5}) holds identically for all$\,\chi(x)\in D_{b^{+}(w_{0})}$,
which implies that%

\begin{equation}
\overline{\hat{a}}(w_{0})=\hat{b}^{+}(w_{0}),\,
D_{\bar{a}(w_{0})}=\mathrm{\,}D_{b^{+}(w_{0})}\,\,\text{given by
\,a copy\,of\,}\,(\ref{2.3.4.5}),\label{2.5.4.6}%
\end{equation}
in particular, the asymptotic behavior at the origin of the functions
$\chi(x)$ belonging to the domain of the operator $\overline{\hat{a}}(w_{0})$
is estimated by a copy of (\ref{2.5.4.4}),%
\begin{equation}
\chi(x)=O(x^{1/2}\ln^{1/2}x),\;x\rightarrow0,\,\label{2.5.4.6a}%
\end{equation}
and
\begin{equation}
\ker\overline{\hat{a}}(w_{0})=\{c\left(  \phi(w_{0};x)=(\upsilon
x)^{1/2}\mathrm{e}^{-(\upsilon x)^{2}/2}\right)  \}.\label{2.5.4.6b}%
\end{equation}

iv) It directly follows from (\ref{2.5.4.6}) by taking the adjoints that
\begin{equation}
\overline{\hat{b}}(w_{0})=\hat{a}^{+}(w_{0}),\text{\textrm{\ }}D_{\bar
{b}(w_{0})}=D_{a^{+}(w_{0})}\,\,\text{given by a copy of}\,\,
(\ref{2.3.4.1}).\label{2.5.4.7}%
\end{equation}

By arguments similar to those in the end of subsubsec. 2.2.4, it is easy to
prove that there is no other pair $\hat{g}=\overline{\hat{g}}$ and
$\,\hat{g}^{+}$of closed mutually adjoint operators that are extensions of the
respective $\hat{a}(w_{0})$ and $\hat{b}(w_{0})$, $\hat{a}(w_{0})\subset
\hat{g}\,,\hat{b}(w_{0})\subset\hat{g}\,^{+}$.

We thus show that the pair $\check{a}(w_{0}),\check{b}(w_{0})$ of
mutually adjoint\ by Lagrange differential operations given by a copy of
(\ref{2.3.1.2}) with $\phi(w_{0};x)$ (\ref{2.5.1.1}) and providing unique
generalized oscillator representation (\ref{2.5.1.2}) for $\check{H}%
$\ (\ref{1.1}) with $g_{1}=-1/4$, $g_{2}>0$ generates a unique pair
$\overline{\hat{a}}(w_{0})=\hat{b}^{+}(w_{0})$, $\ \hat{a}^{+}(w_{0}%
)=\overline{\hat{b}}(w_{0})\,$of closed mutually adjoint\textrm{\thinspace}
operators. The operator $\overline{\hat{a}}(w_{0})=$ $\hat{b}^{+}(w_{0})$ is
an extension of the initial operator $\hat{a}(w_{0})$, it is associated with
$\check{a}(w_{0})$, and its domain is given by a copy of (\ref{2.3.4.5}).
The operator $\overline{\hat{b}}(w_{0})=\hat{a}^{+}(w_{0})$\ is an
extension of
the initial operator $\hat{b}(w_{0})$, it is associated with $\check{b}%
(w_{0})$, and its domain is given by a copy of (\ref{2.3.4.1}).

\subsection{Resume}

The final conclusion of this section is that for each pair of coupling
constants $g_{1}\geq-1/4$, $g_{2}>0$, we have two %
two-parameter\textrm{\thinspace}families of generalized Calogero Hamiltonians
in a generalized oscillator form, the family $\{\hat{H}_{\mathfrak{e}a}\}$ of
Hamiltonians
\begin{align}
& \hat{H}_{\mathfrak{e}a}=\left\{
\begin{array}
[c]{l}%
\hat{H}_{\mathfrak{e}a(\mu,w)}=\hat{a}^{+}(\mu,w)\overline{\hat{a}}%
(\mu,w)-4\upsilon^{2}w\,\hat{I},\,\mu\in\lbrack0,\pi/2],\;w\in(w_{0}%
,\infty),\\
\hat{H}_{\mathfrak{e}a(w_{0})}=\hat{a}^{+}(w_{0})\overline{\hat{a}}%
(w_{0})-4\upsilon^{2}w_{0}\,\hat{I},\,\,
\end{array}
\right.  \label{3.1.1}\\
& \upsilon=\sqrt[4]{g_{2}}>0,\,w_{0}=-\frac{1}{2}(1+\varkappa),\,\varkappa
=\sqrt{1/4+g_{1}}\geq0,\, \nonumber%
\end{align}
where the operators $\hat{a}^{+}(\mu,w)$ and $\overline{\hat{a}}(\mu,w)$ are
described in the respective items i) and iii) in
subsubsec. 2.2.4 for the case of $g_{1}>-1/4\,(\varkappa>0)$ and in subsubsec.
2.4.4 for the case of $g_{1}=-1/4\,(\varkappa=0)$, while the operators
$\hat{a}^{+}(w_{0})$ and $\overline{\hat{a}}(w_{0})$ are described in the
respective items i) and iii) in subsubsec. 2.3.4 for the case of
$g_{1}>-1/4\,(\varkappa>0)$ and in subsubsec. 2.5.4 for the case of
$g_{1}=-1/4\,(\varkappa=0)$, and the family $\{\hat{H}_{\mathfrak{e}b}\}$ of
Hamiltonians
\begin{equation}
\hat{H}_{\mathfrak{e}b}=\left\{%
\begin{array}
[c]{l}%
\hat{H}_{\mathfrak{e}b(\mu,w)}=\,\overline{\hat{b}\ }(\mu,w)\hat{b}^{+}%
(\mu,w)-4\upsilon^{2}w\,\hat{I},\mu\in\lbrack0,\pi/2],\;w\in(w_{0},\infty),\\
\hat{H}_{\mathfrak{e}b(w_{0})}=\overline{\hat{b}\ }(w_{0})\hat{b}^{+}%
(w_{0})-4\upsilon^{2}w_{0}\,\hat{I},\,
\end{array}
\right.
\label{3.1.2}%
\end{equation}
where the operators $\hat{b}^{+}(\mu,w)$ and $\overline{\hat{b}}(\mu,w)$ are
described in the respective items ii) and iv) in
subsubsec. 2.2.4 for the case of $g_{1}>-1/4\,(\varkappa>0)$ and in subsubsec.
2.4.4 for the case of $g_{1}=-1/4\,(\varkappa=0)$, while the the operators
$\hat{b}^{+}(w_{0})$ and $\overline{\hat{b}}(w_{0})$ are described in the
respective items ii) and iv) in subsubsec. 2.3.4 for the case of
$g_{1}>-1/4\,(\varkappa>0)$ and in subsubsec. 2.5.4 for the case of
$g_{1}=-1/4\,(\varkappa=0)$.

For some values of coupling constants, these families overlap. In
particular, the Hamiltonians $\hat{H}_{\mathfrak{e}a(\mu,w)}$ and
$\hat{H}_{\mathfrak{e}b(\mu,w)}$ with $g_{1}\geq3/4\,$coincide, see
(\ref{2.2.4.23}) and (\ref{2.2.4.25}), $\hat{H}_{\mathfrak{e}a(\mu
,w)}=\hat{H}_{\mathfrak{e}b(\mu,w)}$, $g_{1}\geq3/4$. The same holds for the
Hamiltonians $\hat{H}_{\mathfrak{e}a(w_{0})}$ and $\hat{H}_{\mathfrak{e}%
b(w_{0})}$ with any $g_{1}\geq-1/4$, see (\ref{2.3.4.10}), (\ref{2.3.4.11})
and (\ref{2.5.4.6}), (\ref{2.5.4.7}), so that $\{\hat{H}_{\mathfrak{e}%
a}\}=\{\hat{H}_{\mathfrak{e}b}\}$ for the case of $g_{1}\geq3/4$.

\section{Oscillator representations for all Hamiltonians with $g_1\geq-1/4$,  $g_2>0$}

\subsection{Preliminaries}

This section is a concluding one. Without being afraid of repeating ourselves,
we give here a full answer to the question on generalized oscillator
representations (\ref{1.8}) or\ (equivalently) (\ref{1.9}) for all generalized
Calogero Hamiltonians $\hat{H}_{\mathfrak{e}}$ associated with generalized
Calogero differential operation $\check{H}$\ (\ref{1.1}) with any coupling
constants $g_{1},g_{2}\in\mathbb{R}^{2}$.

An answer to the question is essentially different for different regions in
the plane $\mathbb{R}^{2}$ of coupling constants, namely, for the open
half-planes $\{g_{1}<-1/4\}$ and $\{g_{2}<0\}$ (these regions are overlapping
\ along the open quadrant $\{g_{1}<-1/4$, $g_{2}<0\}$) and for the
quadrant $\{g_{1}\geq-1/4$, $g_{2}\geq0\}$.

As was already indicated in the beginning of sec. 1 and in subsec. 2.1, any
generalized Calogero Hamiltonian $\hat{H}_{\mathfrak{e}}$ with coupling
constants lying in the half-plane $\{g_{1}<-1/4\}$ or in the half-plane
$\{g_{2}<0\}$ does not allow generalized oscillator representation because
such a representation would imply that $\hat{H}_{\mathfrak{e}}$ is bounded
from below, whereas any generalized Calogero Hamiltonian with such values of
coupling constants is not bounded from below \cite{GitTyV12}. This\
conclusion is in complete agreement\ with that according to subsec. 2.1,
there is no generalized oscillator representation (\ref{1.6a}) for
generalized Calogero differential operation $\check{H}$ (\ref{1.1}) with
$g_{1}<-1/4$ or with $g_{2}<0$.

As to the quadrant $\{g_{1}\geq-1/4$, $g_{2}\geq0\}$, we recall that as was
shown in \cite{TyuVo13}, any Calogero Hamiltonian\footnote{We recall that we
conventionally omit the term ``generalized '' for Calogero Hamiltonians with
$g_{2}=0$.} $\hat{H}_{\mathfrak{e}}$ with coupling constants lying on the
horizontal boundary semiaxis $\{g_{1}\geq-1/4$, $g_{2}=0\}$ of the quadrant,
allows a generalized oscillator representation, moreover, a one- or
two-parameter family of such representations. Thus, only the semiopen quadrant
$\{g_{1}\geq-1/4$, $g_{2}>0\}$ remains. Following the considerations advanced
in the end of sec.1, we show in what follows that any generalized Calogero
Hamiltonian $\hat{H}_{\mathfrak{e}}$ with coupling constants lying in this
quadrant also allows a family of generalized oscillator representations, one-
or two-parameter. For completeness, we repeat an extended version of these
considerations.

A hypothesis is that the two families (\ref{3.1.1}) and (\ref{3.1.2}) of
generalized Calogero Hamiltonians in generalized oscillator form cover all
the set of the known generalized Calogero Hamiltonians with coupling
constants lying in the semiopen quadrant $\{g_{1}\geq-1/4$, $g_{2}>0\}$,
see \cite{GitTyV12}. Namely, each generalized Calogero Hamiltonian with
given $g_{1}\geq-1/4$, $g_{2}>0$ can be identified with one or more members
of family (\ref{3.1.1}) or family (\ref{3.1.2}). This identification is
trivial in the case of $g_{1}\in \lbrack3/4,\infty)$ (the semiopen quadrant
$\{g_{1}\geq3/4$, $g_{2}>0\}$) where there is a unique generalized Calogero
Hamiltonian $\hat{H}_{1}$ with given $g_{1}$, $g_{2}$, so that in this
case, the both $\{\hat{H}_{\mathfrak{e}a}\}$ and
$\{\hat{H}_{\mathfrak{e}b}\}$ are reduced
to a one-point set, $\{\hat{H}_{\mathfrak{e}a}\}=\{\hat{H}_{\mathfrak{e}%
b}\}=\hat{H}_{1}$. In the case of $g_{1}\in\lbrack-1/4,3/4)\,(\varkappa
\in\lbrack0,1))$ (the semiopen strip $\{-1/4\leq g_{1}<3/4,g_{2}>0\}$), the
procedure of identification is more complicated.According to
\cite{GitTyV12}, for each pair
of coupling constants $g_{1}$, $g_{2}$ lying in the open strip $\{-1/4<g_{1}%
<3/4,g_{2}>0\}$, there exists a one-parameter family $\{\hat{H}_{2,\nu},$
$\nu\in\lbrack-\pi/2,\pi/2],-\pi/2\,\sim\pi/2\}$ of generalized Calogero
Hamiltonians differing in their domains $D_{H_{2,\nu}}$, while for each pair
of coupling constants $g_{1}$, $g_{2}$ lying on the boundary open semiaxis
$\{g=-1/4,g_{2}>0\}$, there exists a one-parameter family
$\{\hat{H}_{3,\nu},$ $\nu\in\lbrack-\pi/2,\pi/2],-\pi/2\,\sim\pi/2\}$ of
generalized Calogero Hamiltonians\footnote{We slightly change the notation
in comparison with \cite{GitTyV12} for uniformity: the extension parameter
$\nu$ entering the index of $\hat{H}_{3,\nu}$ was denoted in
\cite{GitTyV12} by $\vartheta$, so that $\hat{H}_{3,\nu}$ here coincides
with $\hat{H}_{3,\vartheta}$ in \cite{GitTyV12}.} differing in their
domains $D_{H_{3,\nu}}$. Namely, the domains $D_{H_{2,\nu}}$ and
$D_{H_{3,\nu}}$ are subspaces of the natural domains $D_{\check{H}}^{n}$
for the respective $\check{H}$, the subspaces that are specified by
different s.a. asymptotic boundary conditions at the origin of the
respective form
\begin{equation}
D_{H_{2,\nu}}\ni\chi(x)=c\left[  (\upsilon x)^{1/2+\varkappa}\sin
\nu+\,(\upsilon x)^{1/2-\varkappa}\cos\nu\right]  +O(x^{3/2}),\,x\rightarrow
0,\,\varkappa\in(0,1),\label{3.1.3}%
\end{equation}
and
\begin{equation}
D_{H_{3,\nu}}\ni\chi(x)=c[\,(\upsilon x)^{1/2}\sin\nu+\,2(\upsilon x)^{1/2}%
\ln(\upsilon x)\cos\nu]+O(x^{3/2}),\,x\rightarrow0\,,\varkappa=0.\label{3.1.4}%
\end{equation}

Therefore, an identification of a given $\hat{H}_{2,\nu}$ with a certain
$\hat{H}_{\mathfrak{e}a}$ or $\hat{H}_{\mathfrak{e}b}\,$goes through
evaluating the asymptotic behavior at the origin of functions belonging to
the domain of $\hat{H}_{\mathfrak{e}a}$ or $\hat{H}_{\mathfrak{e}b}$ and its
identification with the asymptotic boundary conditions for the certain
$\hat{H}_{2,\nu}$; the same holds for $\hat{H}_{3,\nu}$. It may happen, and
that really occurs, that $\hat{H}_{\mathfrak{e}a(\mu,w)}$, or
$\hat{H}_{\mathfrak{e}b(\mu,w)}$, with different $\mu,w$ have the same
asymptotic behavior of functions belonging to their domains and define the
same Hamiltonian.

The asymptotic boundary conditions for Hamiltonians with $g_{1}\in(-1/4,3/4)
$\ and with $g_{1}=-1/4$\ differ widely in their character sending us in
separate consideration of the open strip $\{-1/4<g_{1}<3/4,g_{2}>0\}$\ and the
vertical boundary open semiaxis $\{g_{1}=-1/4,g_{2}>0\}$, as we did beforehand
in sec. 2.

We begin a detailed consideration with the family of Hamiltonians
$\{\hat{H}_{\mathfrak{e}a}\}$ (\ref{3.1.1}).

\subsection{Family $\{\hat{H}_{\mathfrak{e}a}\}$}

\subsubsection{Quadrant $\{g_{1}\geq3/4$ ($\varkappa\geq1$), $g_{2}>0\}$}

According to \cite{GitTyV12}, for any pair of coupling constants $g_{1}$,
$g_{2}$ lying in this quadrant, there exists a unique s.a. generalized
Calogero Hamiltonian $\hat{H}_{1}$ defined on the natural domain
$D_{\check{H}}^{n}$. It follows that the Hamiltonian $\hat{H}_{1}$ with
given $g_{1}\geq3/4$, $g_{2}>0$ must be identified with the whole family\
$\{\hat{H}_{\mathfrak{e}a}\}$ with the same $g_{1}$, $g_{2}$,
$\{\hat{H}_{\mathfrak{e}a}\}=\hat{H}_{1}$, which yields the two types of
generalized oscillator representations for this Hamiltonian:%
\begin{align}
\hat{H}_{1}  & =\hat{H}_{\mathfrak{e}a(\mu,w)}=\hat{a}^{+}(\mu,w)\overline
{\hat{a}}(\mu,w)-4\upsilon^{2}w\,\hat{I},\,\label{3.2.1.1}\\
\forall\mu & \in[0,\pi/2],\;\forall w\in(w_{0},\infty),\,w_{0}=-\frac{1}%
{2}(1+\varkappa),\,\varkappa\geq1,\nonumber%
\end{align}
and
\begin{equation}
\hat{H}_{1}=\hat{H}_{\mathfrak{e}a(w_{0})}=\hat{a}^{+}(w_{0})\overline
{\hat{a}}(w_{0})+2\upsilon^{2}(1+\varkappa),\,\varkappa\geq1.\label{3.2.1.2}%
\end{equation}

The formula (\ref{3.2.1.1}) actually presents a two-parameter family of
different generalized oscillator representations for a unique generalized
Calogero Hamiltonian $\hat{H}_{1}$ with given coupling constants $g_{1}%
\geq3/4$, $g_{2}>0$. According to (\ref{2.2.4.24b}), the kernel of any
operator $\overline{\hat{a}}(\mu,w)$ is trivial. This implies that any of
representations (\ref{3.2.1.1}) is not an optimum one in the sense that it does
not allow determining the ground state and the ground-state energy of
$\hat{H}_{1}$; we can say only that the spectrum of $\hat{H}_{1}$ is bounded
from below by $2\upsilon^{2}(1+\varkappa)$.

In contrast to (\ref{3.2.1.1}), representation (\ref{3.2.1.2}) is an optimum
one: $\ker\overline{\hat{a}}(w_{0})$ is nontrivial, it is given by
(\ref{2.3.4.10b}), which implies that $\ker\overline{\hat{a}}(w_{0})$ is the
one-dimensional ground space of $\hat{H}_{1}$ and the ground-state energy
$E_{0}$ of $\hat{H}_{1}$, which is a lower boundary of its spectrum, is
\begin{equation}
E_{0}^{(1)}=2\upsilon^{2}(1+\varkappa),\,\varkappa\geq1.\label{3.2.1.3}%
\end{equation}

The normalized ground state $U^{(1)}(x)\,$ of $\hat{H}_{1}$ is given by%
\begin{equation}
U^{(1)}(x)=\sqrt{\frac{2\upsilon}{\Gamma(1+\varkappa)}}(\upsilon
x)^{1/2+\varkappa}\mathrm{e}^{-(\upsilon x)^{2}/2},\,\varkappa\geq
1.\label{3.2.1.4}%
\end{equation}

\subsubsection{Strip $\{-1/4<g_{1}<3/4$ $(0<\varkappa<1)$, $g_{2}>0\}$}

A consideration in this subsubsection appears to be similar to that in the
previous subsubsection.

According to \cite{GitTyV12}, for any pair of coupling constants $g_{1}$,
$g_{2}$ lying in this strip, there exists a one-parameter family
$\{\hat{H}_{2,\nu},$ $\nu\in\lbrack-\pi/2,\pi/2],-\pi/2\,\sim\pi/2\}$ of
generalized Calogero Hamiltonians $\hat{H}_{2,\nu}$ specified by s.a.
asymptotic boundary conditions at the origin (\ref{3.1.3}).

An asymptotic behavior at the origin of the functions belonging to the domains
of the Hamiltonians $\hat{H}_{\mathfrak{e}a}$ (\ref{3.1.1}) with coupling
constants $g_{1}$, $g_{2}$ lying in the same strip is estimated as follows. By
definition of any operator $\hat{H}_{\mathfrak{e}a(\mu,w)}$, its domain
belongs to or coincides with the domain of the operator %
$\overline{\hat{a}}(\mu,w)$, $D_{H_{\mathfrak{e}a(\mu,w)}}\subseteq
D_{\bar{a}(\mu,w)}$; the same holds for the operator $\hat{H}_{\mathfrak{e}%
a(w_{0})}$, $D_{H_{\mathfrak{e}a(w_{0})}}\subseteq D_{\bar{a}(w_{0})}$.\ But
according to (\ref{2.2.4.24a}) and (\ref{2.3.4.10a}), the asymptotic behavior
of functions $\chi(x)$ belonging to the respective $D_{\bar{a}(\mu,w)}$ and
$D_{\bar{a}(w_{0})}$, is estimated as $\chi(x)=O(x^{1/2})$, which implies that
the functions belonging to $D_{H_{\mathfrak{e}a(\mu,w)}},\forall\mu,\forall
w\in(w_{0},\infty)$, and $D_{H_{\mathfrak{e}a(w_{0})}}$ tend to zero not
weaker than $x^{1/2}$ as $x\rightarrow0$. A comparison of this estimate with
(\ref{3.1.3}) shows that there is only one s.a. generalized Calogero
Hamiltonian with such an asymptotic behavior of the functions belonging to its
domain, namely, $\hat{H}_{2,\pm\pi/2}$. It follows that the Hamiltonian
$\hat{H}_{2,\pm\pi/2}$ with given coupling constants $g_{1}\in(-1/4,3/4)$,
$g_{2}>0$ must be identified with the whole family\ $\{\hat{H}_{\mathfrak{e}%
a}\}$ with the same $g_{1}$, $g_{2}$ , $\{\hat{H}_{\mathfrak{e}a}%
\}=\hat{H}_{2,\pm\pi/2}$, which yields the two types of generalized oscillator
representations for this Hamiltonian:%
\begin{align}
& \hat{H}_{2,\pm\pi/2}=\hat{H}_{\mathfrak{e}a(\mu,w)}=\hat{a}^{+}%
(\mu,w)\overline{\hat{a}}(\mu,w)-4\upsilon^{2}w\,\hat{I},\,\label{3.2.2.1}\\
& \,\forall\mu\in\lbrack0,\pi/2],\;\forall w\in(w_{0},\infty),\,\varkappa
\in(0,1),\nonumber%
\end{align}
and
\begin{equation}
\hat{H}_{2,\pm\pi/2}=\hat{H}_{\mathfrak{e}a(w_{0})}=\hat{a}^{+}(w_{0}%
)\overline{\hat{a}}(w_{0})+2\upsilon^{2}(1+\varkappa),\,\varkappa
\in(0,1).\label{3.2.2.2}%
\end{equation}

These representations are evident extensions of the previous respective
representations (\ref{3.2.1.1}) and (\ref{3.2.1.2}) to $\,\varkappa\in(0,1)$,
and a comment to them is an extension of the previous one: formula
(\ref{3.2.2.1}) actually presents a two-parameter family of different
generalized oscillator representations for a unique generalized Calogero
Hamiltonian $\hat{H}_{2,\pm\pi/2}$ with given coupling constants $g_{1}%
\in(-1,4,3/4)$, $g_{2}>0$, any of representations (\ref{3.2.2.1}) is not an
optimum representation because of (\ref{2.2.4.24b}), while representation
(\ref{3.2.2.2}) is an optimum one because of (\ref{2.3.4.10b}), the
ground-state energy $E_{0}^{(2)}(\pm\pi/2)$ of $\hat{H}_{2,\pm\pi/2}$, which
is a lower boundary of its spectrum, is
\begin{equation}
E_{0}^{(2)}(\pm\pi/2)=2\upsilon^{2}(1+\varkappa),\,\,\varkappa\in
(0,1),\label{3.2.2.3}%
\end{equation}
which is an extension of (\ref{3.2.1.3}) to $\varkappa\in(0,1)$, and the
normalized ground state $U_{\pm\pi/2}^{(2)}(x)$ of $\hat{H}_{2,\pm\pi/2}$ is
given by%
\begin{equation}
U_{\pm\pi/2}^{(2)}(x)=\sqrt{\frac{2\upsilon}{\Gamma(1+\varkappa)}}(\upsilon
x)^{1/2+\varkappa}\mathrm{e}^{-(\upsilon x)^{2}/2},\,\,\varkappa
\in(0,1),\label{3.2.2.4}%
\end{equation}
which is an extension of (\ref{3.2.1.4}) to $\varkappa\in(0,1)$.

\subsubsection{Semiaxis $\{g_{1}=-1/4$ $(\varkappa=0)$, $g_{2}>0\}$}

A consideration in this case, $\varkappa=0$, is completely similar to that in
the previous case of $\,\varkappa\in(0,1)$.

According to \cite{GitTyV12}, for any pair of coupling constants $g_{1}$,
$g_{2}$ lying on this semiaxis, there exists a one-parameter family
$\{\hat{H}_{3,\nu},$ $\nu\in\lbrack-\pi/2,\pi/2],-\pi/2\,\sim\pi/2\}$ of
generalized Calogero Hamiltonians $\hat{H}_{3,\nu}$ specified by s.a.
asymptotic boundary conditions at the origin (\ref{3.1.4}).

An asymptotic behavior at the origin of the functions belonging to the domains
of the Hamiltonians $\hat{H}_{\mathfrak{e}a}$ (\ref{3.1.1}) with coupling
constants $g_{1}$, $g_{2}$ lying on the same semiaxis is estimated as follows.
By definition of any operator $\hat{H}_{\mathfrak{e}a}$, its domain belongs to
or coincides with the domain of the operator $\overline{\hat{a}}%
$, $\overline{\hat{a}}$ is $\overline{\hat{a}}(\mu,w)$ or
$\overline{\hat{a}}(w_{0})$, i.e., $D_{H_{\mathfrak{e}a}}\subseteq
D_{\bar{a}}$.\ But according to (\ref{2.4.4.20a}) and (\ref{2.5.4.6a}), the
asymptotic behavior of functions $\chi(x)$ belonging to any $D_{\bar{a}}$, is
estimated as $\chi(x)=O(x^{1/2}\ln^{1/2}x)$, which implies that the functions
belonging to any $D_{H_{\mathfrak{e}a}}$ tend to zero not weaker than
$x^{1/2}\ln^{1/2}x$ as $x\rightarrow0$. A comparison of this estimate with
(\ref{3.1.3}) shows that there is only one s.a. generalized Calogero
Hamiltonian with such an asymptotic behavior of the functions belonging to its
domain, namely, $\hat{H}_{3,\pm\pi/2}$. It follows that the Hamiltonian
$\hat{H}_{3,\pm\pi/2}$ with given $g_{1}=-1/4$, $g_{2}>0$ must be identified
with the whole family\ $\{\hat{H}_{\mathfrak{e}a}\}$ with the same $g_{1}$,
$g_{2}$, $\{\hat{H}_{\mathfrak{e}a}\}=$ $\hat{H}_{3,\pm\pi/2}$, which yields
the two types of generalized oscillator representations for this Hamiltonian:%
\begin{align}
& \hat{H}_{3,\pm\pi/2}=\hat{H}_{\mathfrak{e}a(\mu,w)}=\hat{a}^{+}%
(\mu,w)\overline{\hat{a}}(\mu,w)-4\upsilon^{2}w\,\hat{I},\,\label{3.2.3.1}\\
& \,\forall\mu\in\lbrack0,\pi/2],\;\forall w\in(w_{0},\infty),\,w_{0}%
=-\frac{1}{2},\nonumber%
\end{align}
and
\begin{equation}
\hat{H}_{3,\pm\pi/2}=\hat{H}_{\mathfrak{e}a(w_{0})}=\hat{a}^{+}(w_{0}%
)\overline{\hat{a}}(w_{0})+2\upsilon^{2}.\label{3.2.3.2}%
\end{equation}

These representations are evident extensions of the previous respective
representations (\ref{3.2.2.1}) and (\ref{3.2.2.2}) to $\,\varkappa=0$, and a
comment to them is an extension of the previous one: formula (\ref{3.2.3.1})
actually presents a two-parameter family of different generalized oscillator
representations for a unique generalized Calogero Hamiltonian $\hat{H}_{3,\pm
\pi/2}$ with given coupling constants $g_{1}=-1/4$, $g_{2}>0$, any of
representations (\ref{3.2.3.1}) is not an optimum representation because of
(\ref{2.4.4.20b}), while representation (\ref{3.2.3.2}) is an optimum one
because of (\ref{2.5.4.6b}), the ground-state energy $E_{0}^{(3)}(\pm\pi/2)$
of $\hat{H}_{3,\pm\pi/2}$, which is a lower boundary of its spectrum, is
\begin{equation}
E_{0}^{(3)}(\pm\pi/2)=2\upsilon^{2},\label{3.2.3.3}%
\end{equation}
which is an extension of (\ref{3.2.2.3}) to $\varkappa=0$, and the normalized
ground state $U_{\pm\pi/2}^{(3)}(x)$ of $\hat{H}_{3,\pm\pi/2}$ is given by%
\begin{equation}
U_{\pm\pi/2}^{(3)}(x)=\sqrt{2\upsilon}(\upsilon x)^{1/2}\mathrm{e}^{-(\upsilon
x)^{2}/2},\label{3.2.3.4}%
\end{equation}
which is an extension of (\ref{3.2.2.4}) to $\varkappa=0$.

\subsection{Family $\{\hat{H}_{\mathfrak{e}b}\}$}

\subsubsection{Quadrant $\{g_{1}\geq3/4$ $(\varkappa\geq1),$%
\ $g_{2}>0\}$}

For any pair of coupling constants $g_{1}$, $g_{2}$ lying in this quadrant, we
have the identities $\overline{\hat{b}}(\mu,w)=\hat{a}^{+}(\mu,w)$, $\hat
{b}^{+}(\mu,w)=\overline{\hat{a}}(\mu,w)$, see the respective (\ref{2.2.4.25})
and (\ref{2.2.4.23}), and $\overline{\hat{b}}(w_{0})=\hat{a}^{+}(w_{0})$,
$\hat{b}^{+}(w_{0})=\overline{\hat{a}}(w_{0})$, see the respective
(\ref{2.3.4.11}) and (\ref{2.3.4.10}). These identities and the result of
subsubsec. 3.2.1 provide the identities $\{\hat{H}_{\mathfrak{e}%
b}\}=\{\hat{H}_{\mathfrak{e}a}\}=\hat{H}_{1}$ and yield %
equivalent forms of generalized oscillator representations for $\hat{H}_{1}$
presented in\ subsubsec. 3.2.1:
\begin{align}
\hat{H}_{1}  & =\hat{H}_{\mathfrak{e}b(\mu,w)}=\overline{\hat{b}}(\mu
,w)\hat{b}^{+}(\mu,w)-4\upsilon^{2}w\,\hat{I},\,\nonumber\\
\forall\mu & \in[0,\pi/2],\;\forall w\in(w_{0},\infty),\,\,\varkappa
\geq1,\nonumber
\end{align}
which is another, equivalent, form of the known two-parameter family of
nonoptimum generalized oscillator representations (\ref{3.2.1.1}) for
$\hat{H}_{1}$, and
\begin{equation*}
\hat{H}_{1}=\hat{H}_{\mathfrak{e}b(w_{0})}=\overline{\hat{b}}(w_{0})\hat
{b}^{+}(w_{0})+2\upsilon^{2}(1+\varkappa),\,\varkappa\geq1.
\end{equation*}
which is another, equivalent, form of the known optimum generalized oscillator
representation (\ref{3.2.1.2}) for $\hat{H}_{1}$. Of course, the comment
following (\ref{3.2.1.2}) holds including formulas (\ref{3.2.1.3}) and
(\ref{3.2.1.4}).

When proceeding to $\hat{H}_{\mathfrak{e}b}$ with $g_{1},g_{2}$ lying in the
open strip $\{-1/4<g_{1}<3/4,g_{2}>0\}$, we have, in view of (\ref{2.2.4.23}),
(\ref{2.2.4.25}) and (\ref{2.3.4.10}), (\ref{2.3.4.11}), to distinguish the
cases $\mu=\pi/2$, $w>w_{0}\,\,\ $and $w=w_{0}$ from the case $\mu\in
\lbrack0,\pi/2)$, $w>w_{0}$.$\,$

\subsubsection{Strip $\{-1/4<g_{1}<3/4$ ($0<\varkappa<1$), $g_2>0\}$,
cases $\mu=\pi/2$, $w>w_0=-\frac{1}{2}(1+\varkappa)$ and $w=w_{0}$}

A consideration in this subsubsection is completely similar to that in the
previous subsubsection.

For any pair of coupling constants $g_{1}$, $g_{2}$ lying in this strip and
these values of the parameters $\mu,w$, we have the identities $\overline
{\hat{b}}(\pi/2,w)=\hat{a}^{+}(\pi/2,w)$, $\hat{b}^{+}(\pi/2,w)=\overline
{\hat{a}}(\pi/2,w)$, see the respective (\ref{2.2.4.25}) and
(\ref{2.2.4.23}),
and $\overline{\hat{b}}(w_{0})=\hat{a}^{+}(w_{0})$, $\hat{b}^{+}%
(w_{0})=\overline{\hat{a}}(w_{0})$, see the respective (\ref{2.3.4.11}) and
(\ref{2.3.4.10}). These identities and the result of subsubsec. 3.2.2 provide
the identities $\hat{H}_{\mathfrak{e}b(\pi/2,w)}=\hat{H}_{\mathfrak{e}%
a(\pi/2,w)}=\hat{H}_{2,\pm\pi/2}=\hat{H}_{\mathfrak{e}a(w_{0})}%
=\hat{H}_{\mathfrak{e}b(w_{0})}$ and yield equivalent forms of a part of
generalized oscillator representations for $\hat{H}_{2,\pm\pi/2}$ presented
in\ subsubsec. 3.2.2:
\begin{align}
& \hat{H}_{2,\pm\pi/2}=\hat{H}_{\mathfrak{e}b(\pi/2,w)}=\overline{\hat{b}}%
(\pi/2,w)\hat{b}^{+}(\pi/2,w)-4\upsilon^{2}w\,\hat{I},\,\nonumber\\
& \,\forall w\in(w_{0},\infty),\;\varkappa\in(0,1),\nonumber
\end{align}
which is another, equivalent, form of the one-parameter family of nonoptimum
generalized oscillator representations for $\hat{H}_{2,\pm\pi/2\text{ }}$ that
is a restriction of the known two-parameter family of representations
(\ref{3.2.2.1}) to $\mu=\pi/2$\ , and
\begin{equation*}
\hat{H}_{2,\pm\pi/2}=\hat{H}_{\mathfrak{e}b(w_{0})}=\overline{\hat{b}}%
(w_{0})\hat{b}^{+}(w_{0})+2\upsilon^{2}(1+\varkappa),\,\,\varkappa
\in(0,1),
\end{equation*}
which is another, equivalent, form of the known optimum generalized
oscillator representation (\ref{3.2.2.2}) for $\hat{H}_{2,\pm\pi/2}$. Of
course, the comment following (\ref{3.2.2.2}) holds including formulas
(\ref{3.2.2.3}) and (\ref{3.2.2.4}).

\subsubsection{Strip $\{-1/4<g_1<3/4$ ($0<\varkappa<1$), $g_2>0\}$,
case $\mu\in\lbrack0,\pi/2)$, $w>w_{0}=-\frac{1}{2}(1+\varkappa)$}

In this case, we deal only with the operators $\hat{H}_{\mathfrak{e}b(\mu,w)}$
and have to find the asymptotic behavior at the origin of functions belonging
to their domains.

We begin with functions $\phi(\mu,w;x)$ (\ref{2.2.1.2}) with indicated values
of parameters, namely, with representing their asymptotic behavior at the
origin given in (\ref{2.2.1.5}) in a new form:
\begin{align}
&  \phi(\mu,w;x)=c(\mu,w)[(\upsilon
x)^{1/2+\varkappa}\sin\theta(\mu,w)+(\upsilon
x)^{1/2-\varkappa}\cos\theta(\mu,w)]+O(x^{5/2-\varkappa}),\,x\rightarrow
0,\nonumber\\
&  \tan\theta(\mu,w)=\tan\mu-\frac{\Gamma(1-\varkappa)}{\Gamma(1+\varkappa
)}\frac{\Gamma\left(  \frac{1}{2}(1+\varkappa)+w\right)  }{\Gamma\left(
\frac{1}{2}(1-\varkappa)+w\right)  }\mathrm{,}\ \tan\mu\geq0\,,\nonumber\\
&\,c(\mu,w)   =\frac{\cos\mu}{\cos\theta(\mu,w)},\,\theta(\mu,w)\in\left(
-\frac{\pi}{2},\frac{\pi}{2}\right)  ,\nonumber\\
&  \mu\in\lbrack0,\frac{\pi}{2}),\text{\textrm{\ }}w\in(w_{0}\,,\infty
)\,,\;\varkappa\in(0,1).\nonumber
\end{align}

By definition of the operator $\hat{H}_{\mathfrak{e}b(\mu,w)}$, its domain
$D_{H_{\mathfrak{e}b(\mu,)}}$ consists of functions $\chi(x)\in D_{b^{+}%
(\mu,w)}$ such that $\hat{b}^{+}(\mu,w)\chi(x)=\,\check{a}(\mu,w)\chi
(x)=\eta(x)\in D_{\overline{b}(\mu,w)}$. The first condition implies that
$\chi(x)$ allows representation (\ref{2.2.4.13}) with $x_{0}=0$ and, in
general, $D\neq0$, see (\ref{2.2.4.14}), while the second condition implies
that $\eta(x)=O(x^{1/2}),\;x\rightarrow0$, see (\ref{2.2.4.26a}). Estimating
the integral term in (\ref{2.2.4.13}) with such $\eta(x)$, we obtain that the
asymptotic behavior of functions $\chi(x)\in$ $D_{H_{\mathfrak{e}b(\mu,w)}}$,
$\mu\in\lbrack0,\pi/2),\,w\in(w_{0}\,,\infty),\,\varkappa\in(0,1)$, at the
origin is given by%
\begin{equation}
\chi(x)=c[(\upsilon x)^{1/2+\varkappa}\sin\theta(\mu,w)+(\upsilon
x)^{1/2-\varkappa}\cos\theta(\mu,w)]+O(x^{3/2}),\,x\rightarrow
0,\label{3.3.3.2}%
\end{equation}
where $c=D\,c(\mu,w)$ is arbitrary. A comparison of (\ref{3.3.3.2}) and
(\ref{3.1.3}) naturally identifies the parameter\ $\nu$ in (\ref{3.1.3})
with the angle $\theta(\mu,w)$ in (\ref{3.3.3.2}) and establishes that the
Hamiltonian $\hat{H}_{2,\nu}$ with
given $g_{1}\in(-1/4,3/4)$, $g_{2}>0$ and $\nu\in\left(  -\frac{\pi}%
{2},\frac{\pi}{2}\right)  $ must be identified with all those
$\hat{H}_{b(\mu ,w)}$ with the same $g_{1},g_{2}$, for which the parameters
$\mu$ and $w$ are related by$\ \theta(\mu,w)=\nu$, i.e., \
$\{\hat{H}_{b(\mu,w)},\theta (\mu,w)=\nu\}=\hat{H}_{2,\nu}$. It is
convenient to take $\mu$ as an independent parameter, then $w$ is
determined from the relation $\tan
\theta(\mu,w)=\tan\nu$, or%

\begin{align}
& \frac{\Gamma(1-\varkappa)}{\Gamma(1+\varkappa)}\frac{\Gamma\left(
\frac{1}{2}(1+\varkappa)+w\right)  }{\Gamma\left(  \frac{1}{2}(1-\varkappa
)+w\right)  }=\tan\mu-\tan\nu\label{3.3.3.3}\\
& \nu\in(-\pi/2,\pi/2),\mu\in\lbrack0,\frac{\pi}{2}),\text{\textrm{\ }%
}\varkappa\in(0,1),\nonumber%
\end{align}
considered as an equation with respect to $w$ under the additional condition
$w>w_{0}$. It can be shown that the l.h.s. of eq. (\ref{3.3.3.3}) as a
function of $w$ $>w_{0}$ is a continuous monotonically increasing
function\footnote{This actually was shown in \cite{GitTyV12}.} ranging from
$-\infty$ to $\infty$ as $w$ ranges from $w_{0}+0$ to $\infty$. It follows
that eq. (\ref{3.3.3.3}) always has a unique solution $w=w(\mu,\nu)$ and
$w(\mu,\nu)$ as a function of $\mu$ increases monotonically from
$w(0,\nu)>-\frac{1}{2}(1+\varkappa)$ to $\infty$ as $\mu$ ranges from $0$ to
$\frac{\pi}{2}-0$. The previous result then can be written as
$\{\hat{H}_{b(\mu,w(\mu,\nu))}\}=\hat{H}_{2,\nu}$, which yields the
one-parameter family of generalized oscillator representations for the
Hamiltonian $\hat{H}_{2,\nu}$ with given coupling constants $g_{1}%
\in(-1/4,3/4)$, $g_{2}>0$ :%
\begin{align}
\hat{H}_{2,\nu}  & =\hat{H}_{\mathfrak{e}b(\mu,w(\mu,\nu))}=\overline{\hat{b}%
}(\mu,w(\mu,\nu))\hat{b}^{+}(\mu,w(\mu,\nu))-4\upsilon^{2}w\,(\mu
,\nu)\hat{I},\,\label{3.3.3.4}\\
\nu & \in\left(  -\frac{\pi}{2},\frac{\pi}{2}\right)  ,\forall\mu\in
\lbrack0,\pi/2),\varkappa\in(0,1),\nonumber%
\end{align}
the parameter is $\mu$, and $w=w(\mu,\nu)$ is a solution of eq.
(\ref{3.3.3.3}).

According to (\ref{2.2.4.17b}), any of representations (\ref{3.3.3.4}) with
$\mu\in(0,\pi/2)$ is nonoptimum, we can only say that $\hat{H}_{2,\nu}$ is
bounded from below by $-4\upsilon^{2}w\,(0,\nu)$, while representation with
$\mu=0$ is an optimum one, the ground-state energy $E_{0}^{(2)}(\nu)$ of
$\hat{H}_{2,\nu}$, which is a lower boundary of its spectrum, is
\begin{equation*}
E_{0}^{(2)}(\nu)=-4\upsilon^{2}w\,(0,\nu),\,\varkappa\in(0,1),
\end{equation*}
and the normalized ground state $U_{\nu}^{(2)}(x)$ of $\hat{H}_{2,\nu}$ is
given by%
\begin{equation*}
U_{\nu}^{(2)}(x)=Q_{0}(\nu)\mathrm{e}^{-(\upsilon x)^{2}/2}(\upsilon
x)^{1/2+\varkappa\,}\Psi(\frac{1}{2}(1+\varkappa)+w\,(0,\nu),1+\varkappa
;(\upsilon x)^{2}),\,\varkappa\in(0,1),\label{3.3.3.6}
\end{equation*}
$Q_{0}(\nu)$ is a normalization factor.

The ground-state energy $E_{0}^{(2)}(\nu)$ is a unique solution of the
equation
\begin{equation}
\frac{\Gamma(1-\varkappa)}{\Gamma(1+\varkappa)}\frac{\Gamma\left(  \frac{1}%
{2}(1+\varkappa)-E/4\upsilon^{2}\right)  }{\Gamma\left(  \frac{1}%
{2}(1-\varkappa)-E/4\upsilon^{2}\right)  }=-\tan\nu,\,\nu\in(-\pi
/2,\pi/2),\label{3.3.3.7}%
\end{equation}
considered as an equation for $E$ under the additional condition
$E<2\upsilon^{2}(1+\varkappa).$ We are unable to present an explicit
expression for $E_{0}(\nu)$ except $E_{0}(0)=2\upsilon^{2}(1-\varkappa)$, but
we can assert that $E_{0}^{(2)}(\nu)$ monotonically increases from $-\infty$
to $2\upsilon^{2}(1+\varkappa)-0$ as $\nu$ ranges from $-\pi/2+0$ to $\pi
/2-0$. We note that according to \cite{GitTyV12}, eq. (\ref{3.3.3.7})
determines the spectrum of $\hat{H}_{2,\nu}$, which is a discrete
one,\textrm{\ spec}$\hat{H}_{2,\nu}=$ $\{E_{n}(\nu)\}$, $n=0,1,2,...$,
$E_{n+1}(\nu)>E_{n}(\nu)$, and the condition $E<2\upsilon^{2}(1+\varkappa)$
separates out the minimum eigenvalue, the ground-state energy $E_{0}^{(2)}%
(\nu)$.

It remains to consider $\hat{H}_{\mathfrak{e}b}$ with $g_{1},g_{2}$ lying on
the open semiaxis $\{g_{1}=-1/4,g_{2}>0\}$. In view of
(\ref{2.4.4.19}), (\ref{2.4.4.21}) and (\ref{2.5.4.6}), (\ref{2.5.4.7}), we
have to distinguish the cases $\mu=\pi/2$, $w>w_{0}\,=-\frac{1}{2}\ $and
$w=w_{0}=-\frac{1}{2}$ from the case $\mu\in\lbrack0,\pi/2)$, $w>w_{0}$.$\,$

\subsubsection{Semiaxis $\{g_{1}=-1/4$ $(\varkappa=0)$, $g_{2}>0\},$
cases $\mu=\pi/2$, $w>w_{0}=-\frac{1}{2}$ and $w=w_{0}$}

A consideration in this subsubsection is completely similar to that in
subsubsec. 3.3.2.

For any pair of coupling constants $g_{1}$, $g_{2}$ lying on this semiaxis and
these values of the parameters $\mu,w$, we have the identities $\overline
{\hat{b}}(\pi/2,w)=\hat{a}^{+}(\pi/2,w)$, $\hat{b}^{+}(\pi/2,w)=\overline
{\hat{a}}(\pi/2,w)$, see the respective (\ref{2.4.4.21}) and (\ref{2.4.4.19}),
and $\overline{\hat{b}}(w_{0})=\hat{a}^{+}(w_{0})$, $\hat{b}^{+}%
(w_{0})=\overline{\hat{a}}(w_{0})$, see the respective (\ref{2.5.4.7}) and
(\ref{2.5.4.6}). These identities and the result of subsubsec. 3.2.3 provide
the identities $\hat{H}_{\mathfrak{e}b(\pi/2,w)}=\hat{H}_{\mathfrak{e}%
a(\pi/2,w)}=\hat{H}_{3,\pm\pi/2}=\hat{H}_{\mathfrak{e}a(w_{0})}%
=\hat{H}_{\mathfrak{e}b(w_{0})}$ and yield equivalent forms of a part of
generalized oscillator representations for $\hat{H}_{3,\pm\pi/2}$ presented
in\ subsubsec. 3.2.3:
\begin{align}
& \hat{H}_{3,\pm\pi/2}=\hat{H}_{\mathfrak{e}b(\pi/2,w)}=\overline{\hat{b}}%
(\pi/2,w)\hat{b}^{+}(\pi/2,w)-4\upsilon^{2}w\,\hat{I},\,\nonumber\\
& \,\forall w\in(w_{0},\infty),\nonumber
\end{align}
which is another, equivalent, form of the one-parameter family of nonoptimum
generalized oscillator representations for $\hat{H}_{3,\pm\pi/2\text{ }}$, the
parameter is $w$, that is a restriction of the known two-parameter family of
representations (\ref{3.2.3.1}) to $\mu=\pi/2$\ , and
\begin{equation*}
\hat{H}_{3,\pm\pi/2}=\hat{H}_{\mathfrak{e}b(w_{0})}=\overline{\hat{b}}%
(w_{0})\hat{b}^{+}(w_{0})+2\upsilon^{2},\,
\end{equation*}
which is another, equivalent, form of the known optimum generalized oscillator
representation (\ref{3.2.3.2}) for $\hat{H}_{3,\pm\pi/2}$. Of course, the
comment following (\ref{3.2.3.2}) holds including formulas (\ref{3.2.3.3}) and
(\ref{3.2.3.4}).

\subsubsection{Semiaxis $\{g_{1}=-1/4$ $(\varkappa=0),\;$ $g_{2}>0\},$ case $\mu\in\lbrack0,\pi/2)$, $w>w_{0}=-\frac{1}{2}$}

A consideration in this subsubsection is completely similar to that in
subsubsec. 3.3.3.

We begin with functions $\phi(\mu,w;x)$ (\ref{2.4.1.1}) with indicated values
of parameters, namely, with representing their asymptotic behavior at the
origin given in (\ref{2.4.1.3}) in a new form:%
\begin{align}
& \phi(\mu,w;x)=c(\mu,w)[\,(\upsilon x)^{1/2}\sin\theta(\mu,w)+2\,(\upsilon
x)^{1/2}\ln(\upsilon x)\cos\theta(\mu,w)]+O(x^{5/2}\ln x),\,x\rightarrow
0,\nonumber\\
& \tan\ \theta(\mu,w)=\psi(\frac{1}{2}+w)-2\psi(1)-\tan\mu,\mathrm{\;}\tan
\mu\geq0,\nonumber\\
& c(\mu,w)=-\frac{\cos\mu}{\cos\theta(\mu,w)},\,\theta(\mu,w)\in\left(
-\frac{\pi
}{2},\frac{\pi}{2}\right)  ,\nonumber\\
& \mu\in\lbrack0,\frac{\pi}{2}),\text{\textrm{\ }}w\in(w_{0}\,,\infty
)\,,\,w_{0}\,=-\frac{1}{2},\,\varkappa=0.\nonumber
\end{align}

By definition of the operator $\hat{H}_{\mathfrak{e}b(\mu,w)}$, its domain
$D_{H_{\mathfrak{e}b(\mu,)}}$ consists of functions $\chi(x)\in D_{b^{+}%
(\mu,w)}$ such that $\hat{b}^{+}(\mu,w)\chi(x)=\,\check{a}(\mu,w)\chi
(x)=\eta(x)\in D_{\overline{b}(\mu,w)}$. The first condition implies that
$\chi(x)$ allows representation (\ref{2.4.4.9}) with $x_{0}=0$ and, in
general, $D\neq0$, see (\ref{2.4.4.10}), while the second condition implies
that $\eta(x)=O(x^{1/2}),\;x\rightarrow0$, see (\ref{2.4.4.22a}). Estimating
the integral term in (\ref{2.4.4.9}) with such $\eta(x)$, we obtain that the
asymptotic behavior of functions $\chi(x)\in$ $D_{H_{\mathfrak{e}b(\mu,w)}}$,
$\mu\in\lbrack0,\pi/2),\,w\in(w_{0}\,,\infty),\,w_{0}\,=-\frac{1}%
{2},\,\varkappa=0$, at the origin is given by%
\begin{equation}
\chi(x)=c[(\upsilon x)^{1/2}\sin\theta(\mu,w)+2\,(\upsilon x)^{1/2}%
\ln(\upsilon x)\cos\theta(\mu,w)]+O(x^{3/2}),\,x\rightarrow0,\label{3.3.5.2}%
\end{equation}
where $c=D\,c(\mu,w)$ is arbitrary. A comparison of (\ref{3.3.5.2}) and
(\ref{3.1.4}) naturally identifies the parameter $\nu$ in (\ref{3.1.4})
with the angle $\theta(\mu,w)\in\left( -\frac{\pi}{2},\frac{\pi}{2}\right)
$ in (\ref{3.3.5.2}) and establishes that the Hamiltonian $\hat{H}_{3,\nu}$
with given $g_{1}=-1/4,g_{2}>0$ and $\nu \in\left(
-\frac{\pi}{2},\frac{\pi}{2}\right)  $ must be identified with all those
$\hat{H}_{b(\mu,w)}$ with the same $g_{1},g_{2}$, for which the parameters
$\mu$ and $w$ are related by$\ \theta(\mu,w)=\nu$, i.e., \
$\{\hat{H}_{b(\mu,w)},\;\theta(\mu,w)=\nu\}=\hat{H}_{3,\nu}$. It is
convenient to take $\mu$ as an independent parameter, then $w$ is determined
from the relation $\tan\theta(\mu,w)=\tan\nu$, or%
\begin{equation}
\psi(\frac{1}{2}+w)=\tan\nu+\tan\mu+2\psi(1),\;
\nu\in(-\pi/2,\pi/2),\mu\in\lbrack0,\frac{\pi}{2}),\label{3.3.5.3}%
\end{equation}
considered as an equation with respect to $w$ under the additional condition
$w>\,w_{0}\,$. The function $\psi(\frac{1}{2}+w)$ is a continuous
monotonically increasing function of $w$ ranging from $-\infty$ to $\infty$
as $w$ ranges from $w_{0}+0$ to $\infty$. It follows that eq.
(\ref{3.3.5.3}) always has a unique solution $w=w(\mu,\nu)$ and
$w(\mu,\nu)$ as a function of $\mu$ increases monotonically from
$w(0,\nu)>-\frac{1}{2}$ to $\infty$ as $\mu$ ranges from $0$ to
$\frac{\pi}{2}-0$. The previous result then can be written as
$\{\hat{H}_{b(\mu,w(\mu,\nu))}\}=\hat{H}_{3,\nu}$, which yields the
one-parameter family of generalized oscillator representations for the
Hamiltonian $\hat{H}_{3,\nu}$ with given coupling constants $g_{1}=-1/4$,
$g_{2}>0$ :%
\begin{equation}
\hat{H}_{3,\nu}=\hat{H}_{\mathfrak{e}b(\mu,w(\mu,\nu))}=\overline{\hat{b}%
}(\mu,w(\mu,\nu))\hat{b}^{+}(\mu,w(\mu,\nu))-4\upsilon^{2}w\,(\mu
,\nu)\hat{I},\;
\nu\in\left(-\frac{\pi}{2},\frac{\pi}{2}\right),\forall\mu\in
[0,\pi/2),\label{3.3.5.4}%
\end{equation}
the parameter is $\mu$, and $w=w(\mu,\nu)$ is a solution of eq.
(\ref{3.3.5.3}).

According to (\ref{2.4.4.13a}), any of representations (\ref{3.3.5.4}) with
$\mu\in(0,\pi/2)$ is nonoptimum, we can only say that $\hat{H}_{3,\nu}$ is
bounded from below by $-4\upsilon^{2}w\,(0,\nu)$, while representation with
$\mu=0$ is an optimum one, the ground-state energy $E_{0}^{(3)}(\nu)$ of
$\hat{H}_{3,\nu}$, which is a lower boundary of its spectrum, is
\begin{equation*}
E_{0}^{(3)}(\nu)=-4\upsilon^{2}w\,(0,\nu),
\end{equation*}
and the normalized ground state $U_{\nu}(x)$ of $\hat{H}_{3,\nu}$ is given by%
\begin{equation*}
U_{\nu}^{(3)}(x)=Q_{0}(\nu)\mathrm{e}^{-(\upsilon x)^{2}/2}(\upsilon
x)^{1/2\,}\Psi(\frac{1}{2}+w\,(0,\nu),1;(\upsilon x)^{2}),
\end{equation*}
$Q_{0}(\nu)$ is a normalization factor.

The ground-state energy $E_{0}^{(3)}(\nu)$ is a unique solution of the
equation
\begin{equation}
\psi(\frac{1}{2}-\frac{E}{4\upsilon^{2}})=\tan\nu+2\psi(1),\,\nu\in(-\pi
/2,\pi/2),\label{3.3.5.7}%
\end{equation}
considered as an equation for $E$ under the additional condition
$E<2\upsilon^{2}$. We are unable to present an explicit expression for
$E_{0}^{(3)}(\nu)$, but we can assert that $E_{0}^{(3)}(\nu)$ monotonically
decreases from $2\upsilon^{2}-0$ to $-\infty$ as $\nu$ ranges from $-\pi/2+0$
to $\pi/2-0$.

We note that according to \cite{GitTyV12}, eq. (\ref{3.3.5.7}) determines the
spectrum of $\hat{H}_{3,\nu}$, which is a discrete one,\textrm{\ spec}%
$\hat{H}_{3,\nu}=$ $\{E_{n}(\nu)\}$, $n=0,1,2,...$, $E_{n+1}(\nu)>E_{n}(\nu)$,
and the condition $E<2\upsilon^{2}$ separates out the minimum eigenvalue, the
ground-state energy $E_{0}^{(3)}(\nu)$.\\

{\bf Acknowledgement}

I.T. thanks RFBR, grant 11-01-00830, and B.V. thanks RFBR, grant
11-02-00685, for partial support.

\end{document}